\documentclass[journal,english]{IEEEtran}

\usepackage{epsfig}
\usepackage{times}
\usepackage{float}
\usepackage{afterpage}
\usepackage{amsmath,amstext,amssymb,amsthm,mathrsfs}
\usepackage{bm}
\usepackage{latexsym}
\usepackage{color}
\usepackage{pstricks}
\usepackage{graphicx}
\usepackage[center]{caption}
\usepackage{caption}
\usepackage{subcaption}
\usepackage{booktabs}
\usepackage{multicol}
\usepackage[T1]{fontenc}
\usepackage{cite}
\usepackage{hyperref}
\usepackage{aecompl}
\usepackage{enumitem}
\usepackage{comment}

\allowdisplaybreaks

\long\def\comment#1{}

\newcommand{\dv}{{\mathbf d}}

\newcommand{\pv}{{\mathbf p}}

\newcommand{\uv}{{\mathbf u}}

\newcommand{\Fc}{{\mathcal F}}
\newcommand{\Gc}{{\mathcal G}}
\newcommand{\Hc}{{\mathcal H}}

\newcommand{\Lc}{{\mathcal L}}

\newcommand{\Qc}{{\mathcal Q}}

\newcommand{\Sc}{{\mathcal S}}

\newcommand{\Vc}{{\mathcal V}}

\newcommand{\Bsf}{{\mathsf B}}

\newcommand{\Ksf}{{\mathsf K}}

\newcommand{\Msf}{{\mathsf M}}
\newcommand{\Nsf}{{\mathsf N}}

\newcommand{\Rsf}{{\mathsf R}}

\newcommand{\Usf}{{\mathsf U}}

\renewcommand{\arg}{{\hbox{arg}}}

\newtheorem{thm}{Theorem}%[section]
\newtheorem{cor}{Corollary}
\newtheorem{lem}{Lemma}

\newtheorem{rem}{Remark}

\providecommand{\definitionname}{Definition}

\newcommand{\dt}[1]{{\blue #1}}%DT: 

\begin{document}

\title{On the Fundamental Limits of Device-to-Device Private Caching under Uncoded Cache Placement and User Collusion} 

\author{
Kai~Wan,~\IEEEmembership{Member,~IEEE,} 
Hua~Sun,~\IEEEmembership{Member,~IEEE,}
Mingyue~Ji,~\IEEEmembership{Member,~IEEE,}  
Daniela Tuninetti,~\IEEEmembership{Fellow,~IEEE,}
and~Giuseppe Caire,~\IEEEmembership{Fellow,~IEEE}
\thanks{
The results of this paper were presented in parts at 
    the 2020 IEEE International Conference on Communications, Dublin, Ireland,~\cite{wan2019d2dprivate}, 
and the 2020 IEEE  International Symposium on Information Theory, Los Angeles, California, USA,~\cite{wan2020ISITD2D}.}
\thanks{
K.~Wan and G.~Caire are with the Electrical Engineering and Computer Science Department, Technische Universit\"at Berlin, 10587 Berlin, Germany (e-mail:  kai.wan@tu-berlin.de; caire@tu-berlin.de). The work of K.~Wan and G.~Caire was partially funded by the European Research Council  under the ERC Advanced Grant N. 789190, CARENET.}
\thanks{
H.~Sun is with the Department of Electrical Engineering, University of North Texas, Denton, TX 76203 (email: hua.sun@unt.edu). The work of H.~Sun is supported in part by funding from NSF grants CCF-2007108 and CCF-2045656.
}
\thanks{
M.~Ji is with the Electrical and Computer Engineering Department, University of Utah, Salt Lake City, UT 84112, USA (e-mail: mingyue.ji@utah.edu). The work of M.~Ji was supported in part by NSF Awards 1817154 and 1824558.}
\thanks{
D.~Tuninetti is with the Electrical and Computer Engineering Department, University of Illinois Chicago, Chicago, IL 60607, USA (e-mail: danielat@uic.edu). The work of D.~Tuninetti was supported in part by NSF Award 1910309.}
}
\maketitle
%\IEEEpeerreviewmaketitle{}

\begin{abstract}
In the coded caching problem, as originally formulated by Maddah-Ali and Niesen, a server communicates via a noiseless shared broadcast link to multiple users that have local storage capability. In order for a user to decode its demanded file from the coded multicast transmission, the demands of all the users must be globally known, which may violate the privacy of the users. To overcome this privacy problem, Wan and Caire recently proposed several schemes that attain coded multicasting gain while simultaneously guarantee information theoretic privacy of the users' demands.
 In Device-to-Device (D2D) networks, the demand privacy problem is further exacerbated by the fact that each user is also a transmitter, which appears to be needing the knowledge of the files demanded by the remaining users in order to form its coded multicast transmission. This paper shows how to solve this seemingly infeasible problem.   The main contribution of this paper is the development of new achievable and converse bounds for D2D coded caching that are to within a constant factor of one another when privacy of the users' demands must be guaranteed even in the presence of colluding users (i.e., when some users share cached contents and demanded file indices). %In particualr:
First, a D2D private caching scheme is proposed, whose key feature is the addition of virtual users in the system in order to ``hide'' the demands of the real users. By comparing the achievable D2D private load with an existing converse bound for the shared-link model without demand privacy constraint, the proposed scheme is shown to be order optimal, except for the very low memory size regime with more files than users.
Second, in order to shed light into the open parameter regime,  %for the D2D private caching problem with two users, 
a new achievable scheme and a new converse bound under the constraint of uncoded cache placement (i.e., when each user stores directly a subset of the bits of the library) are developed for the case of two users, and shown to be to within a constant factor of one another for all system parameters. 
%\dt{!!!DT's note: removed one sentence as there is another work for K=N=2!!!} %To the best of our  knowledge, this is the first converse bound for caching problems that genuinely accounts for the demand privacy constraint and is the key theoretical novelty of this work.
%
Finally, the two-user converse bound is extended to any number of users by a cut-set type argument. With this new converse bound, the virtual users scheme is shown to be order optimal in all parameter regimes under the constraint of uncoded cache placement and user collusion.
\end{abstract}

\section{Introduction}
\label{sec:intro}

Internet data traffic has grown dramatically in the last decade because of on-demand video streaming.   The users' demands concentrate on a relatively limited number of files (e.g., latest films and shows) and that the price of memory components in the devices is usually  
significantly less than  the price of bandwidth. On the above observation,   caching becomes an efficient and promising technique for future communication systems~\cite{5gcaching}, which leverages the device memory to store data so that future demands can be served faster. 

Coded caching was originally proposed by Maddah-Ali and Niesen (MAN) for shared-link networks~\cite{dvbt2fundamental}. In the MAN model, a server has access to a library of $\Nsf$ equal-length files and is connected to $\Ksf$ users through an error-free broadcast link. Each user can store up to $\Msf$ files in its cache. A caching scheme includes {\it placement} and  {\it delivery} phases that are designed so as to minimize the %worst-case 
{\it load} (i.e., the number of files sent on the shared link that suffices to satisfy every possible demand vector). 
In the original MAN model, no constraint is imposed in order to limit the amount of information that the delivery phase leaks to a user about the demands of the remaining users. Such a privacy constraint is critical in modern broadcast services, such as peer-to-peer networks, and is the focus of this paper. %The MAN model has been extended to include such a `privacy' constraint, as well as to include peer-to-peer communications among users.

In order to appreciate the main contributions of our work, in the next sub-section we briefly review the various models of coded caching  studied in the literature, which will lead to the new problem formulation in this paper.

\subsection{Brief review of coded caching models} 
\label{sub:Brief Description}
Table~\ref{tab:summaryofresults} shows relevant known results and new results for various coded caching models. 
The complete memory-load tradeoff is obtained as the lower convex envelope of the listed points.
These results are valid for any system parameters $(\Nsf,\Ksf)$; other results that may lead to better tradeoffs but only apply to limited parameter regimes are not reported for sake of space.

\begin{table*}[ht]
\caption{Achievable %worst-case 
loads for various coded caching models. Notation: $\dv_{\setminus\{k\}}$ denotes the vector obtained from the demand vector $\dv$ by removing the $k$-th  element, and $N_e\left(\dv_{\setminus\{k\}}\right)$ gives the number of distinct elements in $\dv_{\setminus\{k\}}$. }
\label{tab:summaryofresults}
\centering

\begin{align*}
\begin{array}{|c|c|c|}
\hline
(\Msf,\Rsf)   &\text{No Privacy} & \text{With Privacy} \\
\hline
\text{Shared-link}  & 
%\Rsf^\text{(sl-p)}\left(\Ksf,\Nsf,t; \dv \right) := 
\left( t\frac{\Nsf}{\Ksf}, \frac{\binom{\Ksf}{t+1} - \binom{\Ksf-\min(\Nsf,\Ksf)}{t+1}}{\binom{\Ksf}{t}} \right) &
%\Rsf^\text{(sl-p)}\left(\Ksf\Nsf,\Nsf,t; \widetilde{\dv} \right) = 
\left( t\frac{1}{\Ksf}, \frac{\binom{\Ksf\Nsf}{t+1} - \binom{\Ksf\Nsf-\Nsf}{t+1}}{\binom{\Ksf\Nsf}{t}} \right) \\
  &
  t\in[0:\Ksf], \ \text{from~\cite{exactrateuncoded}} & 
  t\in[\Ksf\Nsf], \ \text{from~\cite{wan2019privatecaching,kamath2019demandpri}} \\ 
\hline
\text{D2D} & 
%\frac{1}{\Ksf}\sum_{k\in[\Ksf]} \Rsf^\text{(sl-p)}\left(\Ksf-1,\Nsf,t-1; \dv\backslash\{d_k\} \right) = 
\left( t\frac{\Nsf}{\Ksf}, \underset{\dv  \in [\Nsf]^{\Ksf}}{\max } \frac{\binom{\Ksf-1}{t} - 
 \frac{1}{\Ksf}\sum_{k\in[\Ksf]} \binom{\Ksf-1-N_e\left(\dv_{\setminus\{k\}}\right)}{t}}{\binom{\Ksf-1}{t-1}} \right) &
\left( \frac{\Nsf+t-1}{\Ksf}, \frac{\binom{\Nsf(\Ksf-1)}{t} - \binom{\Nsf(\Ksf-1)-\Nsf}{t}}{\binom{\Nsf(\Ksf-1)}{t-1}} \right) \\
  &
  t\in[\Ksf], \ \text{from~\cite{Yapard2djournal}} & 
  t\in[\Nsf(\Ksf-1)+1], \ \text{Scheme~A in this paper} \\
\hline
\end{array}
\end{align*}

\end{table*}

\subsubsection{Shared-link networks without privacy constraints}\label{sec:intro:sl-p}
In the MAN placement phase, letting $t=\Ksf\Msf/\Nsf \in [0:\Ksf]$ represent the number of times a file can be copied in the network's aggregate memory (excluding the server), each file is partitioned into $\binom{\Ksf}{t}$ equal-length subfiles, each of which is cached by a different $t$-subset of users. In the MAN delivery phase, each user demands one file. According to the users' demands, the server sends $\binom{\Ksf}{t+1}$ {\it MAN multicast messages}, each of which has the size of a subfile and is useful to $t+1$ users simultaneously.  The load of the MAN coded caching scheme is thus $\Rsf = \frac{\binom{\Ksf}{t+1}}{\binom{\Ksf}{t}} = \frac{\Ksf-t}{t+1}$.\footnote{\label{foot:metadata in MAN}In the MAN caching scheme, in order to allow each user to decode its demanded file, the composition of each coded multicast message sent by the server must be broadcasted along with the multicast message itself. This is akin to the ``header'' in linear network coding, that defines the structure of the linear combination of the files to enable decoding. Such composition requires to broadcast {\it metadata} along the coded multicast messages. Since the size of the metadata does not scale with the file size, the metadata overhead does not contribute to the load in the limit of large file size.  %can be neglected when computing the achievable load.
}
The MAN scheme is said to achieve a {\it global coded caching gain}, also referred to as {\it multicasting gain}, equal to $t+1$ because the load with uncoded caching $\Rsf_\text{uncoded} = \Ksf-t = \Ksf(1-\Msf/\Nsf)$ is reduced by a factor $t+1$.
This gain scales linearly with network's aggregate memory size.
Yu, Maddah-Ali, and Avestimehr (YMA) in~\cite{exactrateuncoded} proved that $\binom{\Ksf-N_e(\dv)}{t+1}$ of the MAN multicast messages are redundant when a file is requested simultaneously by multiple users, where $N_e(\dv) \in[\min(\Nsf,\Ksf)]$ is the number of distinct file requests in the demand vector $\dv \in[\Nsf]^\Ksf$. The YMA scheme is known to be exactly optimal under the constraint of {\it uncoded cache placement}~\cite{exactrateuncoded}, and order optimal to within a factor of $2$ otherwise~\cite{yufactor2TIT2018}, for both worst-case load and average load when files are requested independently and equally likely.
The converse bound under the constraint of uncoded cache placement for the worst-case load was first derived by a subset of the authors in~\cite{ontheoptimality,indexcodingcaching2020} by exploiting the index coding acyclic converse bound in~\cite{onthecapacityindex}.
For the case $\Nsf\geq \Ksf =2$, the exact optimality without constraints on the type of placement was characterized in~\cite{symmetryouterbound} by a non-trivial converse bound leveraging the symmetries in the coded caching problem.

\subsubsection{Shared-link networks with privacy constraints}\label{sec:intro:sl+p}
For the successful decoding of an MAN multicast message, the users need to know the composition of this message (i.e., which subfiles are coded together). As a consequence, users are aware of the demands of other users. In practice, schemes that leak information on the demand of a user  to other users are highly undesirable. For example, this may reveal critical information on user behavior, and allow user profiling by discovering what types of content the users' request. 
Shared-link coded caching with private demands, which aims to preserve the privacy of the users' demands from other users, was originally discussed in~\cite{Engle2017privatecaching} and formally analyzed in an information-theoretical framework by Wan and Caire in~\cite{wan2019privatecaching}. In the private coded caching model, the information about the cached content of each user is unknown to the other users and the   composition of each coded multicast message sent by the server must be broadcasted along with the multicast message itself.  Following the private coded caching model in~\cite{wan2019privatecaching}, various private schemes were proposed in~\cite{wan2019privatecaching,kamath2019demandpri,aravind2020coded,colluding2020yan,Namboodiriprivatecaching,gholami2022coded}. 
Relevant to this paper is the private coded caching scheme based on virtual user proposed in~\cite{wan2019privatecaching}, which operates a MAN scheme as if there were $\Ksf \Nsf$ users in total, i.e., $\Nsf \Ksf-\Ksf$ virtual users in addition to the $\Ksf$ real users, and the demands of the virtual users as set such that each of the $\Nsf$ files is demanded exactly $\Ksf$ times. This choice of demands for the virtual users is such that any real user ``appears'' to have requested equally likely any of the files from the viewpoint of any other user, which guarantees the privacy of the demands.
An  improved private caching scheme based on virtual user strategy was proposed in~\cite{kamath2019demandpri}, which used the YMA delivery instead of the MAN delivery. Compared to converse bounds for the shared-link model without privacy constraint, it can be shown that this scheme based on virtual users is order optimal in all regimes, except for $\Ksf < \Nsf$ and $\Msf < \frac{\Nsf}{\Ksf}$~\cite{wan2019privatecaching}.\footnote{The problem in this regime can be intuitively understood as follows: for $\Msf=0$ the  achievable load in~\cite{wan2019privatecaching} is $\Nsf$ while the converse bound is $\min(\Ksf,\Nsf)=\Ksf$; the ratio of this two numbers  can be unbounded.}
 
%\dt{SUBPACK IS IRRELEVANT HERE \sout{By observing that the private caching schemes in~\cite{wan2019privatecaching,kamath2019demandpri} need high subpacketiation levels (i.e., the number of subfiles into which each file must be partitioned in the placement phase), the authors in~\cite{aravind2019twofiles} proposed a private caching scheme with a coded cache placement for two-user and two-file systems. When $\Msf=1$, the proposed scheme in~\cite{aravind2019twofiles} achieves a load $2/3$ with a subpacketiation level equal to $3$ (i.e., each file is partinioned in three subfiles), and it was proved that any private caching scheme with uncoded cache placement cannot achieve the load $2/3$ with subpacketiation level $3$.}}

To the best of our knowledge, the only converse bound that truly accounts for privacy constraints in the system model of~\cite{wan2019privatecaching} was proposed in~\cite{privatecaching2019KN2} for the case $\Ksf=\Nsf=2$. By combining the novel converse bound in~\cite{privatecaching2019KN2} with existing bounds without privacy constraint, the  exact optimality  was fully characterized in~\cite{privatecaching2019KN2} for $\Ksf=\Nsf=2$.

\subsubsection{D2D networks without privacy constraints}\label{sec:intro:d2d-p}
In practice, the content of the library may have been already distributed across the users' local memories and thus can be delivered through peer-to-peer or Device-to-Device (D2D) communications. The shared-link coded caching model was extended to D2D networks in~\cite{d2dcaching}. In the D2D delivery phase, each user broadcasts packets to all other users as functions of its cached  content and the users' demands. The D2D load is the sum of the bits sent by  all users normalized by the file length.

With the MAN cache placement where each file can be copied $t\in[0:\Ksf]$ times in the aggregate network memory, the D2D coded caching scheme  in~\cite{d2dcaching} further partitions each MAN subfile into $t$ equal-length sub-subfiles. Each user then acts as a shared-link server to convey its assigned sub-subfiles to the remaining users either with the MAN delivery~\cite{d2dcaching} or the YMA delivery~\cite{Yapard2djournal}. This scheme effectively splits the D2D network into $\Ksf$ parallel shared-link models, each having $\Nsf$ files and serving $\Ksf-1$ users with memory parameter $t-1$. Yapar et {\it al.} in~\cite{Yapard2djournal} proved that this scheme is order optimal to within a factor of $4$, and exactly optimal under the constraint of uncoded cache placement and {\it one-shot delivery} (i.e., in a one-shot delivery, any user can recover any requested bit from the content of its own cache and the transmitted messages by at most one other user). %\dt{!!! DT's note: was footnote, now is in main text!!!}

\subsection{New D2D networks with privacy constraints}\label{sec:intro:d2d+p}
  In D2D networks, the demand privacy problem is further exacerbated by the fact that each user is also a transmitter, which 
broadcasts coded multicast transmissions based on its cached content.  Based on the intuition developed from the shared-link model, one is tempted to conclude that it is impossible to guarantee privacy in D2D networks as the demand vector knowledge appears to be necessary to design the coded multicast messages.
Rather surprisingly, in this paper we show that it is possible to guarantee privacy of the users' demands against the other users also in a D2D setting. 
In our new D2D private caching model, the placement phase is similar to the shared-link private coded caching model. 
The delivery phase contains two steps. %as illustrated in Fig.~\ref{fig: system_model}. 
In the first step, each user broadcasts a query to the other users based on its local cached content and its demand;
 since the query size does not scale with the file size, this step does not contribute to the load in the limit for large file size. 
%The queries    telling the users what to send can be seen as protocol information, thus     requiring a communication load negligible with respect to the actual file transmission.  
In the second step, after collecting all the queries from all the users, each user broadcasts coded multicast messages as a function of the queries and its cached content. In the large file size regime, 
the load of the system is defined as the load in the second step of D2D communication. The objective of this paper is to design a D2D private coded caching scheme for $\Ksf$ users, $\Nsf$ files and memory size $\Msf \geq \Nsf/\Ksf$ (so that the aggregate cache in the entire network suffices to store the entire library) with minimum   transmitted load by all users in the delivery phase, while preserving the privacy of the users' demands against the other users.

In the Private Information Retrieval (PIR) problem~\cite{origPIR1995Chor} the privacy of the user's demand  against the servers  has been considered. In the PIR setting, a user wants to retrieve a desired file from some distributed non-colluding databases (servers), and the objective is to prevent any server from retrieving any information about the index of the user's demanded file. Recently, the authors in~\cite{PIR2017Sun} characterized the information-theoretic capacity of the PIR problem by proposing a novel converse bound and a coded PIR scheme based on interference alignment. The $T$-privacy PIR problem with colluding servers were originally considered in~\cite{PIRcludding2017Sun}, where it is imposed that any $T$-subset of queries sent from the user cannot reveal any information about the demand.  The $T$-privacy PIR problem with at most $T$ colluding servers where each server has a local coded storage was considered in~\cite{pirMDScollu2017,PIRcluddingcoded2019taja}. Since D2D communications have not been considered in the PIR literature, the D2D caching problem with private demands treated in this paper is not a special case of any existing PIR problem. %\dt{Our converse bound for the case of two users is inspired by PIR.}  

\subsection{Contributions}  
\label{sub:contributions}
%\dt{!!!DT's note: removed what was the first itemized list and only left what was the second list, with the detailed contributions!!!}
 
We start by giving the first known information-theoretic formulation of the D2D coded caching problem with demand privacy. Then we organize the main contributions of this paper as follows.
\begin{enumerate}[label=\alph*)] %(\roman*)

\item
{\bf Results for general $(\Nsf,\Ksf)$ from non-trivial extensions of past works}:  we prove a constant gap result for all parameter regimes except for $\Nsf> \Ksf$ and $\Msf < 2\Nsf/\Ksf$ (i.e., the small memory regime with more files than users).  More precisely, we propose:
\begin{enumerate}[label=(a.\arabic*)]

\item 
Coded Scheme~A (Theorem~\ref{thm:SchemeA}): This scheme carefully combines the idea of introducing virtual users~\cite{kamath2019demandpri} with that of splitting the D2D network into multiple parallel shared links~\cite{Yapard2djournal}.%d2dcaching,

\item
Optimality (Theorem~\ref{thm:OrderOptimalityA}): By comparing Scheme~A with a converse bound for the shared-link model without the privacy constraint in~\cite{yufactor2TIT2018}, we prove that Scheme~A is order optimal to within a factor of $6$ when $\Nsf\geq \Ksf$ and $\Msf\Ksf/\Nsf \geq 2$, %$\Msf\geq 2\Nsf/\Ksf$, 
and of $12$ when $\Nsf<\Ksf$ and $\Msf\Ksf/\Nsf \geq 1$. %$\Msf\geq \Nsf/\Ksf$. 

\end{enumerate}

\item
{\bf Results specifically for the case $\Ksf=2$ under uncoded cache placement}: 
at this point the regime $\Nsf > \Ksf$ and $\Msf\Ksf/\Nsf \in [1,2)$ %$\Msf\in[\Nsf/\Ksf, 2\Nsf/\Ksf)$ 
is open, which motivates the in-depth study of the simplest open case, namely the two-user case. 
We prove the first known general converse bound under uncoded cache placement that accounts for privacy constraints and  leads to a constant gap result for any number of files and any memory regime. In particular, we propose: 
\begin{enumerate}[label=(b.\arabic*)]

\item %(referred to as {\it Scheme~B} in the following) 
Coded Scheme~B (Theorem~\ref{thm:SchemeB}): This scheme     outperforms Scheme~A for the two-user case.

\item 
New Converse (Theorem~\ref{thm:two user converse}): Inspired by the converse bounds for non-private shared-link caching models under uncoded cache placement from~\cite{indexcodingcaching2020} and for PIR systems from~\cite{PIR2017Sun}, we propose a new converse bound under  uncoded cache placement for the two-user case by fully considering the privacy constraint.\footnote{\label{foot:difference to N=K=2}Our bound is not a generalization of the one for the shared-link private caching model with $\Nsf=\Ksf=2$ in~\cite{privatecaching2019KN2}, because 
the proposed converse bound heavily depends on the fact that the transmission of each user is a function of the queries and cached content of this user.}
 
\item 
Optimality (Theorem~\ref{thm:optimality K2}): With the new converse bound, under the constraint of uncoded cache placement and $\Nsf\geq \Ksf=2$, we show that Scheme~B is exactly optimal when $\Msf\in [\Nsf/2, (\Nsf+1)/2]$ or $\Msf \in \left[\frac{\Nsf(3\Nsf-5)}{2(2\Nsf-3)}, \Nsf \right]$, and is order optimal to within a factor of $3$ (numerical simulations suggest $4/3$) for the remaining memory size regime. 

\end{enumerate}

\item
{\bf Results for general $(\Nsf,\Ksf)$ under uncoded cache placement and user collusion}: we leverage the new converse bound for the two-user case  in a cut-set type bound  and prove a constant gap result for all parameter regime, while at the same time considering 
%the setting of user collusion. 
%With the above results for the two-user case, we can consider
a stronger notion of privacy that allows for colluding users.  
We propose:
\begin{enumerate}[label=(c.\arabic*)]

\item New Converse (Theorem~\ref{thm:K user converse}): We extend the proposed two-user converse bound to the $\Ksf$-user system by dividing the $\Ksf$ users into two groups, and derive a converse bound under uncoded cache placement and user collusion.

\item Optimality (Theorem~\ref{thm:order optimality K users}): Under the constraint of uncoded cache placement and  user collusion, Scheme~A is shown to be order optimal   to  within a factor of $18$ (numerical simulations suggest $27/2$) when $\Nsf > \Ksf$ and $\Msf\Ksf/\Nsf \in [1,2)$. %$\Msf\in[\Nsf/\Ksf, 2\Nsf/\Ksf)$. 
This proves that Scheme~A is order optimal in all memory regimes (that is, also in the regime that was open under the converse bound  for the non-private shared-link model) and it is robust to colluding users.  

\end{enumerate}

\end{enumerate}

\begin{rem}[The powerfulness of the two-user converse bound]\label{rem:K2power}
It is rather surprising and quite remarkable to see that, in the considered D2D private coded caching problem, the converse for the case of $\Ksf = 2$ users 
 combined with a cut-set extension  yields the  order optimality   for any system parameters under the constraint of uncoded cache placement and user collusion. This is in stark contrast  to plenty of well-known multiuser information theory problems where
the optimality results for the $\Ksf = 2$ case do not generalize, and give in fact little or no hint to  the $\Ksf > 2$ case. Paramount examples include the general broadcast channel with degraded message sets~\cite{generalBCwithdegraded,networkinformation},   the  $\Ksf$-user Gaussian  interference channel~\cite{twouserinterference,kuserinterference}, and the non-private shared-link coded caching~\cite{symmetryouterbound}.
%we see the results for the two-user case as the key breakthrough that allows us to derive order optimality results for any system parameters under the constraint of uncoded cache placement and user collusion.}
\hfill $\square$ 
\end{rem}

\begin{rem}[Cost of   D2D]\label{rem:Cost of Privacy}
By using the   result in~\cite{porter2019embeddedindex}, one can immediately infer that, under the constraint of uncoded cache placement and without privacy constraint, the gap between the achieved loads in the shared-link and D2D scenarios is at most $2$. This is no longer the case when privacy is introduced, where the gap between the loads in private shared-link and private D2D scenarios can be arbitrarily large %. However, this paper shows that this is no longer the case with privacy constraints.  Indeed, the gap between the achieved loads in private shared-link and private D2D scenarios can be arbitrarily large  
 (i.e., the gap is larger than $\Nsf/\min(\Nsf,\Ksf)$ when $\Msf=\Nsf/\Ksf$, which can be unbounded). Similar observations were made in the context of secure shared-link pliable index coding~\cite{TangLiuITW2019}, where the authors showed that problems that are feasible without security constraints became infeasible when security is considered  (i.e., there is no constant gap factor independent of the system parameters).
 \hfill $\square$ 
\end{rem}

\subsection{Paper organization}
The rest of this paper is organized as follows.
Section~\ref{sec:model} formulates the D2D private caching model.
Section~\ref{sec:main} provides an overview of all our technical results, and provides some numerical evaluations. 
Sections~\ref{sec:AchievableSchemes} and~\ref{sec:converse} provide proofs of the proposed achievable schemes and converse bounds, respectively.
%%Section~\ref{sec:AchievableSchemes} presents our   proposed achievable schemes and the order optimality results.
%Section~\ref{sec:num} provides some numerical evaluations for the proposed achievable and converse bounds. 
Section~\ref{sec:conclusion} concludes the paper.
Some proofs  (i.e., more technical lemmas and tedious gap derivations)  may be found in the Appendices.

\subsection{Notation convention}
%We use the following notation convention.
Calligraphic symbols denote sets, 
bold symbols denote vectors, 
and sans-serif symbols denote system parameters. 
% In general, lower-case symbols denote realizations of random variables indicated with upper-case symbols. 
We use $|\cdot|$ to represent the cardinality of a set or the length of a vector.
Sets of consecutive integers are denoted as
$[a:b]:=\left\{ a,a+1,\ldots,b\right\}$ and $[n] := [1:n]$. 
The symbol $\oplus$ represents bit-wise XOR. 
$a!=a\times (a-1) \times \cdots \times 1$ represents the factorial of $a$.
%the number of $k$-permutations of $n, n\geq k,$ is indicated as $P(n,k):=n\cdot(n-1)\cdots(n-k+1)$.
We use the convention $\binom{x}{y}=0$ if $x<0$ or $y<0$ or $x<y$.

\section{System Model}
\label{sec:model}

%\dt{!!!DT's note: removed: large B, epsB, and metadata; note ell in~\eqref{eq:load}; note local randomness in~\eqref{eq:cK}!!!}

A $(\Ksf,\Nsf)$ D2D private caching system comprises the following elements.
\begin{itemize}
\item
A library with $\Nsf$ independently generated files, where each file is composed of $\Bsf$ i.i.d. bits.
% where $\Bsf$ is assumed sufficiently large such that any subfile division of the files is possible.
The files are denoted by $(F_{1},F_{2},\dots,F_{\Nsf})$.
\item
$\Ksf$ users, each equipped with a local cache.
%of $\Msf \Bsf$ bits, where  $\Msf \in \left[\frac{\Nsf}{\Ksf} ,\Nsf\right]$. 
%There is a trusted server without access to the library in the system. This server is connected to each user through an individual  secure  link.
\item
An error-free broadcast link from each user to all other users (e.g., a shared medium).\footnote{D2D networks may be implemented at the physical/MAC layer, such that the nodes are physical devices sharing a common transmission medium, or at the logical or ``application'' layer, as for example in current peer-to-peer  file sharing systems such as 
BitTorrent, Gnutella, Kazaa and several others. 
We do not make such distinction here and just compute the load as the sum of all nodes (or ``peers'') transmissions  expressed in bits, necessary to satisfy the users demands. 
This load notion is compliant with the previously defined coded caching models for D2D and shared link systems.}
\end{itemize}

%Let $\epsilon_\Bsf \geq 0$ be a constant. 
The system operates in two phases.
\begin{itemize}
\item Placement Phase.
Note that the placement phase is done without knowledge of later demand. 
Each user $k\in[\Ksf]$ first  generates some local randomness   $ P_{k} $, which is independent of the library $F_1,\ldots,F_{\Nsf}$ and independent across users, and is only known at user $k\in[\Ksf]$.   
Then user $k $ stores $Z_{k}$ in its cache,  
where
\begin{align}
H\big(Z_{k}| P_{k}, \ F_{1},\ldots,F_{N} \big)=0 \ \ \text{(placement constraint)},
\label{eq:cK}
\end{align} 
%\dt{!!!DT's note: no more conditioning on Pk later on; right???}
The vector of all caches is $\mathbf{Z}:=(Z_{1},Z_{2},\dots,Z_{\Ksf})$.

\item Delivery Phase. 
User $k\in [\Ksf]$ demands the file indexed by $d_k\in [\Nsf]$.
The demand vector is $\mathbf{d}:=(d_1,d_2,\ldots,d_{\Ksf})$. 
The delivery phase contains the following two steps. %, as illustrated in Fig.~\ref{fig: system_model}:
\begin{itemize}

\item Step 1: user $k\in [\Ksf]$, given its randomness $P_k$, cached content $Z_{k}$ and   demand $d_k$, broadcasts the query $\ell_{k}$ to the other users. 
\iffalse
 where
\dt{
\begin{align}
H\big(\ell_{k}|Z_k, P_k,  d_{k} \big)=0,  \ \ \text{(query constraint)}.
\label{eq:encoding constraint}
\end{align}
%Denote the query vector by $\mathbf{L}=(\ell_{1},\ldots,\ell_{\Ksf})$.
}
\fi
%%%%\iffalse
%%%%\item Step 2: 
%%%%{\red according to the users' demands and the metadata  of cached content, the trusted server sends back the   metadata $\mathscr{M}(P_k)$  to user $k$, 
%%%%%\dt{IF "broadcasts the metadata vector $(\mathscr{M}(P_1),\ldots,\mathscr{M}(P_K))$ to all users via a shared-link" THEN NO NEED FOR A USER TO SEND ITS METADATA} 
%%%%where the metadata $\mathscr{M}(P_k)$ describes how the coded packets (i.e., payload) $P_k$, to be broadcasted by user $k\in [\Ksf]$, are composed. Note that $\mathscr{M}(P_1),\ldots,\mathscr{M}(P_{\Ksf})$ are random variables over $\Pc_1,\ldots,\Pc_{\Ksf}$, representing all types of transmissions by the $\Ksf$ users.}
%%%%\fi

\item Step 2: after having received all the queries, user $k\in [\Ksf]$ broadcasts the signal $X_{k}$ %=(\mathscr{M}(P_k), P_k)
to the other users, where
\begin{align}
H\big(X_{k}|Z_k, P_k, \ell_{1},\ldots,\ell_{\Ksf}\big)=0,  \  \text{(encoding constraint)}.
\label{eq:encoding constraint}
\end{align}
Note that,  the queries  $\ell_{1},\ldots,\ell_{\Ksf}$ act as the metadata  explained in Footnote~\ref{foot:metadata in MAN}, implying the   composition of each coded multicast message.
%By the definition of metadata, we also have 
%\begin{align}
%&H(X_{k}|\mathscr{M}(P_k), Z_k )=0.
%\label{eq:metadata Xk}
%\end{align}
%Again, we assume that the sizes of the queries  and  metadata    are negligible with respect to the file size $\Bsf$.

\end{itemize}

%{\it Decoding.} 
%Let $\mathbf{X} := (X_j: j\in [\Ksf])$ be the vector of all transmitted signals. 
%The constraints  on the decoding of the demanded file by each user while maintaining the privacy is given as follows.
Successful decoding is guaranteed if
\begin{align}
&H\big(F_{d_k}|Z_k, P_k, d_k, \ \ell_{1},\ldots,\ell_{\Ksf}, \ X_{1},\ldots,X_{\Ksf} \big) = 0, \nonumber\\&  \forall k\in [\Ksf],   
 \ \ \text{(decoding constraint)}.  
\label{eq:decodability}
\end{align}
%In other words, given what user $k$ has at its disposal after the delivery phase, the desired file $F_{d_k}$ is fully determined (conditional entropy equal to zero).  
Demand privacy\footnote{The privacy constraint in~\eqref{eq:privacy}  corresponds to  perfect secrecy in an information theoretic sense (see~\cite[Chapter 22]{networkinformation}).} is guaranteed if %\dt{(recall that the demands are independent across users)}
%given $d_k$ and $Z_k$, user $k$ cannot get any information about the demands of other users  from $(  P_j: j\in [\Ksf] )$.  The privacy constraint  can be   written as %\dv_{\backslash\{k\}}
\begin{align}
& I\big( \dv_{[\Ksf]\setminus\{k\}}; Z_k, P_k, d_k, \ \ell_{1},\ldots,\ell_{\Ksf}, \ X_{1},\ldots,X_{\Ksf}  \big) =0,  \nonumber\\& \text{(privacy  constraint)}, 
\label{eq:privacy}
\end{align}
%\sout{In other words, the mutual information between $\dv_{\backslash\{k\}}$ and all the information known by each user  after the delivery phase, quantifies in precise information theoretic terms the information leakage of the delivery phase on the demands of other users on the perspective of  user $k$.} 
 where $\dv_{\Sc}$ denotes the vector obtained from $\dv$ by retaining only the elements indexed by $\Sc$.

\end{itemize}

 Assume that  the length of $(P_{k}, \ell_k)$, $k\in [\Ksf]$,  does not scale with $\Bsf$.  By the constraint of privacy,  the number of transmissions
in Step 2 of the  delivery for different demand vectors should be the
same.  Thus
\begin{subequations}
a pair $(\Msf,\Rsf)$ is said to be achievable if all the above constraints are satisfied with 
\begin{align}
& \underset{\Bsf\to\infty}{\lim\sup} \frac{H(Z_{k})}{\Bsf} \leq \Msf, \ \forall k\in[\Ksf],  \ \ \text{(cache size)},
\label{eq:memory size}
\\
&\underset{\Bsf\to\infty}{\lim\sup} \frac{\sum_{k\in[\Ksf]} H(X_{k}) }{\Bsf} \leq \Rsf,  \ \ \text{(load)}.
\label{eq:load}
\end{align}
\label{eq:def achievable  M,R}
\end{subequations}
%where  $\Msf \in \left[\frac{\Nsf}{\Ksf} ,\Nsf\right]$ is the cache size. Note that for $\Msf \geq \Nsf$ each user can cache the whole library, thus no 
%
Our objective is to determine
\begin{align}
\Rsf^\star(\Msf) := \inf\{ \Rsf: (\Msf,\Rsf) \ \text{is  achievable as in~\eqref{eq:def achievable  M,R}} \}.
\label{eq:optimal load}
\end{align}

We only consider the case $\min(\Ksf,\Nsf)\geq 2$, since the case $\Ksf=1$, a single node network,  does not  make sense in a D2D network  and when $\Nsf=1$ each user knows the demand of  the other users.
In addition, we only need to consider $\Msf \in \left[\frac{\Nsf}{\Ksf} ,\Nsf\right]$, since for $\Msf \geq \Nsf$ each user can cache the whole library, thus no delivery is needed; and for $\Ksf\Msf < \Nsf$ there is not enough space in the overall network memory to store the whole library, thus the problem is not feasible.

%We say that load $\Rsf$ is achievable if  %(recall that   the size of the metadata is negligible with respect to the file size $\Bsf$)
%\begin{align}
%%\Rsf:=\sum_{k\in [\Ksf]} \Rsf_{k}
%\sum_{k\in[\Ksf]} H(X_{k}) \leq \Bsf(\Rsf+ \epsilon_\Bsf) \ \text{(load)},
%\label{eq:loaddef}
%\end{align}
%while all the above constraints are satisfied and $\lim_{\Bsf\to\infty}\epsilon_\Bsf=0$.
%%Similar to the shared-link private caching problem in~\cite{wan2019privatecaching}, it is obvious that the transmitted loads for different demand vectors should be the same;  otherwise, the transmitted load which can be counted by each user will reveal information about the users' demands.
%The objective is to determine, for a fixed $\Msf \in \left[\frac{\Nsf}{\Ksf} ,\Nsf\right]$, the minimum achievable load, which is indicated by $\Rsf^{\star} (\Msf)$.
%
%{\red Notice that the metadatas introduced in this paper which are random variables,  are used to show that the cache placement functions of the users and the delivery function  of the server are not deterministic, which are crucial for preserving the privacy of the users' demands. }

{\it Uncoded Cache Placement.}
If  each user $k\in [\Ksf]$ directly copies some bits of the files into $Z_k$, the cache placement is said to be {\it uncoded}. The optimal load under the constraint of uncoded cache placement is denoted by $\Rsf^{\star}_{\mathrm{u}}(\Msf)$, which is defined as in~\eqref{eq:load} but with the additional constraints that the cache placement phase is uncoded. 
%The optimal load under the constraint of uncoded cache placement is denoted by $\Rsf^{\star}_{\mathrm{u}}$. Obviously, 
Clearly, $\Rsf^{\star}(\Msf) \leq \Rsf^{\star}_{\mathrm{u}}(\Msf)$.

{\it Colluding Users.} 
We say that the users in the system {\it collude} if they exchange the index of their demanded file and their cached content.  
Collusion is a natural consideration to increase the privacy level and is one of the most widely studied variants in the PIR problem \cite{beimel2005general, PIRcludding2017Sun, tajeddine2017private, banawan2018capacity}. 
Privacy constraint against colluding users is a stronger notion than~\eqref{eq:privacy} and is defined as follows 
\begin{align}
% \ \text{(privacy  constraint against colluding users)}.
 & I\big( \dv_{[\Ksf]\setminus \Sc};  (Z_k, P_k:k\in\Sc), \dv_{\Sc},   \ell_{1},\ldots,\ell_{\Ksf},   X_{1},\ldots,X_{\Ksf}  \big) =0,  \nonumber\\& 
 \forall \Sc \subseteq [\Ksf],  \Sc\not=\emptyset.
\label{eq:colluding privacy}
\end{align} 
The optimal load under  uncoded cache placement and the privacy constraint in~\eqref{eq:colluding privacy} is denoted by $\Rsf^{\star}_{\mathrm{u,c}} (\Msf)$. Clearly, $\Rsf^{\star}_{\mathrm{u,c}}(\Msf) \geq \Rsf^{\star}_{\mathrm{u}}(\Msf) \geq  \Rsf^{\star}(\Msf)$.

\begin{rem}\label{rem:relationloads} 
For $\Ksf=2$, the privacy constraints in~\eqref{eq:privacy} and~\eqref{eq:colluding privacy} are equivalent, and thus we have  $\Rsf^{\star}_{\mathrm{u,c}}(\Msf) = \Rsf^{\star}_{\mathrm{u}}(\Msf) \geq   \Rsf^{\star}_{}(\Msf)$. 

%Our converse results with $K > 2$ users require this constraint of colluding users. 
\hfill $\square$ 
\end{rem}
 
%In the rest of the paper, to simplify the notations  we use $ \Rsf^{\star}$, $\Rsf^{\star}_{\mathrm{u}}$, and $\Rsf^{\star}_{\mathrm{u,c}}$ to represent  $ \Rsf^{\star}(\Msf)$, $\Rsf^{\star}_{\mathrm{u}}(\Msf)$, and $\Rsf^{\star}_{\mathrm{u,c}}(\Msf)$, respectively.

\section{Main Results}
\label{sec:main}
In this section, we summarize all the new results in this paper and provide the main ingredients on how the bounds are derived. 
%\dt{!!!Dt's note: moved examples in proof sections!!!}
%In particular, we   provide an example to illustrate the high-level idea to derive the novel converse bound. The detailed proofs of  the proposed achievable   and   converse bounds are   given in Sections~\ref{sec:AchievableSchemes} and~\ref{sec:converse}, respectively.

\subsection{Results for general $(\Nsf,\Ksf)$ by non-trivial extensions of known schemes}
\label{sub:list of achiev}

%%%\iffalse
%%%{\red For the sake of   comparison, we first propose a trivial D2D private caching scheme, which lets each user recover the whole library in order to hide its demanded file.}  %\dt{WHY DO WE NEED THIS SCHEME?}
%%%\begin{thm}[Uncoded Scheme]
%%%\label{thm:UncodedScheme}
%%%For the $(\Ksf,\Nsf)$ D2D private caching system \dt{with a trusted server} %, $\Rsf^{\star} $, $\Rsf^{\star}_{\mathrm{u }}$, and  $\Rsf^{\star}_{\mathrm{u,c}}$ are upper bounded by   
%%%\begin{align}
%%%%\Rsf^{\star}
%%% \Rsf^{\star}_{\mathrm{u,c}} 
%%%\leq \Rsf_{\text{\rm uncoded}} = \frac{\Ksf}{\Ksf-1}(\Nsf-\Msf). 
%%%\label{eq:UncodedScheme}
%%%\end{align}
%%%\hfill $\square$ 
%%%\end{thm} 
%%%\fi

%\dt{I CHANGED SOMEWHAT THE FOLLOWING, AS "subpacketization is different from the MAN cache placement" IS TOO MUCH IMHO; AS WE DISCUSSED, WHAT MAKES IT LOOK DIFFERENT IS THE WAY WE INDEXED THE SUBFILES IN THIS WORK.}

Inspired by the virtual-user strategy in~\cite{kamath2019demandpri}, we propose a private coded caching scheme (referred to as Scheme~A in the following) with a cache placement inspired by the   D2D strategy~\cite{d2dcaching}. More precisely, our scheme effectively divides the D2D network into $\Ksf$ independent shared-link models, each of which serves $\Usf:= (\Ksf-1)\Nsf$ effective users, where $(\Ksf-1)(\Nsf-1)$ users are virtual. 
The achieved load is given in the following theorem; an example that highlights the main ingredients in Scheme~A can be found in Section~\ref{ex:example for Scheme A} and the detailed general description on Scheme~A can be found in Section~\ref{sub:SchemeA}.
\begin{thm}[Scheme~A]
\label{thm:SchemeA}
For the $(\Ksf,\Nsf)$ D2D private caching system, %$\Rsf^{\star}$, $\Rsf^{\star}_{\mathrm{u }}$, and  
$\Rsf^{\star}_{\mathrm{u,c}}$ is upper bounded by  the lower convex envelope of %$(\Msf,\Rsf_{\mathrm{A}})=\left(\Nsf/\Ksf,\Nsf \right)$ and 
the following points
\begin{align}
&(\Msf,\Rsf_{\mathrm{A}})= \left( 
\frac{\Nsf+t-1}{\Ksf},
\frac{\binom{\Usf}{t}-\binom{\Usf-\Nsf}{t}}{ \binom{\Usf}{t-1}}
%\frac{\Usf-t+1}{t}
\right), \ \forall  t\in [ \Usf+1].
\label{eq:extended scheme}
\end{align}
\hfill $\square$ 
\end{thm}
%Notice that when $t=\Usf+1$ in~\eqref{eq:extended scheme}, we have the trivial corner point $(\Msf,\Rsf_{\mathrm{A}})=(\Nsf, 0)$.

Note that Scheme~A satisfies the robust privacy constraint in~\eqref{eq:colluding privacy} against colluding users.
By comparing Scheme~A in Theorem~\ref{thm:SchemeA} and the converse bound for the shared-link caching problem without privacy constraint in~\cite{yufactor2TIT2018}, we have the following order optimality results, whose  proof can be found in Appendix~\ref{sec:proof of order optimality}.
\begin{thm}[Order optimality of Scheme~A]
\label{thm:OrderOptimalityA}
For  the $(\Ksf,\Nsf)$ D2D private caching system, Scheme~A in Theorem~\ref{thm:SchemeA} is order optimal to within a factor of $6$ if $\Nsf\geq \Ksf$ and $\Msf \geq 2\Nsf/\Ksf$, and $12$ if $\Nsf \leq  \Ksf$.
\hfill $\square$ 
\end{thm}

\begin{rem}[Reduction of Subpacketization for Scheme~A]
\label{rem:subpacketization}
%It will be clarified that the virtual-user scheme (i.e., 
Scheme A in  Theorem~\ref{thm:SchemeA} divides each file into   $\Ksf\binom{\Usf}{t-1}$ equal-length subfiles, thus the subpacketization is  $\Ksf\binom{\Usf}{t-1} \approx \Ksf 2^{\Usf \Hc\left( \frac{t-1}{\Usf} \right)} $, where    $\Hc(p)=-p \log_2(p)-(1-p)\log_2(1-p) $ is the binary entropy function. Hence,   the maximal  subpacketization of the virtual-user scheme (when $\frac{t-1}{\Usf} = \frac{1}{2}$) is  
exponential in $\Usf$, which is much higher than the maximal subpacketization  of the $\Ksf$-user MAN coded caching scheme  (which is exponential in $\Ksf$).
Very recently, after the original submission of this paper, the authors in~\cite{colluding2020yan} proposed a shared-link private coded caching scheme based on the cache-aided linear function retrieval~\cite{arxivfunctionretrieval}, which can significantly reduce the subpacketization of the shared-link virtual-user  private caching schemes  in~\cite{wan2019privatecaching,kamath2019demandpri}. 
In addition to the cached  content 
by the MAN placement, the authors let each user privately cache some
linear combinations of uncached subfiles in the MAN placement
which are regarded as keys. In such way, the effective demand
of each user in the delivery phase becomes the sum of these
linear combinations and the subfiles of its desired file, such
that the real remand is concealed. We can directly use the extension strategy in~\cite{d2dcaching} to extend this shared-link private caching scheme to our D2D setting  to obtain   {\it Scheme C}, which achieves
the lower convex envelope of  $\left(\frac{\Ksf}{\Nsf},\Nsf \right)$ and 
the following points
\begin{align}
&(\Msf,\Rsf_{\mathrm{C}})= \left( 
\frac{t(\Nsf-1)}{\Ksf}+1,
\frac{\binom{\Ksf-1}{t}-\binom{\Ksf-1-\Nsf}{t}}{ \binom{\Ksf-1}{t-1}}
%\frac{\Usf-t+1}{t}
\right), \ \forall  t\in [\Ksf].
\label{eq:less subpacket scheme}
\end{align}
The subpacketization of the scheme in~\eqref{eq:less subpacket scheme} is $\Ksf\binom{\Ksf}{t}  \approx \Ksf 2^{\Ksf \Hc\left(t/\Ksf \right)}$, which is the same as the $\Ksf$-user non-private D2D coded caching scheme in~\cite{d2dcaching}. As the shared-link private caching scheme in~\cite{colluding2020yan}, Scheme C also satisfies the robust privacy constraint in~\eqref{eq:colluding privacy} against colluding users. 
\hfill $\square$ 
\end{rem}

\subsection{Results for $\Ksf=2$: new converse bound to truly account for privacy constraints}
%\subsection{Novel D2D Private Caching Converse Bound under the Constraint of Uncoded Cache Placement}
\label{sub:list of converse}

The order optimality results in Theorem~\ref{thm:OrderOptimalityA} is derived from an existing converse bound without privacy constraint and does not cover the regime $\Nsf > \Ksf$ and $\Msf\in[\Nsf/\Ksf, 2\Nsf/\Ksf)$. Hence, we are motivated to derive a new converse bound by fully incorporating the privacy constraint for the simplest open case, that is, for a two-user system. %To the best of our knowledge, in the literature, there does not exist any caching converse bound which considers the privacy constraint. 

When $\Ksf=2$, we observe that in Scheme~A some cached  content is redundant. By removing those redundancies we derive a new scheme (referred to as Scheme~B in the following) whose achieved load is given in the following theorem; an example that highlights the main ingredients in Scheme~B can be found in Section~\ref{ex:example for Scheme B} and the detailed general description on Scheme~B can be found in Section~\ref{sub:SchemeB}.
\begin{thm}[Scheme~B]
\label{thm:SchemeB}
For the $(\Ksf,\Nsf)=(2,\Nsf)$  D2D private caching system,  
$\Rsf^{\star}_{\mathrm{u}}=\Rsf^{\star}_{\mathrm{u,c}}$ is upper bounded by  the lower convex envelope
of $(\Msf,\Rsf_{\mathrm{B}})=\left(\Nsf , 0 \right)$ and the following points
\begin{align}
& (\Msf,\Rsf_{\mathrm{B}})= \left(\frac{\Nsf}{2} + \frac{\Nsf t^{\prime}}{2(\Nsf+t^{\prime}-1)}, \frac{\Nsf(\Nsf-1)}{(t^{\prime}+1)(\Nsf+t^{\prime}-1)}\right), \nonumber\\ &  \forall  t^{\prime}\in [0:\Nsf-1].
\label{eq:second scheme}
\end{align}
\hfill $\square$ 
\end{thm}
 
In Appendix~\ref{sec:proof of strictly better} we prove the following corollary.
\begin{cor}\label{cor:proof of strictly better}
By comparing Scheme~A in Theorem~\ref{thm:SchemeA} for $\Ksf=2$ and Scheme~B in Theorem~\ref{thm:SchemeB}, we find $\Rsf_{\mathrm{B}} \leq \Rsf_{\mathrm{A}}$. 
\hfill $\square$ 
\end{cor}

 Next we turn our attention to converse bounds that truly incorporate the privacy constraint.
The following converse bound is one of the key novelties of this paper. It truly accounts for the privacy constraint in the general setting $\Nsf \geq 2$.  The main idea   is to derive  several bounds that contain a ``tricky'' entropy term that needs to be bounded in a non-trivial way; in some bounds this entropy term appears with a positive sign and in others with a negative sign; by linearly combining the bounds, the ``tricky'' entropy term cancels out. Different from the converse bound in~\cite{privatecaching2019KN2} for the shared-link caching with private demands for $\Nsf=\Ksf=2$, our converse bound focuses on uncoded cache placement and works for any $\Nsf \geq \Ksf=2$. 
 Theorem~\ref{thm:two user converse} is proved in full generality in Section~\ref{sub:two user converse}. For the sake of clarity,  an example of the key steps in the proof is provided Section~\ref{ex:converse K2N2} for the case of $\Nsf=2$ files.
%In Section~\ref{sub:two user converse} we show how to generalize Example~\ref{ex:converse K2N2} to the case where $\Ksf=2$ and $\Nsf\geq 2$, so as to arrive at the following theorem. 
%
\begin{thm}[New converse bound for the two-user system] 
\label{thm:two user converse}
For the $(\Ksf,\Nsf)=(2,\Nsf)$ D2D private caching system  where $\Nsf \geq  \Ksf=2 $, assuming $\Msf=\frac{\Nsf}{2}+y$ where $y\in \left[0,\frac{\Nsf}{2}\right]$, we have the following bounds %for $\Rsf^{\star}_{\mathrm{u}}$
\begin{align}
&\Rsf^{\star}_{\mathrm{u}} \geq \Nsf-2y-\frac{4y+(\Nsf- \Ksf/2 )h}{h+2} \nonumber\\& +\frac{h^2(\Nsf- \Ksf/2 )-\Nsf(2\Nsf/\Ksf-3 )+h(\Nsf+ \Ksf/2 ) }{(h+1)(h+2)}\frac{2y}{\Nsf}, \notag \\& \forall h\in[0:\Nsf-3], \ \  \text{only active for $\Nsf\geq 3$},
\label{eq:K2 converse 1}
\\
&\Rsf^{\star}_{\mathrm{u}} \geq  \Ksf \left(1-\frac{3y}{\Nsf}\right),
\label{eq:K2 converse 2}
\\
&\Rsf^{\star}_{\mathrm{u}} \geq  \Ksf \left(\frac{1}{2}-\frac{y}{\Nsf}\right).
\label{eq:K2 converse 3}
\end{align}
\hfill $\square$ 
\end{thm}
%\dt{REMOVED RED COMMENT AND ADDED SOMTHING IN EQ~\eqref{eq:K2 converse 1}}
%{\red  Note that when $\Nsf=2$, we have $\Nsf-3<0$ and then  the  inequalities in~\eqref{eq:K2 converse 1}  do  not exist; thus the converse bound for this case is $\Rsf^{\star}_{\mathrm{u}} \geq \max\left( \Ksf \left(1-\frac{3y}{\Nsf}\right), \Ksf \left(\frac{1}{2}-\frac{y}{\Nsf}\right)\right)$.}

By comparing the new converse bound in Theorem~\ref{thm:two user converse} and Scheme~B in Theorem~\ref{thm:SchemeB}, we have the following optimality result under the constraint of uncoded cache placement (the proof can be found in Appendix~\ref{sec:optimality K2}).
\begin{thm}[Optimality for the two-user system] 
\label{thm:optimality K2}
For the $(\Ksf,\Nsf)=(2,\Nsf)$ D2D private caching system  where $\Nsf \geq \Ksf=2$, Scheme~B in Theorem~\ref{thm:SchemeB} is exactly optimal under the constraint of uncoded cache placement when $\frac{\Nsf}{2}\leq \Msf \leq \frac{\Nsf+1}{2}$ or $\frac{\Nsf(3\Nsf-5)}{2(2\Nsf-3)} \leq \Msf \leq \Nsf$. Otherwise, Scheme~B is order optimal to within a factor of $3$ (numerical simulations suggest $4/3$).

\hfill $\square$ 
\end{thm}
 
%%% I DO NOT LIKE THE `MULTIPLICATION OF THE THEOREMS' !!! 
%%%By comparing the novel converse bound in Theorem~\ref{thm:two user converse} and Scheme~B in Theorem~\ref{thm:SchemeB}, we can also have the following  order  optimality results under the constraint of uncoded cache placement, whose proof is in Appendix~\ref{sec:order optimality K2}.
%%% \begin{thm}[Order optimality for two-user systems] 
%%% \label{thm:order optimality K2}
%%%  For  the $(\Ksf,\Nsf)$ D2D private caching system \dt{with a trusted server} where $\Nsf \geq \Ksf=2$, Scheme~B is order optimal under the constraint of uncoded cache placement  within a factor of $3$ (numerical simulations suggest $4/3$).
%%%  \hfill $\square$ 
%%% \end{thm}

From Theorem~\ref{thm:optimality K2}, we can directly derive the following corollary.
\begin{cor}
\label{cor:optimality K2 special case}
For the $(\Ksf,\Nsf)=(2,\Nsf)$ D2D private caching system Scheme~B in Theorem~\ref{thm:SchemeB} is exactly optimal under the constraint of uncoded cache placement  in all memory regimes  when $\Nsf \in \{2,3\}$.
\hfill $\square$ 
\end{cor}

\subsection{Order optimality results for any system parameter when users may collude} 
In Section~\ref{sub:K user converse} we extend Theorem~\ref{thm:two user converse} to any $\Ksf \geq 2$ with the consideration of the privacy constraint against colluding users in~\eqref{eq:colluding privacy}.  The main idea is to divide the users into two groups in a cut-set-like fashion and generate a powerful aggregate user whose cache contains the caches of all users in each group (implying collusion). The derived converse bound is as follows.
\begin{thm}[New converse bound for the $\Ksf$-user system] 
\label{thm:K user converse}
For the $(\Ksf,\Nsf)$ D2D private caching system  where $\Nsf \geq \Ksf \geq 3$, assuming $\Msf=\frac{\Nsf}{\Ksf}+\frac{2 y }{\Ksf}$ where $y\in \left[0,\frac{\Nsf}{2}\right]$, we have 
\begin{align}
&\Rsf^{\star}_{\mathrm{u,c}} \geq \frac{\left\lfloor \Ksf/2 \right\rfloor }{ \left\lceil\Ksf/2 \right\rceil  } \frac{\left\lfloor 2\Nsf/\Ksf \right\rfloor  }{  2\Nsf/\Ksf }   \times \text{\rm RHS~eq\eqref{eq:K2 converse 1}} 
, \nonumber\\ &\forall h\in \left[0:  \left\lfloor 2\Nsf/\Ksf -3 \right\rfloor \right], \ \ \text{only active for $  \Nsf/\Ksf \geq 3/2 $},
\label{eq:K user converse first segment}
\\
&\Rsf^{\star}_{\mathrm{u,c}} \geq \frac{\left\lfloor \Ksf/2 \right\rfloor }{ \left\lceil\Ksf/2 \right\rceil  } \times \text{\rm RHS~eq\eqref{eq:K2 converse 2}},
%\left\{\Ksf-\frac{3\Ksf y}{\Nsf} \right\},
\label{eq:K user converse second segment}
\\
&\Rsf^{\star}_{\mathrm{u,c}} \geq \frac{\left\lfloor \Ksf/2 \right\rfloor }{ \left\lceil\Ksf/2 \right\rceil }  \times \text{\rm RHS~eq\eqref{eq:K2 converse 3}}. 
% \left\{ \frac{\Ksf}{2}-\frac{y \Ksf}{\Nsf} \right\}.
\label{eq:K user converse third segment}
\end{align}
\hfill $\square$ 
\end{thm}
%\dt{REMOVED RED COMMENT AND ADDED SOMTHING IN EQ~\eqref{eq:K user converse first segment}}
%{\red  Note that when $  2\Nsf/\Ksf -3   < 0$,    the inequalities in~\eqref{eq:K user converse first segment}  do  not exist; thus the converse bound for this case is $\Rsf^{\star}_{\mathrm{u}} \geq \frac{\left\lfloor \Ksf/2 \right\rfloor }{ \left\lceil\Ksf/2 \right\rceil  } \times \max\left(\text{\rm RHS~eq~\eqref{eq:K2 converse 2}},   \text{\rm RHS~eq~\eqref{eq:K2 converse 3}}\right)$.}

By comparing Scheme~A in Theorem~\ref{thm:SchemeA} and the combination of the new converse bound in Theorem~\ref{thm:K user converse} and the converse bound for shared-link caching without privacy in~\cite{indexcodingcaching2020}, we can characterize the order optimality of Scheme~A under the constraint of uncoded cache placement and user collusion in all parameter regimes (the proof can be found in Appendix~\ref{sec:order optimality K users}).
\begin{thm}[Order optimality for the $\Ksf$-user system] 
\label{thm:order optimality K users}
For  the $(\Ksf,\Nsf)$ D2D private caching system where $\Nsf \geq \Ksf $, Scheme~A in Theorem~\ref{thm:SchemeA} is order optimal to within a factor of $18$ (numerical simulations suggest $27/2$) under the constraint of uncoded cache placement and  user collusion.

\hfill $\square$ 
\end{thm}

Note that when $\Nsf<\Ksf$, Theorem~\ref{thm:OrderOptimalityA} shows that Scheme~A is generally order optimal to within a factor of $12$. Hence, from Theorems~\ref{thm:OrderOptimalityA} and~\ref{thm:order optimality K users}, we can directly have the following conclusion.
\begin{cor}
\label{cor:general order optimality uncoded cache}
For  the $(\Ksf,\Nsf)$ D2D private caching system, Scheme~A in Theorem~\ref{thm:SchemeA} is order optimal to within a factor of $18$  under the constraint of uncoded cache placement and  user collusion.
\hfill $\square$ 
\end{cor}

%\dt{???DT: restate here that this cut-set-like result is ratter surprising???} 

\begin{rem}[Coded vs Uncoded Cache Placement]
\label{rem:uncoded vs coded}
 For the  non-private shared-link coded caching problem in~\cite{dvbt2fundamental},  by comparing the optimal coded caching scheme with uncoded cache placement in~\cite{exactrateuncoded} and the general converse bound in~\cite{yufactor2TIT2018}, it was proved that the gain of coded cache placement is at most $2$. Similarly, 
 for the  non-private D2D coded caching problem in~\cite{d2dcaching},  by comparing the   coded caching scheme with uncoded cache placement in~\cite{Yapard2djournal} and the general converse bound in~\cite{yufactor2TIT2018}, it was proved that the gain of coded cache placement is at most $4$.
 However, for the considered D2D private coded caching problem, by comparing the proposed converse bounds under uncoded cache placement and Scheme C (which is with coded cache placement), it is interesting to find that the gain of coded cache placement is not always within a constant gap. More precisely, let us focus on the two-user system and consider $\Msf=\frac{\Nsf+1}{2}$. By letting $y=\frac{1}{2}$ and $h=0$ in~\eqref{eq:K2 converse 1}, we have 
$
 \Rsf^{\star}_{\mathrm{u}} \geq \frac{\Nsf-1}{2}. %\label{eq:coded vs uncoded, uncoded}
$
By letting $t=2$  in~\eqref{eq:less subpacket scheme},   Scheme C achieves the memory-load pair $(\Msf,\Rsf_{\mathrm{C}})= \left(\frac{\Nsf+1}{2}, 1 \right)$. Hence, we have $\frac{\Rsf^{\star}_{\mathrm{u}}}{\Rsf_{\mathrm{C}}} \geq \frac{\Nsf-1}{2}$, which can be unbounded (in the sense that it can be made larger than any constant by choosing a sufficiently large $\Nsf$).
\hfill $\square$ 
\end{rem}

\subsection{Numerical evaluations}
\label{sec:num}
We conclude the overview of our main results with some numerical evaluations.
%We   provide numerical evaluations of the proposed achievable and converse bounds for the $(\Ksf,\Nsf,\Msf ) $ D2D caching system with private   demands. 
For the achievable schemes, we plot   Scheme~A in Theorem~\ref{thm:SchemeA}, Scheme~B in Theorem~\ref{thm:SchemeB} (for the two-user system), and Scheme~C in Remark~\ref{rem:subpacketization} (with coded cache placement).
We also plot the converse bound under uncoded cache placement in Theorem~\ref{thm:two user converse} for $\Ksf=2$ and the converse bound under uncoded cache placement and user collusion  in Theorem~\ref{thm:K user converse} for $\Ksf\geq 3$. 
For sake of comparison, we also plot the converse bound in~\cite{yufactor2TIT2018} and  the   converse bound under the constraint of uncoded cache placement in~\cite{indexcodingcaching2020} for shared-link caching without privacy. 

 In Fig.~\ref{fig:numerical 1a}, we consider the case where  $\Ksf=2$ and  $\Nsf=8$.  Here the converse bounds in~\cite{indexcodingcaching2020} and~\cite{yufactor2TIT2018} are the same.  It can be seen in Fig.~\ref{fig:numerical 1a} that, Scheme~B and the proposed converse bound meet for all memories except    $4.5 \leq \Msf\leq 6$. When $4<\Msf \leq 5.8$, Scheme~C, with coded cache placement, achieves a lower load than the converse bound under uncoded cache placement  in Theorem~\ref{thm:two user converse}. 

In Fig.~\ref{fig:numerical 1b}, we consider the case where  $\Ksf=10$ and  $\Nsf=40$.   It can be seen in Fig.~\ref{fig:numerical 1b} that compared  to the converse bound   in~\cite{indexcodingcaching2020}, the proposed converse bound is  tighter   when $\Msf$ is small. This is mainly because in the proposed converse bound we treat $\Ksf/2=5$ users as a powerful super-user, which loosens the converse bound when $\Msf$ grows. However, for the low memory size regime,  this strategy performs well and gives the order optimality result of Scheme~A, while the gap between  the converse bound   in~\cite{indexcodingcaching2020} and Scheme~A is not a constant. 
 Hence, combining the proposed converse bound and the converse bound  in~\cite{indexcodingcaching2020}, we can obtain the order optimality results of Scheme~A for any memory size.   
 
In Fig.~\ref{fig:numerical 1c}, we consider the case where  $\Ksf=40$ and  $\Nsf=10$.  It can be seen  that the multiplicative gap between  Scheme~A  and the converse bounds for non-private shared-link coded caching problem is to within a constant. In addition, Scheme A outperforms Scheme C for any $\Msf \in [0,\Nsf]$.

 \iffalse
\begin{figure}
    \centering
    \begin{subfigure}[t]{0.5\textwidth}
        \centering
        \includegraphics[scale=0.6]{revised_D2Dprivate_k2n8} 
        \caption{\small $\Ksf=2$, $\Nsf=8$.}
        \label{fig:numerical 1a}
    \end{subfigure}%
    ~ 
    \begin{subfigure}[t]{0.5\textwidth}
        \centering
        \includegraphics[scale=0.6]{revised_D2Dprivate_k10n40} 
        \caption{\small $\Ksf=10$, $\Nsf=40$.}
        \label{fig:numerical 1b}
    \end{subfigure}
    \\
        \begin{subfigure}[t]{0.5\textwidth}
        \centering
        \includegraphics[scale=0.6]{revised_D2Dprivate_k40n10} 
        \caption{\small $\Ksf=40$, $\Nsf=10$.}
        \label{fig:numerical 1c}
    \end{subfigure}
    \caption{\small The memory-load tradeoff for the D2D private caching system.}
    \label{fig:num}
\end{figure} 
 \fi
 
 \begin{figure}%[ht]
%\vspace{-2mm}
\centerline{\includegraphics[scale=0.6]{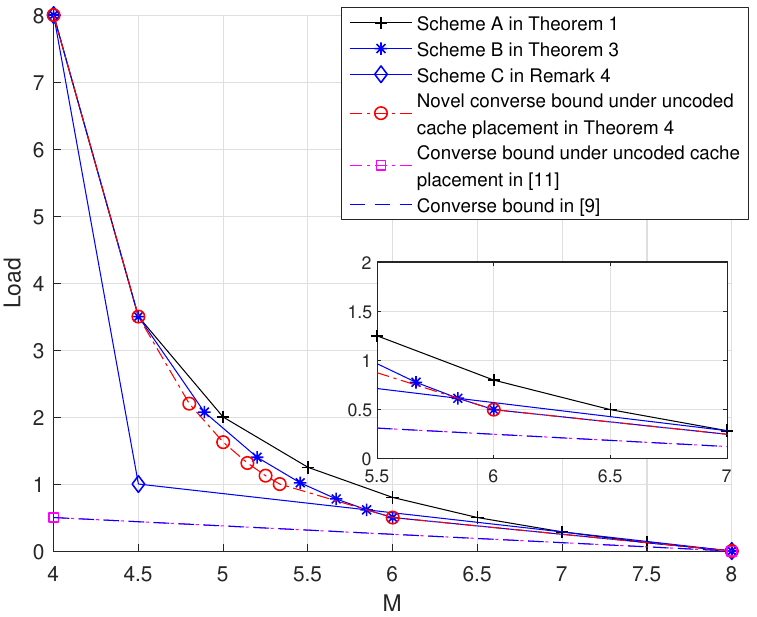}}
\caption{\small The memory-load tradeoff for the D2D private caching system, where $\Ksf=2$ and $\Nsf=8$.}
\label{fig:numerical 1a}
%\vspace{-5mm}
\end{figure}

 \begin{figure}%[ht]
%\vspace{-2mm}
\centerline{\includegraphics[scale=0.6]{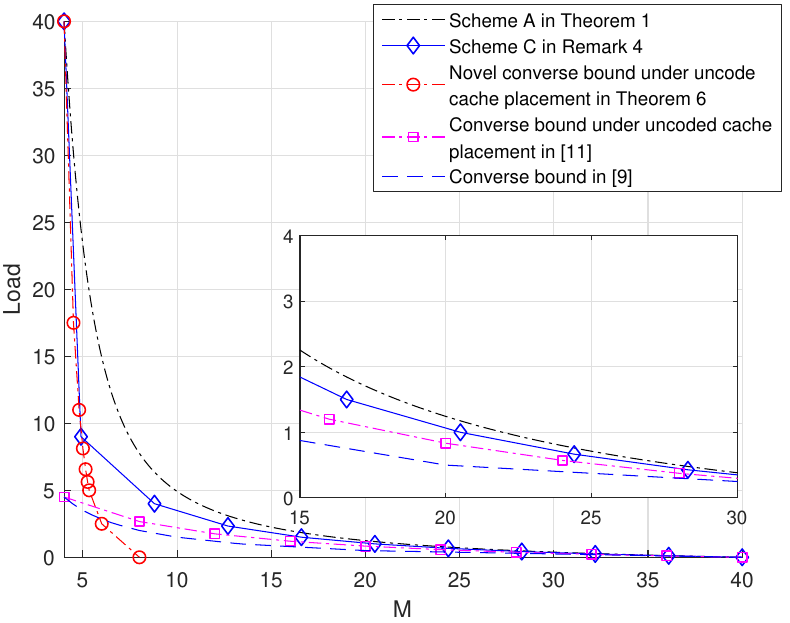}}
\caption{\small The memory-load tradeoff for the D2D private caching system, where $\Ksf=10$ and $\Nsf=40$.}
\label{fig:numerical 1b}
%\vspace{-5mm}
\end{figure}
  
   \begin{figure}%[ht]
%\vspace{-2mm}
\centerline{\includegraphics[scale=0.6]{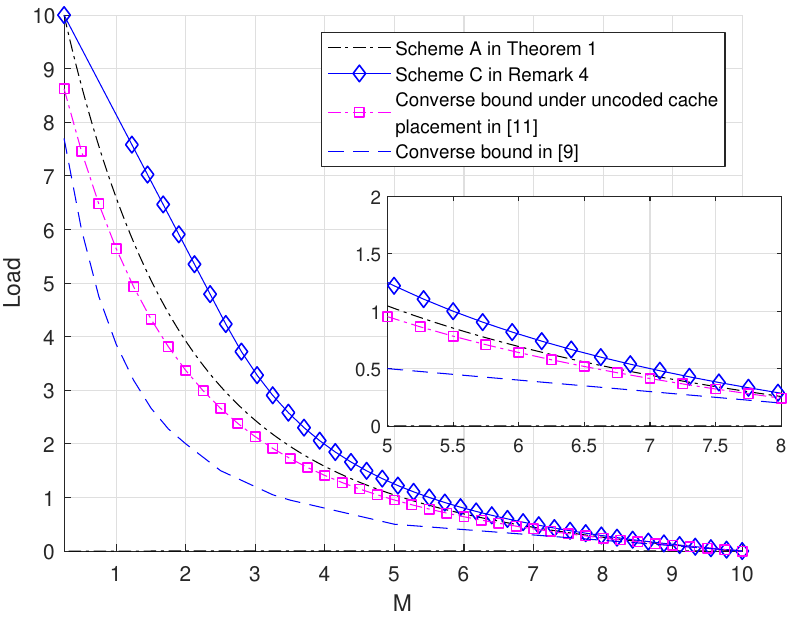}}
\caption{\small The memory-load tradeoff for the D2D private caching system, where $\Ksf=40$ and $\Nsf=10$.}
\label{fig:numerical 1c}
%\vspace{-5mm}
\end{figure}

\section{Achievable Schemes}%{D2D Caching Schemes with Private  Demands}
\label{sec:AchievableSchemes}
In this section we provide the details of the achievable schemes together with illustrative examples.

%\dt{!!!DT's note: moved the examples in this section; they were in the one above!!!}

%Note that in the rest of the paper, when we introduce achievable schemes, we directly provide the construction of the payloads and skip the description on their metadata
%\dt{!!!DT's: remove metadata, or rephrase as no more metadata in model definition!!!}

%%%%\begin{example}\rm
\subsection{Example of Scheme~A}\label{ex:example for Scheme A}
Before introducing Scheme~A in full generality, we present an example to illustrate the main idea for the D2D private system with $\Ksf=2$ users, $\Nsf=3$ files, and  $t=2$ (corresponding to cache size $\Msf=\frac{5}{2}$).

%there are $\Ksf(\Nsf-1)=4$ virtual users, in addition to the $\Ksf=2$ real users; 
 At a high level, we aim to create a ``virtual users''-system with a total $\Ksf\Nsf=6$ effective (i.e., real or virtual) users. We then effectively divide the ``virtual users''-system into $\Ksf=2$ independent shared-link models, in each of which a real user broadcasts coded multicast packets to $(\Ksf-1)\Nsf=3$ effective users (including $\Ksf-1=1$ real users and $(\Ksf-1)(\Nsf-1)=2$ virtual users). The demand vector of the effective users served on each independent shared-link model is such that each file is requested exactly $\Ksf-1=1$ times, thereby guaranteeing privacy.

%\dt{??DT's question: is Sk the local randomness in the model definition??}

{\it File Partitioning.}
Each file is partitioned into $6$ equal-length subfiles as
\begin{align}
  F_i=\{F^1_{i,\{4,5\}},F^1_{i,\{4,6\}},F^1_{i,\{5,6\}}, \ F^2_{i,\{1,2\}},F^2_{i,\{1,3\}}, F^2_{i,\{2,3\}} \}, 
  \label{eq:schemeA:ex:filesplit}
\end{align}
where $  i\in[3].$ 
Each subfile contains $\Bsf/6$ bits.  
The subfiles $(F^1_{i,\{4,5\}},F^1_{i,\{4,6\}},F^1_{i,\{5,6\}}: i\in [3])$ are to be delivered in the first independent shared-link model by real user $1$ to the effective users indexed by $[\Nsf+1:2\Nsf]=[4:6]$.
Similarly, the subfiles $(F^2_{i,\{1,2\}},F^2_{i,\{1,3\}},F^2_{i,\{2,3\}}: i\in [3])$ are to be delivered in the second independent shared-link model by real user $2$ to the effective users indexed by $[\Nsf]=[3]$.

{\it Placement Phase.}
Real user $1$ stores all the subfiles with superscript $1$ (which it is charged to deliver in the delivery phase), and similarly, real user $2$ must store all subfiles with superscript $2$. In addition, each real user also stores other sub-files as follows.
Real user $k\in[2]$ selects $P_k$ uniformly i.i.d. over $[3]$.
%and let real user $k$ ``pretend'' to be effective user  $\theta_k := \Nsf(k-1)+ P_k = 3(k-1)+ P_k$ among the $\Nsf\Ksf=6$ effective users. 
The realization of $P_1$ is unknown to real user $2$, and similarly $P_2$ is unknown to real user $1$. 
Real user $k\in [2]$  impersonates effective user  $\theta_k  = 3(k-1)+ P_k$. Thus, the actual cache content of each real user $k\in [2]$ is
\begin{align}
  Z_{k} = \{ F^k_{i,\Vc} : i\in[3], \forall \Vc \}
  \bigcup_{j\in[\Ksf]\setminus\{k\}} \{ F^j_{i,\Vc} : i\in[3], \theta_k \in\Vc \}. 
  \label{eq:schemeA:ex:Zreal}
\end{align}
For example, if we assume $P_1=1$ (real user $1$  impersonates  effective user $1$) and $P_2=1$ (real user $2$  impersonates   effective user $4$), then real users' cached contents are
\begin{align}
  &Z_1=(F^1_{i,\{4,5\}},F^1_{i,\{4,6\}},F^1_{i,\{5,6\}}, \ F^2_{i,\{1,2\}},F^2_{i,\{1,3\}}:i\in [3]),\label{eq:schemeA:ex:Z1}
\\&Z_2=(F^1_{i,\{4,5\}},F^1_{i,\{4,6\}}, \ F^2_{i,\{1,2\}},F^2_{i,\{1,3\}},F^2_{i,\{2,3\}}:i\in [3]),\label{eq:schemeA:ex:Z2}
\end{align}
each of $\Msf = 3\frac{5}{6}$ files.

Thus in the first shared-link model served by real user $1$ with the library $(F^1_{i,\{4,5\}},F^1_{i,\{4,6\}},F^1_{i,\{5,6\}}:i\in [3])$, each effective user $k\in [4:6]$ caches $( F^1_{i,\Vc} :  \Vc \in \{\{4,5\},\{4,6\}, \{5,6\}\}, k\in \Vc )$. In the second shared-link model served by real user $2$ with the library $(F^2_{i,\{1,2\}},F^2_{i,\{1,3\}},F^2_{i,\{2,3\}}:i\in [3])$, each effective user $k\in [3]$ caches $( F^2_{i,\Vc} :  \Vc \in \{\{1,2\},\{1,3\}, \{2,3\}\}, k\in \Vc )$.

\iffalse
For the delivery phase, for each $k\in [2]$, we consider that the cached content of 
we consider that the cached contents of the effective users is
\begin{align}
  \widetilde{Z}_{j} = \{ F^k_{i,\Vc} : k\in[2], i\in[3], j \in \Vc\}, \ j \in [6]. \dt{RIGHT?}
\end{align}
\fi

{\it Delivery Phase.}
In order to guarantee privacy, we want that each file is demanded the same number of times by the effective users served in each independent shared-link model. Therefore, we let real user $k\in[\Ksf]$, who wants to retrieve the file indexed by $d_k$, choose uniformly i.i.d. at random one permutation among all permutations of $[\Nsf]$ with $P_k$-th entry equal to $d_k$.

Assume that the demand vector is $(d_1,d_2)=(1,1)$.  Denote the demand of effective user $k$ by $q_k$.  
Real user $1$, who  impersonates  effective user $1$ with demand $q_1=1$,   randomly chooses $(q_2,q_3)$  to be either   $(2,3)$ or $(3,2)$, with equal probability. Real user $1$ sends $\ell_1=(q_1,q_2,q_3)$ as a query to real user $2$.
Similarly, real user $2$, who  impersonates    effective user $4$ with demand $q_4=1$,   randomly chooses $(q_5,q_6)$  to be either   $(2,3)$ or $(3,2)$, with equal probability. Real user $2$ sends $\ell_2=(q_4,q_5,q_6)$ as a query to real user $1$.
It can be seen that in each  independent shared-link model each file  is demanded exactly once.
%the union of the demanded files of effective users  in $[3]$ is $[3]$, while the union of the demanded files of effective users  in $[4:6]$ is also $[3]$. 

%By assuming $(q_5,q_6)=(2,3)$, 
Real user $1$ then sends  
\begin{align}
X_1=F^1_{q_4,\{5,6\}} \oplus F^1_{q_5,\{4,6\}} \oplus F^1_{q_6,\{4,5\}};\label{eq:schemeA:ex:X1}
\end{align} 
thus real user $2$, who   has cached $F^1_{q_5,\{4,6\}} , F^1_{q_6,\{4,5\}} $, can recover $F^1_{q_4,\{5,6\}}$.
%By assuming $(q_2,q_3)=(2,3)$, 
Similarity, real user $2$ then sends  
\begin{align}
X_2= F^2_{q_1,\{2,3\}} \oplus F^2_{q_2,\{1,3\}} \oplus F^2_{q_3,\{1,2\}};\label{eq:schemeA:ex:X2}
\end{align}
thus real user $1$, who   has cached  $F^2_{q_2,\{1,3\}}$ and $F^2_{q_3,\{1,2\}}$, can recover $F^2_{q_1,\{2,3\}}$.
%Thus real user $1$ who caches $F^2_{2,\{1,3\}} , F^2_{3,[2]} $, can recover $F^2_{1,\{2,3\}}$.

{\it Performance.}
In the delivery phase, the load is $2\frac{1}{6}$, which coincides with~\eqref{eq:extended scheme}.
%For the privacy constraint, let us focus on real user $1$. Intuitively,  each file is demanded by exactly $1$ effective user in $[4:6]$. 
Privacy is guaranteed as, from the viewpoint of real user $1$, who does not know the realization of $P_2$, all the effective users in $[4:6]$ are equivalent; similarly for real user $2$.  The information theoretic proof on the privacy will be provided later for the general case. 
Note that $|P_k|, |\ell_k|$ where $k\in [2]$ do not scale with $\Bsf$, satisfying our assumption in Section~\ref{sec:model}.
In conclusion, the proposed scheme is decodable and secure.

%%%\hfill$\square$
%%%\end{example}

\subsection{Proof of Theorem~\ref{thm:SchemeA}: Description of Scheme~A}
\label{sub:SchemeA} 
We are now ready to generalize the example in Section~\ref{ex:example for Scheme A}.

Recall that $\Usf= (\Ksf-1)\Nsf$ denotes the number of virtual users.  
%We focus on the memory size $\Msf=\frac{\Nsf+t-1}{\Ksf}$, where 
Let $t\in [\Usf+1]$. 
%We focus on the memory size $\frac{(\Ksf-1)(t-1)+\Usf }{\Ksf\Usf}\Nsf$, where $t\in [ \Usf]$.  
Similar  to the ``virtual users'' scheme for the shared-link model in~\cite{kamath2019demandpri}, we aim to contact a D2D system with $\Ksf(\Nsf-1)$ virtual users (in addition the $\Ksf$ real users) and divide it into $\Ksf$ independent shared-link models, each of which serves $\Usf$ effective users, where $(\Ksf-1)(\Nsf-1)$ are virtual users.

{\it File Partitioning.}
Each file is partitioned into $\Ksf \binom{\Usf}{t-1}$ equal-length subfiles as 
%(please note the `interval' for user k that is removed when the superscript is k)
\begin{align}
F_i = &\{ F_{i,\Vc}^{k} : k\in[\Ksf], \Vc\subseteq [\Ksf\Nsf] \setminus [(k-1)\Nsf+1:k\Nsf], \nonumber\\ & |\Vc|=t-1 \}, \
\forall i\in[\Nsf],
\label{eq:NEWsplit}
\end{align}
where each subfile contains  $\frac{\Bsf}{\Ksf \binom{\Usf}{t-1}}$ bits.
%each piece of size $\frac{B}{K \binom{(\Ksf-1)\Nsf}{t-1}}$ bits.
Note that, for each $k\in [\Ksf]$,  in~\eqref{eq:NEWsplit} we have eliminated the index interval  $[(k-1)\Nsf+1:k\Nsf]$, which is associated with real user $k$, from the set of all effective users $[\Ksf\Nsf]$. %This is the `trick' that allows us to solve the challenge identified in Section~\ref{sec:related:NEW}.

{\it Placement Phase.}
Each real user $k\in [\Ksf]$ selects $P_k\in[\Nsf]$ uniformly at random and independently across users.
We let real user $k\in [\Ksf]$  impersonate  effective user $\theta_k := (k-1)\Nsf+ P_k$ among the $\Ksf\Nsf$ effective users. 
The realization of $P_k, k\in [\Ksf],$ %is stored in the cache of user $k$ as the metadata of its cache content, and 
is unknown to   all the other real users, 
%In other words,  we divide the index set of the effective users  $[\Nsf\Ksf]$ into  index intervals, where each real user $k\in [\Ksf]$ belongs to one distinct index interval, which is  $[(k-1)\Nsf+1:k\Nsf]$. However, except real user $k$, 
that is, the other real users do not know the realization of $\theta_k \in [(k-1)\Nsf+1:k\Nsf]$.

%We let real user $k\in [\Ksf]$ cache all sub-files $F_{i,\Vc}^{k}$ such that $\Vc\subseteq [\Ksf\Nsf] \setminus [(k-1)\Nsf+1:k\Nsf]$; \dt{ real user $k$ will be delivering those sub-files in the independent shared-link model. In addition, we then let each real user $k\in [\Ksf]$ also cache all sub-files $F_{i,\Vc}^{j}$ where $\Vc\subseteq [\Ksf\Nsf] \setminus [(j-1)\Nsf+1:j\Nsf], \ \theta_{k} \in \Vc$ and $j\in[\Nsf]\setminus\{k\}$}.

%For each real user $k\in [\Ksf]$, we aim to generate the subfiles to be delivered in the $k^{\text{th}}$ shared-link model, in which real user $k$ broadcasts packets as the server and there are $\Ksf-1$ real user and $(\Ksf-1)(\Nsf-1)$ virtual users to  be served.  More precisely, there are in total $(\Ksf-1)(\Nsf-1)+\Ksf-1=\Usf$   effective users to be served by real user $k$, whose union set is  $[\Nsf\Ksf]\setminus [(k-1)\Nsf+1:k\Nsf]$.

Each real user $k\in [\Ksf]$ caches all sub-files $F_{i,\Vc}^{j}$ for which either $k=j$ or $\theta_k\in\Vc$, for all files $i\in[\Nsf]$, requiring
\begin{align}
\Msf 
&= \Nsf \frac{ \binom{\Usf}{t-1} +(\Ksf-1) \binom{\Usf-1}{t-2} }{ \Ksf \binom{\Usf}{t-1} }
 = \Nsf\frac{1+(t-1)/\Nsf}{\Ksf}.
\label{eq:NEWcache}
\end{align}
%satisfying the memory size constraint. 

{\it Delivery Phase.}
In the first step, each real user $k\in [\Ksf]$ who demands $d_k\in[\Nsf]$,  uniformly and independently selects  a  vector $ \ell_k =(q_{(k-1)\Nsf+1}, \ldots, q_{ k \Nsf } ) $  among all permutations of  $[\Nsf]$ %where the $q^{\text{th}}_{\theta_k}$ 
 whose $P_k$-th element equals $d_k$. Then real user $k\in [\Ksf]$ broadcasts  $\ell_k$ to all the other real users. 
Thus, from the viewpoint of each of the other real users,  the union of the demands of the effective users in  $ [(k-1)\Nsf+1:k\Nsf]$ is always $[\Nsf]$, which is key to guarantee privacy.

In the second step of the delivery phase, each  real user $k \in[\Ksf]$ performs a YMA delivery on the $k$-th  shared-link model with sub-files 
\begin{align}
(F_{i,\Vc}^{k} : i\in[\Nsf], \Vc\subseteq [\Ksf\Nsf] \setminus [(k-1)\Nsf+1:k\Nsf], |\Vc|=t-1 ),
\end{align}
 for effective users $[\Ksf\Nsf] \setminus [(k-1)\Nsf+1:k\Nsf]$.
More precisely, for each   $i\in[N]$, the effective user with the smallest index in $[\Ksf\Nsf] \setminus [(k-1)\Nsf+1:k\Nsf]$  which requires $F_i$ is chosen a leader for   $F_i$.
%we randomly and uniformly choose an effective user in $[\Ksf\Nsf] \setminus [(k-1)\Nsf+1:k\Nsf] $ demanding this file as a leader user. 
The leader set for the  $k $-th shared-link model is denoted by $\Lc_k$.
For each $\Sc \subseteq [\Ksf\Nsf] \setminus [(k-1)\Nsf+1:k\Nsf]$ where 
$|\Sc|=t$,   we let 
\begin{align}
 W^k_{\Sc}= \underset{j\in \Sc}{\oplus } F^{k}_{ q_{j}, \Sc  \setminus \{q_{j}\}}.
 \label{eq:Scheme A broadcast}
\end{align}
Then real user $k$ broadcasts 
\begin{align}
X_{k}=& \big( W^{k}_{\Sc}: \Sc \subseteq [\Ksf\Nsf] \setminus [(k-1)\Nsf+1:k\Nsf] ,  \nonumber\\& |\Sc|=t, \Sc \cap \Lc_k\neq \emptyset  \big).
\label{eq:Scheme A X_k}
\end{align}

{\it Decodability.}
We focus on  real user $k\in [\Ksf]$.
From $X_{j}$ where $j\in [\Ksf]\setminus \{k\}$, it was shown in~\cite[Lemma 1]{exactrateuncoded}, real user $k$ can reconstruct each multicast message   $W^{j}_{\Sc}$ where $\Sc \subseteq ([\Ksf\Nsf] \setminus [(j-1)\Nsf+1:j\Nsf])$ and $|\Sc|=t$.
Then real user $k$ can recover each $F^{j}_{d_k, \Vc}$ where $\Vc \subseteq ([\Ksf\Nsf] \setminus [(j-1)\Nsf+1:j\Nsf])$, $|\Vc|=t-1$, and $\theta_k \notin \Vc$ from
$W^{j}_{\Vc \cup \{\theta_k\} }$, since real user $k$ caches all the subfiles  in $W^{j}_{\Vc \cup \{\theta_k\} }$  except $F^{j}_{d_k, \Vc}$.
In conclusion, real user $k$ can recover all the uncached subfiles of $F_{d_k}$ from $(X_j: j\in [\Ksf]\setminus \{k\})$.

{\it Privacy.}
%The privacy constraints holds, by the similar reason as the   shared-link private caching scheme in~\cite{kamath2019demandpri}. 
We  will prove that the privacy constraint in~\eqref{eq:privacy} holds.\footnote{\label{foot:no need uniformly}Note that the privacy proof in~\cite{kamath2019demandpri} needs the constraint that the demand of each real user is uniformly i.i.d. over $[\Nsf]$. In the following, we will show that this condition is not necessary.}
%Let us first  focus on  real user $k\in [\Ksf]$.
By our construction, the cached content of each effective user  is fixed. Hence, $(X_1,\ldots, X_{\Ksf})$ only depends on the demands of the effective users. 
Since $P_{j}$, $j\in [\Ksf]$,  %\setminus \{k\}
    is chosen uniformly i.i.d over $[\Nsf]$,    $\theta_{j}$ is uniformly 
 i.i.d. over $[(j-1)\Nsf+1:j\Nsf]$. 
 Hence,  for any permutation of $[\Nsf]$ denoted by $\uv$, any $i\in [\Nsf]$, and any $(j,k)\in [\Ksf]^2$ where $j\not= k$,  (assume that  the $p$-th element of $\uv$ is $i$)
\begin{subequations}
\begin{align}
& \Pr \{ (q_{(j-1)\Nsf+1}, \ldots, q_{ j  \Nsf } ) =\uv | d_j=i,  d_k , Z_k  \} \nonumber \\
&= \Pr \{ (q_{(j-1)\Nsf+1}, \ldots, q_{ j  \Nsf } ) =\uv | d_j=i  \} \label{eq:indep of other users infor}\\
&=\Pr\{P_j= p | d_j =i\} \Pr\{(q_{(j-1)\Nsf+1}, \ldots, q_{p-1}, \nonumber\\& q_{p+1},\ldots, q_{ j  \Nsf }  | P_j= p, d_j =i\} \\
&=  \frac{1}{\Nsf}  \Pr\{(q_{(j-1)\Nsf+1}, \ldots, q_{p-1},q_{p+1},\ldots, q_{ j  \Nsf }  | P_j= p, d_j =i\} \label{eq:Sk uniform}\\
&=  \frac{1}{\Nsf} \frac{1}{(\Nsf-1)!}, \label{eq:other demand uniform}
\end{align}
\end{subequations}
where
~\eqref{eq:indep of other users infor} follows since, given $d_j$,  the demands of the effective users in $  [\Ksf\Nsf] \setminus [(j-1)\Nsf+1:j\Nsf] $ are independent of the cached content, queries, and demands of other effective users;
~\eqref{eq:Sk uniform} follows since  $P_j$ is chosen uniformly over $[\Nsf]$ independent of $d_j$; and
~\eqref{eq:other demand uniform} follows since, given $P_j$ and $d_j$, the demand vector of the   effective users in  $ [(j-1)\Nsf+1:j\Nsf]$ is  chosen uniformly among all permutations of $[\Nsf]$ where the $S_j$-th element is $d_j$.
From~\eqref{eq:other demand uniform}, it can be seen that $\Pr \{ (q_{(j-1)\Nsf+1}, \ldots, q_{ j  \Nsf } )  | d_j ,  d_k , Z_k   \} $ does not depend on $(d_j ,  d_k , Z_k  )$; thus 
\begin{align}
 I( q_{(j-1)\Nsf+1}, \ldots, q_{ j  \Nsf } ;d_j|  d_k , Z_k )=0.  \label{eq:j mutual}
 \end{align}
Hence, from~\eqref{eq:j mutual} and the fact that given $d_j$, the demands of the effective users in $  [\Ksf\Nsf] \setminus [(j-1)\Nsf+1:j\Nsf] $ are independent of the cached content, queries, and demands of other effective users, we have    
 \begin{align}
I( q_{1}, \ldots, q_{\Ksf \Nsf } ; \dv|  d_k , Z_k  )=0.  \label{eq:user k privacy}
 \end{align}
Recall that $(X_1,\ldots, X_{\Ksf})$ only depends on the demands of the effective users; thus we can prove~\eqref{eq:privacy}. 
Similarly, we can also prove the privacy constraint against colluding users in~\eqref{eq:colluding privacy}.

{\it Performance.}  
Each real user $k\in [\Ksf]$ broadcasts $\binom{\Usf}{t}-\binom{\Usf-\Nsf}{t}$ multicast  messages, each of which contains  $\frac{\Bsf}{\Ksf\binom{\Usf}{t-1}}$ bits. Hence, the achieved  load  is given by~\eqref{eq:extended scheme}. 
Note that $|P_k|, |\ell_k|$ where $k\in [\Ksf]$ do not scale with $\Bsf$, satisfying our assumption in Section~\ref{sec:model}.

%%%\begin{example}[$\Ksf=2$, $\Nsf=3$, and $\Msf=\frac{9}{4}$]
%%%\label{ex:example for Scheme B}
%%%\rm
\subsection{Example of Scheme~B}\label{ex:example for Scheme B}

We now focus on  the case of $\Ksf=2$ user  and propose a scheme that does not introduce virtual users and removes the redundancy in the placement phase of Scheme A.  Let us return to the example in Section~\ref{ex:example for Scheme A}  but with $\Msf=\frac{9}{4}$ to illustrate the key insights. % on the improvement.

Let us first go back to Scheme~A. Recall that in Scheme~A, each file is split as in~\eqref{eq:schemeA:ex:filesplit}, 
the cached contents of the real users are given by~\eqref{eq:schemeA:ex:Z1} and~\eqref{eq:schemeA:ex:Z2}.
and the transmitted signals are given by~\eqref{eq:schemeA:ex:X1} and~\eqref{eq:schemeA:ex:X2}.
%$F_i$, where $i\in [3]$, is partitioned into $6$ subfiles as 
%$$F_i=\{F^1_{i,\{4,5\}},F^1_{i,\{4,6\}},F^1_{i,\{5,6\}},F^2_{i,\{1,2\}},F^2_{i,\{1,3\}}, F^2_{i,\{2,3\}} \}.$$
%Real user $1$ caches $ F^1_{i,\{4,5\}},F^1_{i,\{4,6\}},F^1_{i,\{5,6\}},F^2_{i,\{1,2\}},F^2_{i,\{1,3\}} $, and real user~$2$ caches $F^1_{i,\{4,5\}},F^1_{i,\{4,6\}},$ $ F^2_{i,\{1,2\}},F^2_{i,\{1,3\}}, F^2_{i,\{2,3\}} $.
 Assume that the demand vector is $(d_1,d_2)=(1,1)$ and the queries are  $\ell_1=\ell_2=(1,2,3)$. Thus the transmitted signals are 
\begin{align}
&X_1=F^1_{1,\{5,6\}} \oplus F^1_{2,\{4,6\}} \oplus F^1_{3,\{4,5\}}, 
\\
&X_2=F^2_{1,\{2,3\}} \oplus F^2_{2,\{1,3\}} \oplus F^2_{3,\{1,2\}}. 
\end{align}
Note that real user~$2$  caches $(F^1_{2,\{4,5\}},F^1_{2,\{4,6\}})$ but only uses $F^1_{2,\{4,6\}}$ in the decoding procedure. 
Similarly, real user~$2$ caches  $(F^1_{3,\{4,5\}},F^1_{3,\{4,6\}})$ but only uses $ F^1_{3,\{4,5\}} $ in the decoding procedure. In other words, the cached subfiles $F^1_{2,\{4,5\}}$ and $F^1_{3,\{4,6\}}$ are redundant for user $2$. Similarly, the cached  subfiles $F^2_{2,\{1,2\}}$ and $F^2_{3,\{1,3\}}$ are redundant for user $1$.
The same is true for any demand vector. 

We propose Scheme~B to remove this cache redundancy as follows.

{\it File Partitioning.}
We partition each file into $4$ subfiles as 
\begin{align}
F_i=\{F^1_{i,1},F^1_{i,2}, F^2_{i,1},F^2_{i,2}  \}, \ i\in [3],
\end{align}
where each subfile contains $\Bsf/4$ bits.  

{\it Placement Phase.}
  User $1$ selects $P_1=(p_{1,2},p_{2,2},p_{3,2})$ uniformly i.i.d. over $[2]^3$; user $2$ selects $P_2=(p_{1,1},p_{2,1},p_{3,1})$ uniformly i.i.d. over $[2]^3$. Then user $1$
caches $Z_1=(F^1_{i,1},F^1_{i,2}, F^2_{i,p_{i,2}}:i\in [3])$, and user $2$ caches $Z_2=(F^1_{i,p_{i,1}} , F^2_{i,1},F^2_{i,2}:i\in [3])$.
Hence, $\Msf = \frac{9}{4}$ files. 

{\it Delivery Phase.}
In the delivery phase, we assume that the demand vector is $(d_1,d_2)=(1,1)$. 
  User $1$ sends query $\ell_1=(q,p_{2,2},p_{3,2})$ to user $2$, where  $q \in [2]\setminus\{p_{1,2}\}$. After receiving $\ell_1$, user $2$
 responds by transmitting
\begin{align}
X_2=F^2_{1,q} \oplus F^2_{2,p_{2,2}} \oplus F^2_{3,p_{3,2}} . 
\end{align} 
User $2$ sends query $\ell_2=(q^{\prime},p_{2,1},p_{3,1})$ to user $1$, where  $q^{\prime} \in [2]\setminus\{p_{1,1}\}$. After receiving $\ell_2$, user $1$  responds by transmitting
\begin{align}
X_1= F^1_{1,q^{\prime}} \oplus F^1_{2,p_{2,1}} \oplus F^1_{3,p_{3,1}}. 
\end{align}
The same can be done for any demand vector.

{\it Performance.}
Similar  to the analysis of Scheme~A, Scheme~B is decodable and private.
In this scheme,  $|P_k|, |\ell_k|$ where $k\in [2]$ do not scale with $\Bsf$ neither.
In this example, Scheme~B achieves the memory-load pair $\left(\frac{9}{4}, \frac{1}{2} \right)$.  
Scheme~A achieves the  memory-load pairs $\left(2, 1\right)$  for $t=1$, and $\left(\frac{5}{2}, \frac{1}{3} \right)$ for $t=2$; 
hence, by memory-sharing Scheme A achieves the load $\frac{2}{3}$ when $\Msf = \frac{9}{4}$.  
Therefore, Scheme~B outperforms Scheme~A.
%%%\hfill$\square$
%%%\end{example} 

\subsection{Proof of Theorem~\ref{thm:SchemeB}: Description of Scheme~B}
\label{sub:SchemeB} 
We now ready to provide the general description of Scheme B.

{\it Placement Phase.}
%We also use the  private placement precoding strategy   proposed in~\cite{wan2019privatecaching}. Fix an integer $t^{\prime} \in [0:\Nsf-1]$.
Each file $F_i$, where $i\in [\Nsf]$, is partitioned in two equal-length parts, denoted as $F_{i}= \{ F^1_{i} , F^2_i \}$ where $|F^1_{i}|=|F^2_{i}|=\Bsf/2$. For each $k\in [2]$, 
we further partition $F^k_i$ into $\binom{\Nsf-1}{t^{\prime}}+\binom{\Nsf-2}{t^{\prime}-1} $ equal-length subfiles, denoted by $F^k_{i,1},\ldots,F^k_{i, \binom{\Nsf-1}{t^{\prime}}+\binom{\Nsf-2}{t^{\prime}-1} }$, where each subfile has $\frac{\Bsf}{2 \left( \binom{\Nsf-1}{t^{\prime}}+\binom{\Nsf-2}{t^{\prime}-1}\right) }$ bits. 
We randomly generate a permutation of $\left[\binom{\Nsf-1}{t^{\prime}}+\binom{\Nsf-2}{t^{\prime}-1} \right] $, denoted by $\pv_{i,k}=\left(p_{i,k}[1],\ldots,p_{i,k}\left[ \binom{\Nsf-1}{t^{\prime}}+\binom{\Nsf-2}{t^{\prime}-1} \right] \right)$, independently and uniformly over the set of all possible permutations.  

  We let $P_1=(\pv_{i,2}:i\in [\Nsf])$ and $P_2=(\pv_{i,1}:i\in [\Nsf])$.  Then, we let user $k$ cache  all subfiles of $F^k_i$. In addition, we let the other user (i.e., the user in $[2]\setminus\{k\}$) cache $F^k_{i,p_{i,k}[1]},\ldots,  F^k_{i,p_{i,k}\left[\binom{\Nsf-2}{t^{\prime}-1} \right]}$.

Considering all the files,
each user in total caches $\left(\binom{\Nsf-1}{t^{\prime}}+2\binom{\Nsf-2}{t^{\prime}-1}\right)\Nsf$ subfiles, requiring memory 
\begin{align}
\Msf 
= \frac{\left(\binom{\Nsf-1}{t^{\prime}}+2\binom{\Nsf-2}{t^{\prime}-1}\right)\Nsf}{2 \left( \binom{\Nsf-1}{t^{\prime}}+\binom{\Nsf-2}{t^{\prime}-1}\right)}
%= \frac{\Nsf}{2} +\frac{  \binom{\Nsf-2}{t^{\prime}-1} \Nsf}{2 \left( \binom{\Nsf-1}{t^{\prime}}+\binom{\Nsf-2}{t^{\prime}-1}\right)} 
= \frac{\Nsf}{2} + \frac{\Nsf t^{\prime}}{2(\Nsf+t^{\prime}-1)}.
\end{align}

{\it Delivery Phase.}
%Consider any demand vector $ (d_1,\ldots,d_{\Ksf}) \in [\Nsf]^{\Ksf}$.
We first focus on the transmission by user~$1$, in charge of delivery the subfiles with superscirpt~1.
For each subset $\Sc \subseteq [\Nsf]$ where $|\Sc|=t^{\prime}+1$, we generate an XOR message containing exactly one subfile of each file in $\Sc$. 
More precisely, for each subset $\Sc \subseteq [\Nsf]$ where $|\Sc|=t^{\prime}+1$,
\begin{itemize}
\item If  $d_2 \in \Sc$, we pick a non-picked subfile among $F^1_{d_2,p_{d_2,1}\left[\binom{\Nsf-2}{t^{\prime}-1}+ 1 \right]},\ldots,  F^1_{d_2,p_{d_2,1}\left[\binom{\Nsf-1}{t^{\prime}}+\binom{\Nsf-2}{t^{\prime}-1} \right]}$. In addition, for each $i \in \Sc \setminus \{d_2\}$, we pick a non-picked subfile among $F^1_{i,p_{i,1}[1]},\ldots,  F^1_{i,p_{i,1}\left[\binom{\Nsf-2}{t^{\prime}-1} \right]}$.
\item If  $d_2 \notin \Sc$, for each $i\in \Sc$, we pick a non-picked subfile among $F^1_{i,p_{i,1}\left[\binom{\Nsf-2}{t^{\prime}-1}+ 1 \right]},\ldots,  F^1_{i ,p_{i,1}\left[\binom{\Nsf-1}{t^{\prime}}+\binom{\Nsf-2}{t^{\prime}-1} \right]}$.
\end{itemize}
We let $W^1_{\Sc}$ be the XOR of the picked $t^{\prime}+1$ subfiles, where $|W^1_{\Sc}|=  \frac{\Bsf}{2 \left( \binom{\Nsf-1}{t^{\prime}}+\binom{\Nsf-2}{t^{\prime}-1}\right) }$.

%Similarly, consider transmission by user~$2$. For each subset $\Sc \subseteq [\Nsf]$ where $|\Sc|=t^{\prime}+1$,   
%\begin{itemize}
%\item If  $d_1 \in \Sc$, we pick a non-picked subfile among $F^2_{d_1,p_{d_1,2}\left[\binom{\Nsf-2}{t^{\prime}-1}+ 1 \right]},\ldots,  F^2_{d_1,p_{d_1,2}\left[\binom{\Nsf-1}{t^{\prime}}+\binom{\Nsf-2}{t^{\prime}-1} \right]}$. In addition, for each $i \in \Sc \setminus \{d_1\}$, we pick a non-picked subfile among $F^2_{i,p_{i,2}[1]},\ldots,  F^2_{i,p_{i,2}\left[\binom{\Nsf-2}{t^{\prime}-1} \right]}$.
%\item If  $d_1 \notin \Sc$, for each $i\in \Sc$, we pick a non-picked subfile among $F^2_{i,p_{i,2}\left[\binom{\Nsf-2}{t^{\prime}-1}+ 1 \right]},\ldots,  F^2_{i ,p_{i,2}\left[\binom{\Nsf-1}{t^{\prime}}+\binom{\Nsf-2}{t^{\prime}-1} \right]}$.
%\end{itemize}
We proceed similarly for user~$2$. We let $W^2_{\Sc}$ be the binary sum of the picked $t^{\prime}+1$ subfiles, where $|W^2_{\Sc}|=  \frac{\Bsf}{2 \left( \binom{\Nsf-1}{t^{\prime}}+\binom{\Nsf-2}{t^{\prime}-1}\right) }$.

Finally,  user $1$ asks user $2$ to transmit $X_1=(W^1_{\Sc}:\Sc \subseteq [\Nsf], |\Sc|=t^{\prime}+1)$, and user $2$ asks user $1$ to transmit 
  $X_2=(W^2_{\Sc}:\Sc \subseteq [\Nsf], |\Sc|=t^{\prime}+1)$.\footnote{\label{foot:ell}In other words, the query $\ell_k$, $k\in [2]$, represents the indices of the subfiles in $W^k_{\Sc}$, where $\Sc \subseteq [\Nsf]$ and $|\Sc|=t^{\prime}+1$.}

{\it Decodability.}
We focus on user~$1$. In each message $W^2_{\Sc}$ where $\Sc \subseteq [\Nsf]$, $|\Sc|=t^{\prime}+1$,  and $d_1 \in \Sc$, user~$1$ caches all subfiles except one subfile from $F_{d_1}$, so user~$1$ can recover this subfile. Hence, user~$1$ in total recovers $ \binom{\Nsf-1}{t^{\prime}}$ uncached subfiles of $F_{d_1}$, and thus can recover $F_{d_1}$. Similarly, user~$2$ can also recover $F_{d_2}$.

{\it Privacy.}
Let us focus on user~$1$. Since user~$1$ does not know the random permutations generated in the placement phase, from its viewpoint, all subfiles in $F^1_i$ where $i\in [\Nsf]$ are equivalent.\footnote{\label{foot:equivalence}In our paper, the statement that from the viewpoint of a  user  $A$ and $B$ are equivalent, means that given the known information of this user, $A$ and $B$ are identically distributed.}   
$X_1$ contains $\binom{\Nsf}{t^{\prime}}$ messages, each of which corresponds to a different $(t^{\prime}+1)$-subset of $[\Nsf]$ and contains exactly one subfile of each file in the subset. 
Hence, the compositions of $X_1$ for different demands of user~$2$ are equivalent from the viewpoint of user~$1$. In addition, $X_2$ is generated independent of $d_2$, and thus $X_2$ cannot reveal any information of $d_2$. 
As a result, the demand of user~$2$ is private against user~$1$. Similarly, the demand of user~$1$ is private against user~$2$.

{\it Performance.} 
Each user broadcasts $\binom{\Nsf}{t^{\prime}+1}$    messages, each of which contains  $\frac{\Bsf}{2 \left( \binom{\Nsf-1}{t^{\prime}}+\binom{\Nsf-2}{t^{\prime}-1}\right) }$ bits. Hence, the achieved  load  is
\begin{align}
\Rsf = \frac{2 \binom{\Nsf}{t^{\prime}+1} }{2 \left( \binom{\Nsf-1}{t^{\prime}}+\binom{\Nsf-2}{t^{\prime}-1}\right) } 
= \frac{\Nsf(\Nsf-1)}{(t^{\prime}+1)(\Nsf+t^{\prime}-1)}.
\end{align}

Note that $|P_k|, |\ell_k|$ where $k\in [2]$ do not scale with $\Bsf$, satisfying our assumption in Section~\ref{sec:model}.
%coinciding with~\eqref{eq:second scheme}.

\section{New Converse Bounds Under the Constraint of Uncoded Cache Placement and User Collusion}  
\label{sec:converse}

In this section, we provide the proofs of our new converse bounds in Theorems~\ref{thm:two user converse} and~\ref{thm:K user converse}. 
We first introduce the proposed converse bound for the two-user system and  then extend it to the $\Ksf$-user system. 
We start by introducing an example to illustrate in the simplest possible case the new ideas needed to derive our new converse bound.

\subsection{Example of converse} 
%\begin{example}[D2D private caching system with $(\Ksf,\Nsf,\Msf)=(2,2,6/5)$]
%\rm
\label{ex:converse K2N2}

We consider the D2D private system with $(\Ksf,\Nsf)=(2,2)$ and $\Msf=6/5$, for which the achieved load by both Scheme~A and Scheme~B is $\Rsf=7/5$. The converse bound under the constraint of uncoded cache placement and one-shot delivery for D2D caching without privacy  in~\cite{Yapard2djournal} gives $\Rsf^{\star}_{\mathrm{u}}(6/5)\geq4/5$.\footnote{\label{foot:one shot 2}For $\Ksf=2$, any D2D caching scheme is one-shot.} 
In the following, we prove that $\Rsf^{\star}_{\mathrm{u}}(6/5)=7/5$.
%For the sake of simple illustration, we assume that each user uses the same cache size to cache each file, and that the cache of each user is full. In other words, each user caches $\Msf \Bsf/\Nsf=3 \Bsf / 5$ bits of each file. It will be explained in Remark~\ref{rem:wlog} that these two assumptions can be relaxed. %are also 

Assume we have a working system, that is, a system where all encoding, decoding and privacy constraints listed in Section~\ref{sec:model} are met.
%{\blue Notice that in~\eqref{eq:K2N2 privacy}, when we derive the converse bound on $H(X_1)$, we omit the metadatas of cache configurations, whose lengths are negligible compared to the file size. In the rest of this paper, when we derive the converse bounds on the numbers of transmitted bits, to avoid heavy notation, we also omit the metadatas, since they yield terms $o(\Bsf)$, which will disappear when normalized by the file size $\Bsf \to \infty$.}
%\dt{!!!DT: WE NO LONGER HAVE METADATA IN THE MODEL; REMOVE THIS With a slight abuse of notation, a set operation over cache configurations is meant to represents the set operation over the cached information bits only, i.e., excluding metadata.  In addition, each notation of a set or a vector of bits also includes the metadata for these bits. !!!}
 In the following, in order not to clutter the derivation with unnecessary ``epsilons'', we shall neglect the terms $P_k,\ell_k$ where $k\in [\Ksf]$ that contribute $\epsilon_\Bsf = o(\Bsf)$ when $\Bsf\to\infty$ to bounds like the one in~\eqref{eq:K2N2 privacy}.  
%In the following, in order not to clutter the derivation with unnecessary `epsilons and deltas', we shall neglect the terms (such as metadata, queries, etc.) that contribute $\epsilon_\Bsf = o(\Bsf)$ when $\Bsf\to\infty$ to a bounds like the one in~\eqref{eq:K2N2 privacy}. 
Finally,  without loss of generality  (see Remark~\ref{rem:wlog}), each user caches a fraction $\Msf/\Nsf = 3/5$ of each file
and each bit in the library is cached by at least one user.

Assume that the cache configurations of the two users  are $Z^1_1$ and $Z^1_2$, where  $Z^1_1 \cup Z^1_2=\{F_1,F_2\}$.
For the demand vector $(d_1,d_2)=(1,1)$, any working scheme must produce transmitted signals $(X_1,X_2)$ such that the demand vector $(d_1,d_2)=(1,1)$ can be satisfied. 
The following observation is critical:  because of the privacy constraint, % in~\eqref{eq:privacy}
from the viewpoint of user~$1$, there must exist a cache configuration of user~$2$, denoted by $Z^2_2$, such that $Z^1_1 \cup Z^2_2 =\{F_1,F_2\}$,  $H(X_2|Z^2_2)=0$, and $F_2$ can be decoded from $(X_1,Z^2_2)$. If such a cache configuration $Z^2_2$ did not exist, then user~$1$ would know that the demand of user~$2$ is $F_1$ from $(Z^1_1,X_1,X_2,d_1)$, which is impossible in a working private system. Similarly, from the viewpoint of user~$2$, there must exist a cache configuration of user~$1$, denoted by $Z^2_1$, such that $Z^2_1 \cup Z^1_2 =\{F_1,F_2\}$,  $H(X_1|Z^2_1)=0$, and $F_2$ can be decoded from $(X_2,Z^2_1 )$.

%Recall that each user caches $3 \Bsf / 5$ of the bits of each file, and 
From $(Z^1_1,Z^1_2)$, because of Remark~\ref{rem:wlog}, for each file $F_i, \ i\in[2]$, we have\footnote{Intuitively, with uncoded cache placement, each file is split into disjoint pieces as $F_i=(F_{i,\{1\}},F_{i,\{2\}},F_{i,\{1,2\}}), i\in[2],$ and the users cache $Z_1=\cup_{i=1}^{2}(F_{i,\{1\}}, F_{i,\{1,2\}}), Z_2=\cup_{i=1}^{2}( F_{i,\{2\}},  F_{i,\{1,2\}})$; by symmetry, let $x\in[0,1]$ with $|F_{i,\{1\}}|=|F_{i,\{2\}}|=\Bsf x/2$ and $|F_{i,\{1,2\}}|=\Bsf (1-x)$ such that $x/2 + 1-x = \Msf/\Nsf=3/5 \to x=2(1-\Msf/\Nsf)= 4/5.$ In the proof, one can think of different cache configurations as different ways to split the  files. }  
\begin{subequations}
\begin{align}
&|F_i \cap  Z^1_1| %\dt{=|F_{i,2}^1|}
=\frac{\Bsf\Msf}{\Nsf}=\frac{3\Bsf}{5},
\\
&|F_i \setminus  Z^1_1| %\dt{=|F_{i,2}^1|}
=|F_i \setminus  Z^1_2| %\dt{=|F_{i,1}^1|}
=\Bsf- \frac{3\Bsf}{5}
=\frac{2\Bsf}{5},
\\
&|F_i \cap Z^1_1 \cap Z^1_2| %\dt{=|F_{i,12}^1|}
=\frac{\Bsf}{5}.
\end{align} 
%\end{subequations}
Similarly, since $Z^1_1\cup Z^1_2=Z^1_1\cup Z^2_2=\{F_1,F_2\}$, we also must have 
%\begin{subequations}
\begin{align}
%&|F_i \setminus  Z^1_1| %\dt{=|F_{i,2}^1|}
%=|F_i \setminus  Z^2_2| %\dt{=|F_{i,1}^2|}
%=\frac{2\Bsf}{5},
%\\
&|F_i \cap Z^1_1 \cap Z^2_2| %\dt{=| \underbrace{(F_{i,1}^1 \cup F_{i,12}^1)}_{has F_{i,1}^2} \cap \underbrace{(F_{i,2}^2 \cup F_{i,12}^2)}_{has F_{i,2}^1} | }
= \frac{\Bsf}{5},
\label{eq:example intersection}
\\
&F_i \setminus Z^1_1 %\dt{=F_{i,2}^1}
\subseteq  F_i \cap  Z^1_2 \cap Z^2_2. %\dt{= (F_{i,2}^1 \cup F_{i,12}^1) \cap Z^2_2 = F_{i,2}^1 \cup (F_{i,12}^1 \cap Z^2_2)}
\label{eq:contain complement}
\end{align}
\end{subequations}

%Recall that $F_1$ can be reconstructed from $(X_1,Z^1_2)$, and $F_2$ can be reconstructed from $(X_1,Z^2_2)$.

Inspired by the genie-aided converse bound for shared-link caching networks without privacy in~\cite{indexcodingcaching2020,exactrateuncoded}, we construct a genie-aided super-user with cached content
\begin{align}
Z^{\prime} =\big(Z^1_2, \  Z^2_2\setminus (F_1\cup Z^1_2) \big),
\end{align}
who is able to recover the whole library from $(X_1,Z^{\prime})$.
Indeed, after file $F_1$ is reconstructed from $(X_1,Z^1_2)$, the combination of $(F_1\cup Z^1_2)$ and $Z^2_2\setminus (F_1\cup Z^1_2)$ gives $Z^2_2$;
now, file $F_2$ can be reconstructed from $(X_1,Z^2_2)$.
Therefore, we have
\begin{subequations}
\begin{align}
2 \Bsf 
&=H(F_1,F_2) \leq  H\big(X_1,Z^{\prime} \big) \\
&= H\big(X_1,Z^1_2, Z^2_2\setminus (F_1\cup Z^1_2)  \big) \\
&=H\big(X_1,Z^1_2 \big) + H\big(Z^2_2\setminus (F_1\cup Z^1_2) | X_1,Z^1_2, F_1\big) \\  
&\leq H(X_1)+ H(Z^1_2) +H\big(Z^2_2|  Z^1_2 , F_1 \big) \label{eq:K2N2 privacy step1}  \\
&=H(X_1)+ H(Z^1_2)  +H (F_2 \cap Z^2_2 \cap  Z^1_1   |  Z^1_2  ) \label{eq:K2N2 privacy step4}  \\
&=H(X_1)
+ \underbrace{H(Z^1_2)}_{\leq \Msf \Bsf}
+ \underbrace{H (F_2 \cap Z^2_2 \cap  Z^1_1 )}_{\leq \Bsf/5} \nonumber\\& 
- H ( \underbrace{F_2 \cap Z^2_2 \cap  Z^1_1  \cap Z^1_2}_{:= \Qc}  ), \label{eq:K2N2 privacy step5} 
\end{align}
\label{eq:K2N2 privacy}
\end{subequations}
where~\eqref{eq:K2N2 privacy step4} follows since, from~\eqref{eq:K2N2 privacy step1}, only the bits in $F_2$ are left, and   $Z^2_2 \setminus  Z^1_2 = (Z^2_2 \cap  Z^1_1) \setminus Z^1_2$ following the reasoning leading to~\eqref{eq:contain complement}; the last step in~\eqref{eq:K2N2 privacy step5} follows since the bits in a file are independent.

\begin{figure}%[ht]
%\vspace{-2mm}
\centerline{\includegraphics[scale=0.32]{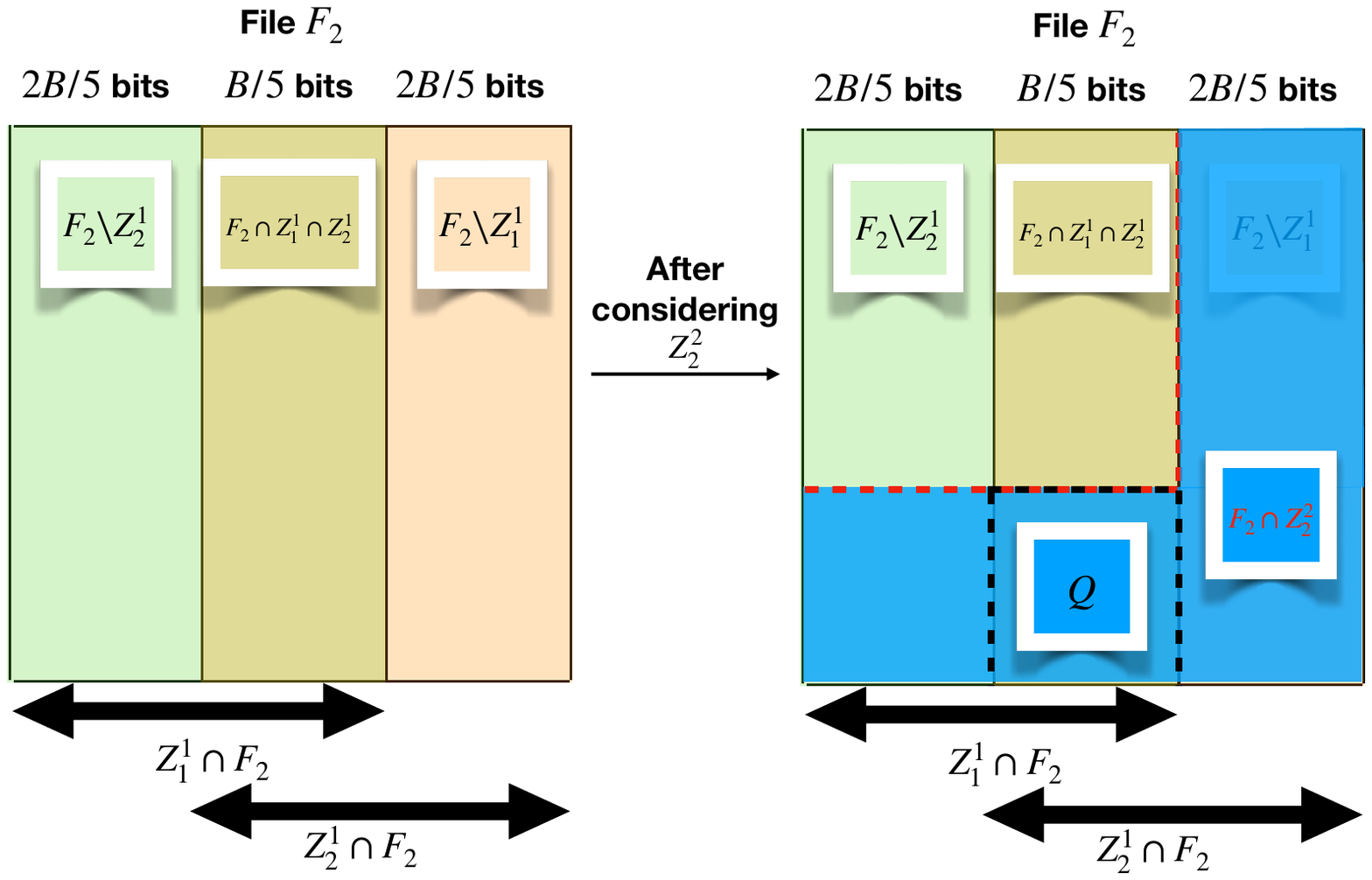}}
\caption{\small Illustration of the composition of $\Qc := F_2 \cap  Z^1_1  \cap Z^1_2  \cap  Z^2_2$.}
\label{fig:illustrationQ}
%\vspace{-5mm}
\end{figure}

At this point, we need a bound that can be combined with the one in~\eqref{eq:K2N2 privacy} such that it contains on the right hand side the term $H(X_2)$, so that  $H(X_1)+H(X_2)$ can be bounded by $\Bsf \Rsf_{\mathrm{u}}$, and a term that allows one to get rid of the negative entropy of the random variable 
\begin{align}
\Qc := F_2 \cap  Z^1_1  \cap Z^1_2  \cap  Z^2_2,  
\label{eq:def of Q in the example DT}
\end{align}
which is illustrated in Fig.~\ref{fig:illustrationQ}.

In the next step, we will introduce another approach to construct a genie-aided  super-user, in order to derive an inequality eliminating $\Qc$ in~\eqref{eq:K2N2 privacy step5}. 
We then focus on the cache configurations $Z^1_1$ and $Z^2_1$, and the transmitted packets $X_2$. 
Recall that $F_1$ can be reconstructed from  $(Z^1_1,X_2)$, and $F_2$ can be reconstructed from  $(Z^2_1,X_2)$.
Furthermore, by recalling the definition of $\Qc$ in~\eqref{eq:def of Q in the example DT}, it can be seen that the bits in $(F_2 \cap Z^1_1) \setminus \Qc$ are independent of $X_2$. Thus $F_1$ can be reconstructed from  $(Z^1_1\cap F_1, \Qc, X_2)$. 
Hence, we can construct a super-user with cached content 
\begin{align}
Z^{\prime\prime}=(Z^1_1\cap F_1, Z^2_1\cap F_2, \Qc),
\end{align}
who can decode both files. Thus
%%%This virtual user can first reconstruct $F_1$ from  $(Z^1_1\cap F_1, \Qc, X_2)$. 
%%%Then it can   reconstruct $F_2$ from $(F_1, Z^2_1\cap F_2,X_2)$. In other words,  
\begin{subequations}
\begin{align}
2\Bsf &= H(F_1,F_2) \leq H(X_2, Z^{\prime\prime})
\\&\leq H(X_2) + \underbrace{H(Z^1_1\cap F_1)}_{\leq 3\Bsf/5}  + \underbrace{H(Z^2_1\cap F_2)}_{\leq 3\Bsf/5} +  H(\Qc) . \label{eq:K2N2 from index coding DT step2}
\end{align}
\label{eq:K2N2 from index coding DT}
\end{subequations}
%%%Hence, we have 
%%%\begin{align}
%%%H(X_2) &\geq H(F_1,F_2)-H(Z^{\prime})
%%%\\& \geq H(F_1,F_2)-H(Z^1_1\cap F_1)-H(\Qc)-H(Z^2_1\cap F_2)
%%%\\&= \frac{4\Bsf}{5}-q.
%%%\label{eq:K2N2 from index coding}
%%%\end{align}
Finally, by summing~\eqref{eq:K2N2 privacy step5} and~\eqref{eq:K2N2 from index coding DT step2}, we have that any achievable rate under uncoded cache placement must satisfy
\begin{align}
\Rsf_{\mathrm{u}}\geq \frac{H(X_1)+H(X_2)}{\Bsf}\geq  \frac{7 }{5}.
\label{eq:K2N2 final result}
\end{align}
The bound in~\eqref{eq:K2N2 final result} shows that Scheme~A and Scheme~B are indeed optimal for the considered memory point.

\begin{rem}[A high-level explanation of the converse technique]
The key take-away points in the example in Section~\ref{ex:converse K2N2} are as follows.
\begin{itemize}

\item By exploiting the privacy constraints, we note that from the viewpoint of each user $k$ (i.e., given cache $Z_k$ and transmitted packets $(X_1,X_2)$), any demand of the other user is equally possible.  Hence, there must exist a cache configuration of the other user that allow for the decoding of any file using the same $(X_1,X_2)$. 

\item 
We introduce an auxiliary random variable $\Qc$ to represents the set of  bits  $ F_2 \cap  Z^1_1  \cap Z^1_2  \cap  Z^2_2 $.  We then use two different approaches to construct genie-aided super-users to decode the whole library, in such a way that we can get rid of ``tricky'' entropy term when the various bounds are summed together:
\begin{enumerate}

\item In the first approach, we focus on $(X_1, Z^1_2, Z^2_2)$ and construct a genie-aided super-user 
%with cache $\big( Z^1_2,    Z^2_2\setminus (F_1\cup Z^1_2) \big)$, 
who can reconstruct the whole library by receiving $X_1$.
The bits in $\Qc$ belong to the overlap of $Z^1_2$ and $Z^2_2$. 
%  These $|\Qc|$ bits appearing in $X_2$ should be contained by both  $Z^1_2$ and  $Z^2_2$, and thus should belong to the common bits of $Z^1_2$ and  $Z^2_2$. 
Hence, the size of the genie-aided super-user's cache decreases when $|\Qc|$ increases. 
In other words, the needed transmitted load increases when $|\Qc|$ increases (see~\eqref{eq:K2N2 privacy step5}). 

\item In the second approach, we focus on $(X_2, Z^1_1, Z^2_1)$ and construct a genie-aided super-user 
%with cache $ ( F_1 \cap Z^1_1, \Qc,   F_2\cap Z^2_1    )$, 
who can  reconstruct the whole library by receiving $X_2$. 
Now the bits in $\Qc$   are in the cache  of the super-user. 
Hence, the size of the genie-aided super-user's cache increases when $|\Qc|$ increases. In other words, the needed transmitted load decreases when $|\Qc|$ increases (see~\eqref{eq:K2N2 from index coding DT step2}). 

\end{enumerate} 
Finally, by summing~\eqref{eq:K2N2 privacy step5} and~\eqref{eq:K2N2 from index coding DT step2}, the effect of $\Qc$ is fully cancelled,  such that we derive~\eqref{eq:K2N2 final result}.
\end{itemize}
\hfill $\square$ 
\end{rem}

\begin{rem}[On Optimality of Symmetric Placement]
\label{rem:wlog}
To derive the converse bound under the constraint of uncoded cache placement in the above example, we assumed that every user caches a fraction $\Msf/\Nsf$ of each file.  This assumption is without loss of generality. Assume that there exists a caching scheme where users %different memory sizes to cache files.
 cache different fraction of the files.  By taking a permutation of $[\Nsf]$ and by using the same strategy to fill the users' caches, we can get another caching scheme. By symmetry, these two caching schemes have the same load. 
Hence, by considering all possible permutations and taking memory-sharing among all such cache schemes, we have constructed a scheme where every user caches the same fraction of each file, with the same achieved load as the original caching scheme.

In addition, in the example, we also assumed the total number of cached bits by each user is exactly $\Msf \Bsf$, i.e., the cache of each user is full. 
Assume that the total number of cached bits by user $k$ is $\Msf_k \Bsf$.
By reasoning as above, we can prove that for any caching scheme, there must exist a caching scheme where $\Msf_1=\cdots=\Msf_{\Ksf}$ and with the same load as the above scheme.  Furthermore, the converse bounds in Theorem~\ref{thm:two user converse} and Theorem~\ref{thm:K user converse}  derived under the assumption that $\Msf_1=\cdots=\Msf_{\Ksf}=\Msf$, are non-increasing with the increase of $\Msf$. Hence, the assumption that the total number of cached bits by each user is exactly $\Msf \Bsf$ bits, is also without loss of generality. 

Hence, in the proof of our new converse bounds, without loss of generality, we can assume each uses caches a fraction $\frac{\Msf}{\Nsf}$ of each file. 
\hfill $\square$ 
\end{rem}

\subsection{Proof of Theorem~\ref{thm:two user converse}: Two-user system}
\label{sub:two user converse}

We focus on uncoded cache placement.  
Without loss of generality, each uses caches a fraction $\frac{\Msf}{\Nsf}$ of each file (as explained in Remark~\ref{rem:wlog}). 
Let 
\begin{align}
\Msf=\frac{\Nsf}{2}+y,
\end{align}
where $y\in \left[0, \frac{\Nsf}{2}\right]$.

Assume the cache configurations of the two users are $(Z^1_1,Z^1_2)$, where $Z^1_1 \cup Z^1_2=\{F_1,\ldots,F_{\Nsf}\}$.
For the demand vector  $(d_1,d_2)=(1,1)$,
any achievable scheme must produce transmitted packets $(X_1,X_2)$,
such that the demand vector  $(d_1,d_2)=(1,1)$ can be satisfied. 
  By the privacy constraint in~\eqref{eq:privacy}, %by the same reasoning used in Example~\ref{ex:converse K2N2}, 
from the  viewpoint of user $1$ with cache configuration $Z^1_1$,  there must exist   some cache configuration $Z^j_{2}$ such that  $Z^1_1\cup Z^j_{2}=\{F_1,\ldots,F_{\Nsf}\}$, $H(X_2|Z^{j}_{2})=0$, and $H(F_{ j}|X_1,Z^{j}_{2})=0$, for any $j\in [\Nsf]$; otherwise,    user $1$  will know that the demand of user $2$ is not $F_j$.  Similarly,
we have the following lemmas.
\begin{lem}
 \label{lem:tree demand}
For any  $i\in [\Nsf]$ and  $j\in [\Nsf]$, there must exist some cache configurations  $Z^{i}_{1}$   and $Z^j_{2}$, such that 
\begin{subequations} 
\begin{align}
&Z^{i}_{1}\cup Z^1_2=Z^1_1\cup Z^j_{2}=\{F_1,\ldots,F_{\Nsf}\}; \label{eq:lib tree}\\
&H(X_1|Z^{i}_{1})=H(X_2|Z^{j}_{2})=0; \label{eq:transmit tree}\\
&H(F_{ i}|X_2,Z^{i}_{1})=H(F_{ j}|X_1,Z^{j}_{2})=0.\label{eq:decode tree}
\end{align} 
\end{subequations} 
\end{lem}
%%%\begin{IEEEproof}
%%%Obviously, when $i=j=1$, Lemma~\ref{lem:tree demand} holds because of the definition of  $(Z^1_1,Z^1_2,X_1,X_2)$.
%%%
%%% By the privacy constraint in~\eqref{eq:privacy}, from the viewpoint of user~$1$ (whose cache is $Z^1_1$), 
%%%for any $j\in [2:\Nsf]$, 
%%% there must be a cache configuration of user~$2$, denoted by $Z^j_2$, such that $Z^1_1 \cup Z^j_2=\{F_1,\ldots,F_{\Nsf}\}$,  $ X_2 $ can be transmitted from $Z^j_2$ (i.e., $  H(X_2|Z^{j}_{2},\mathscr{M}(P_2))=0$), and $F_j$ can be decoded from $(X_1,Z^j_2)$ (i.e., $ H(F_{ j}|X_1,Z^{j}_{2})=0$).  
%%%  
%%% Similarly,  from the viewpoint of user~$2$ (whose cache is $Z^1_2$), 
%%%for any $i\in [2:\Nsf]$, 
%%%there must be a cache configuration of user~$1$, denoted by $Z^i_1$, such that $Z^i_1 \cup Z^1_2=\{F_1,\ldots,F_{\Nsf}\}$,  $ X_1 $ can be transmitted from $Z^i_1$ (i.e., $  H(X_1|Z^{i}_{1},\mathscr{M}(P_1))=0$), and $F_i$ can be decoded from $(X_2,Z^i_1)$ (i.e., $ H(F_{i}|X_2,Z^{i}_{1})=0$).  
%%%  
%%%Hence, Lemma~\ref{lem:tree demand} is proved. 
%%%\end{IEEEproof} 

%After choosing $Z^{i}_{1}$   and $Z^j_{2}$ where $i,j \in [\Nsf]$ by Lemma~\ref{lem:tree demand}, using the similar proof as Lemma~\ref{lem:tree demand}, we obtain the following lemma.
\begin{lem}
\label{lem:tree demand 2}
From $Z^{i}_{1}$ and $Z^j_{2}$ where $i,j \in [\Nsf]$ as in Lemma~\ref{lem:tree demand}, it must hold
\begin{itemize}

\item consider  $Z^{i}_{1}$ where $i\in[\Nsf]$. For any $j^{\prime}\in [\Nsf]$, there must exist a cache configuration denoted by $Z^{(i ,j^{\prime})}_{2}$ such that  $Z^{i}_{1}\cup Z^{(i,j^{\prime})}_{2}=\{F_1,\ldots,F_{\Nsf}\}$,  $ H(X_2|Z^{(i,j^{\prime})}_{2})=0$, and $H(F_{j^{\prime}}|X_1,Z^{(i,j^{\prime})}_{2})=0$; and

\item consider  $Z^j_{2}$ where $j\in[\Nsf]$. For any $i^{\prime}\in [\Nsf]$, there must exist a cache configuration denoted by $Z^{(i^{\prime},j)}_{1}$ such that  $Z^{(i^{\prime},j)}_{1}\cup Z^j_{2}=\{F_1,\ldots,F_{\Nsf}\}$,  $ H(X_1|Z^{(i^{\prime},j)}_{1})=0$, and $H(F_{i^{\prime}}|X_2,Z^{(i^{\prime},j)}_{1})=0$.

\end{itemize}
In addition,   by definition of  Lemma~\ref{lem:tree demand},   % the following construction is possible,
\begin{itemize}

\item
when $i=1$,   we have $Z^{(1,j^{\prime})}_{2}=Z^{j^{\prime}}_{2}$ for each $j^{\prime}\in [\Nsf]$; when $j=1$, we have $Z^{(i^{\prime},1)}_{1}=Z^{i^{\prime}}_{1}$ for each $i^{\prime}\in [\Nsf]$; and

\item
when $j^{\prime}=1$,    we have $Z^{(i,1)}_{2}=Z^{1}_{2}$ for each $i\in [\Nsf]$; when $i^{\prime}=1$, we have $Z^{(1,j)}_{1}=Z^{1}_{1}$  for each $j\in [\Nsf]$.

\end{itemize}
\end{lem}

\begin{figure}%[ht]
%\vspace{-2mm}
\centerline{\includegraphics[scale=0.2]{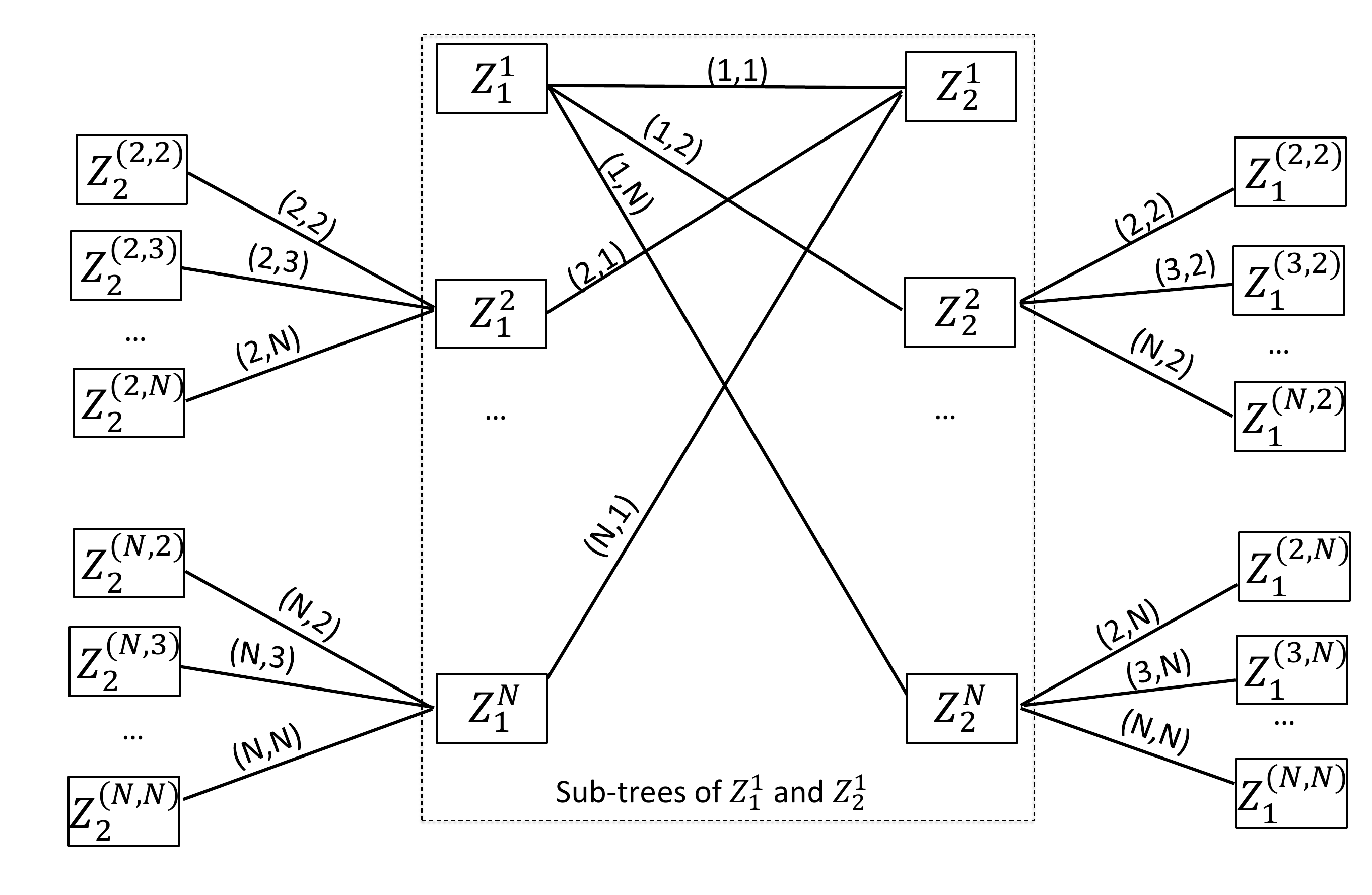}}
\caption{\small Construction of cache configurations   in Lemmas~\ref{lem:tree demand} and~\ref{lem:tree demand 2}.}
\label{fig:tree}
%\vspace{-5mm}
\end{figure}

We can represent the construction of the cache configurations in Lemmas~\ref{lem:tree demand} and~\ref{lem:tree demand 2} by an $\Nsf$-ary  tree, as illustrated in Fig.~\ref{fig:tree}. 
\begin{itemize}
\item Two vertices (assumed to be represented by cache configurations $Z^{\prime}_1$ and $Z^{\prime}_2$)   are connected by an edge with superscript $(i,j)$, if  $Z^{\prime}_1 \cup Z^{\prime}_2=\{F_1,\ldots,F_{\Nsf}\}$, 
$H(X_1|Z^{\prime}_1 )=H(X_2|Z^{\prime}_2)=0$, and $H(F_i|X_2,  Z^{\prime}_1 )=H(F_j|X_1, Z^{\prime}_2)=0$.
\item For each     $i  \in [\Nsf]$, $Z^{i}_{1}$ is connected to  exactly $\Nsf$ vertices, which are $Z^{(i,j^{\prime})}_2$ where $j^{\prime} \in [\Nsf]$.
\item   For each     $j \in [\Nsf]$, $Z^{j}_{2}$ is connected to exactly $\Nsf$ vertices, which are $Z^{(i^{\prime},j)}_1$ where $i^{\prime} \in [\Nsf]$.
\end{itemize}
 
Consider $Z^{i}_1$ where $i\in [\Nsf]$. Recall that $\Msf=\Nsf/2+y$, and that for each $j^{\prime} \in [\Nsf]$, we have $Z^i_1 \cup Z^{(i ,j^{\prime})}_{2}= \{F_1,\ldots,F_{\Nsf}\}$. For each file $F_{p}$ where $p\in [\Nsf]$, by defining
\begin{subequations} 
\begin{align}
Z^{i}_{1,p}:=Z^{i}_1 \cap F_p, \   Z^{(i ,j^{\prime})}_{2,p}:=Z^{(i ,j^{\prime})}_{2} \cap F_p, \ \forall j^{\prime} \in [\Nsf],\label{eq:def of Z1p}
\end{align}
we have
\begin{align}
&|F_p \setminus Z^{i}_{1,p}|= |F_p \setminus Z^{(i ,1)}_{2,p}|=\cdots=|F_p \setminus Z^{(i ,\Nsf)}_{2,p}|=\frac{\Bsf}{2}-\frac{y \Bsf}{\Nsf};\label{eq:unknown size i}\\
&|Z^{i}_{1,p} \cap  Z^{(i ,j^{\prime})}_{2,p}|= \frac{2y\Bsf}{\Nsf}, \ \forall j^{\prime} \in [\Nsf];\label{eq:all known size i}\\
&(F_p \setminus Z^{i}_{1,p}) \subseteq  Z^{(i ,j^{\prime})}_{2,p}, \ \forall j^{\prime} \in [\Nsf].\label{eq:all contain i}
\end{align}
\end{subequations} 
For each file $p\in [\Nsf]$, we define that 
\begin{align}
\Qc^{i}_{1,p} =  Z^i_{1,p} \cap Z^{(i ,1)}_{2,p}\cap \cdots \cap Z^{(i ,\Nsf)}_{2,p},\label{eq:Qi1p}
\end{align}
and that  $q^{i}_{1,p}=|\Qc^{i}_{1,p} |$. 

%denote  the bits in $Z^{i}_{1,p}$  used to encode $X_2$ by $\Qc^{i}_{1,p}$.  
%In other words, $\Qc^{i}_{1,p}$ is the smallest subset of $Z^{i}_{1,p}$ such that 
%\begin{align}
%H\big(X_2|\{F_1,\ldots,F_{\Nsf}\}\setminus Z^{i}_{1,p}, \Qc^{i}_{1,p},\mathscr{M}(P_2) \big)=0.\label{eq:general def of Qi}
%\end{align}
%We also define   $q^{i}_{1,p}=|\Qc^{i}_{1,p} |$. 
%In addition, by the definition of $\Qc^{i}_{1,p}$, it can be seen that  $ Z^i_{1,p} $ contains $\Qc^i_{1,p}$. Since all bits in $\Qc^i_{1,p}$ are used to encode   $X_2$, we also have $\Qc^{i}_{1,p} \subseteq Z^{(i ,j^{\prime})}_{2,p}$. Hence, we have 
%\begin{align}
%\Qc^{i}_{1,p} \subseteq \left(  Z^i_{1,p} \cap Z^{(i ,1)}_{2,p}\cap \cdots \cap Z^{(i ,\Nsf)}_{2,p} \right).\label{eq:Qi1p}
%\end{align}

Similarly, focus on $Z^{j}_2$ where $j\in [\Nsf]$, and 
we have
\begin{subequations} 
\begin{align}
&Z^{j}_{2,p}:=Z^{j}_2 \cap F_p, \   Z^{(i^{\prime} ,j )}_{1,p}:=Z^{(i^{\prime} ,j )}_{1} \cap F_p, \ \forall i^{\prime} \in [\Nsf]; \label{eq:def of Z2p}\\
&|F_p \setminus Z^{j}_{2,p}|= |F_p \setminus Z^{(1 ,j)}_{1,p}|=\cdots=|F_p \setminus Z^{(\Nsf ,j)}_{1,p}|\nonumber\\& =\frac{\Bsf}{2}-\frac{y \Bsf}{\Nsf};\label{eq:unknown size j}\\
&|Z^{j}_{2,p} \cap  Z^{(i^{\prime} ,j )}_{1,p}|= \frac{2y\Bsf}{\Nsf}, \ \forall i^{\prime} \in [\Nsf];\label{eq:all known size j}\\
&(F_p \setminus Z^{j}_{2,p}) \subseteq Z^{(i^{\prime} ,j )}_{1,p}, \ \forall i^{\prime} \in [\Nsf].\label{eq:all contain j}
\end{align}
\end{subequations} 
For each file $p\in [\Nsf]$, we define that 
\begin{align}
\Qc^{j}_{2,p} = Z^j_{2,p} \cap Z^{(1 ,j)}_{2,p}\cap \cdots \cap Z^{(\Nsf , j)}_{2,p}  ,\label{eq:Qj2p}
\end{align}
and that  $q^{j}_{2,p}=|\Qc^{j}_{2,p} |$.

After the above definitions, we are ready to prove Theorem~\ref{thm:two user converse}. As illustrated in the example in Section~\ref{ex:converse K2N2}, we will use two different approaches to construct powerful super-users.

\paragraph*{First approach}
Consider $Z^{i}_1$ where $i\in [\Nsf]$. We then focus the connected vertices of $Z^i_1$ in Fig.~\ref{fig:tree}, i.e., $Z^{(i ,j^{\prime})}_{2}$ where $j^{\prime} \in [\Nsf]$.
By the construction, from $(X_1, Z^{(i ,j^{\prime})}_{2})$, we can reconstruct $F_{j^{\prime}}$.
The first approach is inspired by the acyclic index coding converse bound in~\cite{indexcodingcaching2020,exactrateuncoded} for shared-link caching without privacy.
We pick a permutation of $[\Nsf]$, assumed to be  $\uv=(u_1,\ldots,u_{\Nsf})$, where $u_1=i$.
We can construct a genie-aided  super-user with the cache 
\begin{align}
\cup_{p\in [\Nsf ]}  Z^{(i ,u_p)}_{2}\setminus & \big(F_{u_1}\cup \cdots \cup F_{u_{p-1}} \cup Z^{(i ,u_1)}_{2} \cup \nonumber\\&  \cdots \cup  Z^{(i ,u_{p-1})}_{2} \big). \label{eq:virtual user cache i1}
\end{align}
The genie-aided  super-user can successively decode the whole library from its cache and $X_1$. More precisely, it can first decode  $F_{u_1}$ from $(X_1,Z^{(i ,u_1)}_{2})$. From $(X_1, F_{u_1},Z^{(i ,u_1)}_{2}, Z^{(i ,u_2)}_{2}\setminus (F_{u_1} \cup Z^{(i ,u_1)}_{2})$, then it can  decode $F_{u_2}$. By this way, the genie-aided super-user can decode the whole library.  Hence,   we have
\begin{subequations}
\begin{align}
 &H(F_1,\ldots, F_{\Nsf})  \nonumber \\
 &\leq H (X_1) +H\Big(\cup_{p\in [\Nsf ]}  Z^{(i ,u_p)}_{2}\setminus \big(F_{u_1}\cup \cdots \cup F_{u_{p-1}} \nonumber\\& \cup Z^{(i ,u_1)}_{2} \cup \cdots \cup Z^{(i ,u_{p-1})}_{2} \big) \Big) \label{eq:cut-set first approach}\\ 
 &\leq H(X_1)+ H(Z^{(i ,u_1)}_{2})+H( Z^{(i ,u_2)}_{2}| F_{u_1}, Z^{(i ,u_1)}_{2})+\cdots \nonumber\\& +H(Z^{(i ,u_{\Nsf })}_{2}|F_{u_1},\ldots, F_{u_{\Nsf-1}}, Z^{(i ,u_1)}_{2}, \ldots,   Z^{(i ,u_{\Nsf-1})}_{2}   )\\
 &=H(X_1)+ H(Z^{(i ,i)}_{2})+H( Z^{(i ,u_2)}_{2}| F_{i}, Z^{(i ,i)}_{2})+\cdots  + \nonumber\\&  H(Z^{(i ,u_{\Nsf })}_{2} |F_{i}, F_{u_{2}},\ldots, F_{u_{\Nsf-1}}, Z^{(i ,i)}_{2}, Z^{(i ,u_2)}_{2}, \ldots,   Z^{(i ,u_{\Nsf-1})}_{2}   ) \label{eq:approache 1 step 1}\\
 &=H(X_1)+ H(Z^{(i ,i)}_{2})\nonumber\\& + \Big( H( Z^{(i ,u_2)}_{2,u_2}|  Z^{(i ,i)}_{2,u_2})+\cdots+  H( Z^{(i ,u_{\Nsf})}_{2,u_{\Nsf}}|  Z^{(i ,i)}_{2,u_{\Nsf}}) \Big)+\cdots \nonumber\\& + \Big(H(Z^{(i ,u_{\Nsf })}_{2,u_{\Nsf}}|  Z^{(i ,i)}_{2,u_{\Nsf}}, Z^{(i ,u_2)}_{2,u_{\Nsf}}, \ldots,   Z^{(i ,u_{\Nsf-1})}_{2,u_{\Nsf}}) \Big) \label{eq:approache 1 step 1.2}\\
 &=H(X_1)+ H(Z^{(i ,i)}_{2}) + H(  Z^{(i ,u_2)}_{2,u_2}|   Z^{(i ,i)}_{2,u_2} )+H (  Z^{(i ,u_2)}_{2,u_3} , \nonumber\\& Z^{(i ,u_3)}_{2,u_3}| Z^{(i ,i)}_{2,u_3} )+\cdots+H( Z^{(i ,u_2)}_{2,u_{\Nsf}} ,\ldots, Z^{(i ,u_{\Nsf})}_{2,u_{\Nsf}}|   Z^{(i ,i)}_{2,u_{\Nsf}}  ), \label{eq:approache 1 step 2}
\end{align}
\end{subequations}
where~\eqref{eq:approache 1 step 1} follows since $u_1=i$,~\eqref{eq:approache 1 step 1.2} follows since   all bits in the library are independent,~\eqref{eq:approache 1 step 2} comes from   the chain rule of the entropy.%\footnote{\label{foot:chain rule}$H(Y|X)+H(Z|X,Y)=H(Y,Z|X)$. }.
 
From~\eqref{eq:approache 1 step 2}, it will be proved in Appendix~\ref{sub:proof of first approach bound 1} and Appendix~\ref{sub:proof of first approach bound 2} that (recall $y=\Msf-\Nsf/2$), 
\begin{align}
 H (X_1)&\geq \frac{\Bsf}{2}-\frac{y \Bsf}{\Nsf};  
 \label{eq:proof of first approach bound 1}
 \\
 H (X_1)&\geq \Bsf -\frac{4y\Bsf}{\Nsf}+q^{i}_{1,u_{2}}. 
 \label{eq:proof of first approach bound 2} 
\end{align} 
 
In addition, by considering all permutations of    $[\Nsf]$  where the first element is $i$, we can list all  $(\Nsf-1)!$ inequalities as in~\eqref{eq:approache 1 step 2}. By summing all these   $(\Nsf-1)!$ inequalities, we can obtain the following inequality, which will be proved in    Appendix~\ref{sub:proof of first approach bound 3}, 
  %\begin{subequations}
\begin{align} 
H (X_1) &\geq \frac{\Nsf\Bsf}{2}- y \Bsf -\frac{4(\Nsf-1)y\Bsf}{(h+2)\Nsf}+\frac{2}{h+2}\sum_{p\in[\Nsf]\setminus \{i\}} q^i_{1,p} \nonumber\\
&-\sum_{p\in[\Nsf]\setminus \{i\}} \left\{  \frac{\Nsf-2}{(h+1)(h+2)}\left(\frac{2y\Bsf}{\Nsf}-q^i_{1,p} \right) \right. \nonumber\\& \left. + \frac{h}{h+2}\left(\frac{\Bsf}{2}-\frac{y\Bsf}{\Nsf} \right)       \right\}, \ \forall h\in [0:\Nsf-3]. \label{eq:proof of first approach bound 3}
\end{align}
%\label{eq:proof of first approach step 1}
 %\end{subequations}

By considering all $i\in [\Nsf]$, we can list all  $\Nsf$ inequalities as in~\eqref{eq:proof of first approach bound 3}. By summing all these $\Nsf$ inequalities, we obtain 
  \begin{subequations}
\begin{align}
H (X_1) &\geq \frac{\Nsf\Bsf}{2}- y \Bsf -\frac{4(\Nsf-1)y\Bsf}{(h+2)\Nsf} \nonumber\\& +\frac{2}{(h+2)\Nsf} \sum_{i\in [\Nsf]}\sum_{p\in[\Nsf]\setminus \{i\}} q^i_{1,p} \nonumber\\
&-\sum_{i\in [\Nsf]}\sum_{p\in[\Nsf]\setminus \{i\}} \left\{  \frac{\Nsf-2}{(h+1)(h+2)\Nsf}\left(\frac{2y\Bsf}{\Nsf}-q^i_{1,p} \right) \right. \nonumber\\& \left. + \frac{h}{(h+2)\Nsf}\left(\frac{\Bsf}{2}-\frac{y\Bsf}{\Nsf} \right)       \right\}, \ \forall h\in [0:\Nsf-3]. \label{eq:proof of first approach bound 3 step 2}
\end{align} 

\end{subequations}

We now consider $Z^{j}_2$ where $j \in [\Nsf]$. By the similar step as above to derive~\eqref{eq:proof of first approach bound 3 step 2}, we obtain
 % \begin{subequations}
\begin{align}
H (X_2) &\geq \frac{\Nsf\Bsf}{2}- y \Bsf -\frac{4(\Nsf-1)y\Bsf}{(h+2)\Nsf}+ \nonumber\\& \frac{2}{(h+2)\Nsf} \sum_{j\in [\Nsf]}\sum_{p\in[\Nsf]\setminus \{j\}} q^j_{2,p} \nonumber\\
&-\sum_{j\in [\Nsf]}\sum_{p\in[\Nsf]\setminus \{j\}} \left\{ \frac{\Nsf-2}{(h+1)(h+2)\Nsf}\left(\frac{2y\Bsf}{\Nsf}-q^j_{2,p} \right) \right. \nonumber\\& \left. + \frac{h}{(h+2)\Nsf}\left(\frac{\Bsf}{2}-\frac{y\Bsf}{\Nsf} \right)       \right\}, \ \forall h\in [0:\Nsf-3]. \label{eq:proof of first approach bound 3 step 2j}
\end{align} 
%\end{subequations}

By  summing~\eqref{eq:proof of first approach bound 3 step 2} and~\eqref{eq:proof of first approach bound 3 step 2j}, we obtain
  \begin{subequations}
\begin{align} 
&\Rsf^{\star}_{\mathrm{u}} \Bsf  \geq H(X_1)+H(X_2) \\& \geq \negmedspace \Nsf \Bsf \negmedspace - 2y \Bsf -\frac{8(\Nsf-1)y\Bsf}{(h+2)\Nsf}  -\frac{\Nsf-2}{(h+1)(h+2)\Nsf} 4y (\Nsf-1)\Bsf   \nonumber\\
&-\frac{h(\Nsf-1)}{(h+2) }\left( \Bsf -\frac{2y\Bsf}{\Nsf} \right)
 \negmedspace +\negmedspace \left( \negmedspace \frac{2}{(h+2)\Nsf} \negmedspace + \negmedspace \frac{\Nsf-2}{(h+1)(h+2)\Nsf} \negmedspace\right) \nonumber\\& \left( \sum_{i\in [\Nsf]}\sum_{p\in[\Nsf]\setminus \{i\}} q^i_{1,p}+\sum_{j\in [\Nsf]}\sum_{p\in[\Nsf]\setminus \{j\}} q^j_{2,p} \right),   \forall h\in [0:\Nsf-3]. \label{eq:proof of first approach bound 3 step 3}
\end{align}
%\label{eq:proof of first approach}
\end{subequations}

\paragraph*{Second approach}
We then use the second approach to construct genie-aided  super-users.   
We first consider $X_2$. By the construction, from $(X_2, Z^{i}_{1})$ where $i\in [\Nsf]$, we can reconstruct $F_{i}$. 

Now we fix an integer $i\in [\Nsf]$. We pick a permutation of $[\Nsf]$, assumed to be  $\uv=(u_1,\ldots,u_{\Nsf})$, where $u_1=i$.
We can construct a genie-aided  super-user with the cache 
\begin{align}
\cup_{p\in [\Nsf ]}  \left( Z^{u_p}_{1,u_p} \cup \Qc^{u_1}_{1,u_p} \cup \cdots \cup \Qc^{u_{p-1}}_{1,u_p} \right). \label{eq:virtual user cache j2}
\end{align}
Now we prove that the genie-aided super-user can successively decode the whole library from its cache and $X_2$. 
Note that   from  $(Z^{u_1}_1, X_2)$, we can reconstruct $F_{u_1}$. Furthermore, 
for each file $F_{p_1}$ where $p_1 \in [\Nsf]\setminus \{u_1\}$, by recalling the definition of $ \Qc^{u_1}_{1,p_1}$ in~\eqref{eq:Qi1p}, it can be seen that 
 the bits in $Z^{u_1}_{1,p_1} \setminus \Qc^{u_1}_{1,p_1} $ are independent of $X_2$.
Hence, it is enough to reconstruct $F_{u_1}$ from  $(X_2, Z^{u_1}_{1,u_1}, \Qc^{u_1}_{1,u_2},\ldots, \Qc^{u_1}_{1,u_{\Nsf}})$, and thus the super-user can reconstruct $F_{u_1}$. 
After recovering $F_{u_1}$, the super-user can reconstruct $F_{u_2}$ from  $(X_2, F_{u_1},  Z^{u_2}_{1,u_2}, \Qc^{u_2}_{1,u_3},\ldots, \Qc^{u_2}_{1,u_{\Nsf}})$.
By this way, the genie-aided super-user can decode the whole library.  Hence,   we have
 \begin{subequations}
\begin{align}
 H(X_2)&\geq H(F_1,\ldots,F_{\Nsf}) \nonumber\\& - H\left( \cup_{p\in [\Nsf ]}  \big( Z^{u_p}_{1,u_p} \cup \Qc^{u_1}_{1,u_p} \cup \cdots \cup \Qc^{u_{p-1}}_{1,u_p} \big) \right)\\
&\geq \left(H(F_{u_1})-H(Z^{u_1}_{1,u_1})\right) \nonumber\\& + \big(H(F_{u_2})  -H(Z^{u_2}_{1,u_2},  \Qc^{u_1}_{1,u_2}  ) \big)+\cdots  \nonumber\\& + \left(H(F_{u_{\Nsf}})-H(Z^{u_{\Nsf}}_{1,u_{\Nsf}},  \Qc^{u_1}_{1,u_{\Nsf}},\ldots,  \Qc^{u_{\Nsf-1}}_{1,u_{\Nsf}} ) \right).\label{eq:approache 2 step 1}
\end{align}
 \end{subequations}
 
 From~\eqref{eq:approache 2 step 1}, it will be proved in Appendix~\ref{sub:proof of second approach bound 1} and Appendix~\ref{sub:proof of second approach bound 2} that, 
\begin{align}
 H (X_2)& \geq \frac{\Bsf}{2}-\frac{y \Bsf}{\Nsf}; \label{eq:proof of second approach bound 1}\\
  H (X_2) &\geq \Bsf -\frac{2y\Bsf}{\Nsf}-q^{i}_{1,u_{2}}. \label{eq:proof of second approach bound 2}
\end{align}
 
By letting the two permutations to derive~\eqref{eq:approache 1 step 2} and~\eqref{eq:approache 2 step 1} be the same,
we now sum~\eqref{eq:proof of first approach bound 1} and~\eqref{eq:proof of second approach bound 1} to obtain
\begin{align}
\Rsf^{\star}_{\mathrm{u}} \Bsf  \geq H(X_1)+H(X_2) \geq  \Bsf -\frac{2y \Bsf }{\Nsf} , \label{eq:proof of two approach bound 1}
\end{align}
which coincides with the proposed converse bound in~\eqref{eq:K2 converse 3}. Similarly, by summing~\eqref{eq:proof of first approach bound 2} and~\eqref{eq:proof of second approach bound 2}, we obtain
\begin{align}
 \Rsf^{\star}_{\mathrm{u}} \Bsf \geq H(X_1)+H(X_2)\geq  2 \Bsf-\frac{6y \Bsf }{\Nsf}, \label{eq:proof of two approach bound 2}
\end{align}
which coincides with the proposed converse bound in~\eqref{eq:K2 converse 2}.

 In addition, by considering all permutations of    $[\Nsf]$  where the first element is $i$, we can list all  $(\Nsf-1)!$ inequalities as in~\eqref{eq:approache 2 step 1}. By summing all these   $(\Nsf-1)!$ inequalities, we can obtain the following inequalities, which will be proved in  Appendix~\ref{sub:proof of second approach bound 3},  
\begin{align}
 H (X_2) &\geq \frac{\Nsf\Bsf}{2}-y  \Bsf- \frac{2}{h+2}\sum_{p\in[\Nsf]\setminus \{i\}} q^i_{1,p} \nonumber\\
&-\negmedspace\negmedspace \sum_{p\in[\Nsf]\setminus \{i\}}\negmedspace\negmedspace \left\{  \frac{  \sum_{n\in [\Nsf]\setminus \{i,p\}} q^{n}_{1,p} }{(h+1)(h+2)} + \frac{h}{h+2}\left(\frac{\Bsf}{2}-\frac{y\Bsf}{\Nsf} \right)       \right\}, \nonumber\\& \forall h\in [0:\Nsf-3]. \label{eq:proof of second approach bound 3} 
\end{align}
By considering all $i\in [\Nsf]$, we can list all  $\Nsf$ inequalities as in~\eqref{eq:proof of second approach bound 3}. By summing all these $\Nsf$ inequalities, we obtain 
\begin{align}
 H (X_2) &\geq \frac{\Nsf\Bsf}{2}-y  \Bsf- \frac{2}{(h+2)\Nsf}\sum_{i\in [\Nsf]} \sum_{p\in[\Nsf]\setminus \{i\}} q^i_{1,p} \nonumber\\
&-\sum_{i\in [\Nsf]}\sum_{p\in[\Nsf]\setminus \{i\}} \left\{  \frac{  \sum_{n\in [\Nsf]\setminus \{i,p\}} q^{n}_{1,p} }{(h+1)(h+2)\Nsf}  \right. \nonumber\\& \left. + \frac{h}{(h+2)\Nsf}\left(\frac{\Bsf}{2}-\frac{y\Bsf}{\Nsf} \right)       \right\}, \ \forall h\in [0:\Nsf-3].    \label{eq:proof of second approach bound 3 step 2}
\end{align}

 We now consider $X_1$. By the similar steps as above to derive~\eqref{eq:proof of second approach bound 3 step 2}, we obtain
\begin{align}
  H (X_1) &\geq \frac{\Nsf\Bsf}{2}-y \Bsf- \frac{2}{(h+2)\Nsf}\sum_{j\in [\Nsf]} \sum_{p\in[\Nsf]\setminus \{j\}} q^j_{2,p} \nonumber\\
&-\sum_{j\in [\Nsf]}\sum_{p\in[\Nsf]\setminus \{j\}} \left\{  \frac{  \sum_{n\in [\Nsf]\setminus \{j,p\}} q^{n}_{2,p} }{(h+1)(h+2)\Nsf} \right. \nonumber\\& \left. + \frac{h}{(h+2)\Nsf}\left(\frac{\Bsf}{2}-\frac{y\Bsf}{\Nsf} \right)       \right\}, \ \forall h\in [0:\Nsf-3]. \label{eq:proof of second approach bound 3 step 2i}
\end{align}

By   summing~\eqref{eq:proof of second approach bound 3 step 2} and~\eqref{eq:proof of second approach bound 3 step 2i}, we obtain
  \begin{subequations}
\begin{align}
&\Rsf^{\star}_{\mathrm{u}} \Bsf  \geq H(X_1)+H(X_2) \\&  \geq   \Nsf\Bsf -2y  \Bsf \nonumber\\& -\frac{2}{(h+2)\Nsf}\left(\sum_{j\in [\Nsf]} \sum_{p\in[\Nsf]\setminus \{j\}} q^j_{2,p}+\sum_{i\in [\Nsf]} \sum_{p\in[\Nsf]\setminus \{i\}} q^i_{1,p} \right) \nonumber\\
&-\frac{1}{(h+1)(h+2)\Nsf }\left(\sum_{j\in [\Nsf]}\sum_{p\in[\Nsf]\setminus \{j\}}  \sum_{n\in [\Nsf]\setminus \{j,p\}} q^{n}_{2,p} \right. \nonumber\\ & \left. + \sum_{i\in [\Nsf]}\sum_{p\in[\Nsf]\setminus \{i\}}\sum_{n\in [\Nsf]\setminus \{i,p\}} q^{n}_{1,p} \right)-\frac{h(\Nsf-1)}{(h+2) }\left( \Bsf -\frac{2y\Bsf}{\Nsf} \right) \\
&=\Nsf\Bsf -2y \Bsf \nonumber\\& -\frac{2}{(h+2)\Nsf}\left(\sum_{j\in [\Nsf]} \sum_{p\in[\Nsf]\setminus \{j\}} q^j_{2,p}+\sum_{i\in [\Nsf]} \sum_{p\in[\Nsf]\setminus \{i\}} q^i_{1,p} \right) \nonumber\\
&-\frac{1}{(h+1)(h+2)\Nsf }\left((\Nsf-2)\sum_{j_1\in [\Nsf]}\sum_{p_1\in[\Nsf]\setminus \{j_1\}}  q^{j_1}_{2,p_1}  \right. \nonumber\\& \left. +  (\Nsf-2)\sum_{i_2\in [\Nsf]}\sum_{p_2\in[\Nsf]\setminus \{i_2\}}  q^{i_2}_{1,p_2} \right)\nonumber\\
&-\frac{h(\Nsf-1)}{(h+2) }\left( \Bsf -\frac{2y\Bsf}{\Nsf} \right),\ \forall h\in [0:\Nsf-3],
 \label{eq:proof of second approach bound 3 step 3}
\end{align}
% \label{eq:proof of second approach}
\end{subequations}
where~\eqref{eq:proof of second approach bound 3 step 3} follows since\footnote{\label{foot:merge}In the sum $\sum_{j\in [\Nsf]}\sum_{p\in[\Nsf]\setminus \{j\}}  \sum_{n\in [\Nsf]\setminus \{j,p\}} q^{n}_{2,p}$,  let us compute the coefficient of   term $q^{j_1}_{2,p_1}$  where $j_1\neq p_1$. $q^{j_1}_{2,p_1}$ appears in the sum when $p=p_1$ and $n=j_1$. Hence, there are $\Nsf-2$ possibilities of $j$, which are $[\Nsf]\setminus \{p_1,j_1\}$. So the coefficient of  $q^{j_1}_{2,p_1}$ in the sum is $\Nsf-2$.} $$\sum_{j\in [\Nsf]}\sum_{p\in[\Nsf]\setminus \{j\}}  \sum_{n\in [\Nsf]\setminus \{j,p\}} q^{n}_{2,p}=\sum_{j_1\in [\Nsf]}\sum_{p_1\in[\Nsf]\setminus \{j_1\}}  q^{j_1}_{2,p_1} ,$$ and   $$ \sum_{i\in [\Nsf]}\sum_{p\in[\Nsf]\setminus \{i\}}\sum_{n\in [\Nsf]\setminus \{i,p\}} q^{n}_{1,p}= (\Nsf-2)\sum_{i_2\in [\Nsf]}\sum_{p_2\in[\Nsf]\setminus \{i_2\}}  q^{i_2}_{1,p_2}.$$

Finally, by summing~\eqref{eq:proof of first approach bound 3 step 3} and~\eqref{eq:proof of second approach bound 3 step 3}, we obtain $\forall h\in [0:\Nsf-3]$, 
\begin{subequations}
\begin{align}
\Rsf^{\star}_{\mathrm{u}}  
&\geq  \frac{1}{2}\left\{\Nsf- 2y  -\frac{8(\Nsf-1)y }{(h+2)\Nsf} 
-\frac{(\Nsf-2)(\Nsf-1)4y }{(h+1)(h+2)\Nsf}- \right. \nonumber\\ & \left. \frac{h(\Nsf-1)}{(h+2) }\left(1 -\frac{2y }{\Nsf} \right)\right\} \nonumber\\& +\frac{1}{2}\left\{\Nsf -2y\Nsf  -\frac{h(\Nsf-1)}{(h+2) }\left( 1-\frac{2y }{\Nsf} \right)\right\} \\
&=\Nsf-2y-\frac{4y+(\Nsf-1)h}{h+2}\nonumber\\& +\frac{h^2(n-1)-\Nsf(\Nsf-3)+h(\Nsf+1) }{(h+1)(h+2)}\frac{2y}{\Nsf},
\label{eq:proof of two approach bound 3}
\end{align}
\end{subequations}
which coincides with the proposed converse bound in~\eqref{eq:K2 converse 1}.

\subsection{Proof of Theorem~\ref{thm:K user converse}: $\Ksf$-user System}
\label{sub:K user converse}
 
We extend the proposed converse bound for the two-user system  to $\Ksf$-user system  and consider the privacy constraint against colluding users in~\eqref{eq:colluding privacy}.  In the following, we consider the case where $\Ksf/2$ is an integer and $2\Nsf/\Ksf$ is also an integer.  In Appendix~\ref{sec:general k,n} we generalize the proof  to any $\Ksf$ and $\Nsf$.

%We divide $\Ksf$ users into two non-overlapping groups, denoted by $\Kc_1:=\left[\frac{\Ksf}{2} \right]$ and $\Kc_2:=\left[\frac{\Ksf}{2}+1:\Ksf \right]$. We divide $\Nsf$ files into $2\Nsf/\Ksf$ non-overlapping groups, each of which contains $\Ksf/2$ files, denoted by $\Nc_g:=\left[(g-1)\frac{\Ksf}{2}: g\frac{\Ksf}{2} \right]$. 

%Assume the cache configurations of the $\Ksf$ users are   $(Z^1_1,\ldots,Z^1_{\Ksf})$, where $Z^1_1 \cup \cdots Z^1_{\Ksf}=\{F_1,\ldots,F_{\Nsf}\}$. When the demand vector  is  $\dv=(1,\ldots,\Ksf)$, we assume the transmitted packets are $(X_1,\ldots,X_{\Ksf})$, such that the demand vector  $\dv=(1,\ldots,\Ksf)$ can be satisfied. 

%We first focus on the viewpoint of the users in $\Kc_1$.
 
Let $\Msf=\frac{\Nsf}{\Ksf}+\frac{2 y }{\Ksf}$, where $y\in \left[0,\frac{\Nsf}{2}\right]$.
We use a genie-aided proof by generating two aggregate users, denoted by $k_1$ and $k_2$. 
We assume that the cache size of each aggregate user is $\Msf \Bsf \times \frac{\Ksf}{2}=\frac{\Nsf \Bsf}{2}+y \Bsf$, i.e., the cache size of each aggregate user is the total cache size  of $\Ksf/2 $ users.
%The cache  content of virtual users $k_1$ and $k_2$ are the union sets of the cache  content of users in $[\Ksf/2]$ and   of users in $[\Ksf/2+1:\Ksf]$,   respectively. 
In addition, the demanded files of aggregate users $k_1$ and $k_2$ are the union sets of the demanded files of users in $[\Ksf/2]$ and   of users in $[\Ksf/2+1:\Ksf]$,   respectively. The objective is to design a two-user D2D private caching scheme with minimum load $\Rsf^{\star}_{\mathrm{g}}$, such that each aggregate user can decode its demanded files without knowing anything about the demand of the other aggregate user.

Obviously, for any $\Ksf$-user D2D private caching satisfying the encoding~\eqref{eq:encoding constraint}, decoding~\eqref{eq:decodability}, and privacy constraints~\eqref{eq:colluding privacy}, it must be an achievable scheme  for the above genie system. In other words, 
$\Rsf^{\star}_{\mathrm{u,c}}\geq \Rsf^{\star}_{\mathrm{g}}$.
Hence, in the following we characterize a converse bound for $\Rsf^{\star}_{\mathrm{g}}$, which is also a converse bound for $\Rsf^{\star}_{\mathrm{u,c}}$.

We partition the $\Nsf$ files into $2\Nsf/\Ksf$ equal-size groups, each of which contains $\Ksf/2$ files.  %denoted by $\Nc_g:=\left[(g-1)\frac{\Ksf}{2}: g\frac{\Ksf}{2} \right]$. 
Each aggregate user demands one group of files. Hence, it is equivalent to the two-user D2D private caching problem with $2\Nsf/\Ksf$ files, each of which has $\Ksf\Bsf/2$ bits, and each of the two users caches $\left(\frac{\Nsf \Bsf}{2}+y \Bsf\right)$ bits in its cache and  demands one file. 
  
We assume the caches of aggregate users $k_1$ and $k_2$ are  $A^1_1$ and $A^1_2$.
The transmitted packets by aggregate users $k_1$ and $k_2$ are denoted by $X^{\prime}_1$ and $X^{\prime}_2$, such that from $(X^{\prime}_2, A^1_1)$ aggregate user $k_1$ can decode the files in group $1$ and from $(X^{\prime}_1, A^1_2)$ aggregate user $k_2$ can also  decode the files in group $1$.
We then also construct  the cache configurations of aggregate users $k_1$ and $k_2$ by a $2\Nsf/\Ksf$-ary  tree, as we did in Section~\ref{sub:two user converse}.  
   
%{\red IS THERE A SIMPLER WAY TO DESCRIBE WHAT WE DO? WHY DO WE NEED TO REPEAT ALL THE EQUATIONS?}   
   
By the first approach of constructing converse bound described in Section~\ref{sub:two user converse}, 
when we consider  % $x_1$ (transmitted packets of virtual user~$1$)  and  
 $A^i_1$  where $i\in \left[\frac{2\Nsf}{\Ksf} \right]$ (cache of aggregate user~$k_1$ from which and $X^{\prime}_2$, the files in group $i$ can be reconstructed), with a permutation of $[2\Nsf/\Ksf]$  denoted by $\uv=(u_1,\ldots,u_{2\Nsf/\Ksf})$ where $u_1=i$, we obtain (by the similar derivations of~\eqref{eq:proof of first approach bound 1} and~\eqref{eq:proof of first approach bound 2}),
\begin{align}
 H (X^{\prime}_1)& \geq \left(\frac{\Bsf}{2}-\frac{y \Bsf}{\Nsf}\right) \frac{\Ksf}{2}; \label{eq:proof of first approach bound 1 k user}\\
 H (X^{\prime}_1) &\geq \frac{\Ksf}{2}\Bsf-\frac{\Ksf}{2}\frac{4 y \Bsf}{\Nsf}+  q^i_{1,u_2} , \label{eq:proof of first approach bound 2 k user}
\end{align}
where $q^i_{1,u_2}$ represent the number of bits in $A^i_1 \cap A^{(i,1)}_2 \cap \cdots \cap A^{(i,2\Nsf/\Ksf)}_2$, which are from the files  in group $u_2$.
 
% the number of common bits of $A^i_1$ and the files in group $u_2$, which are used to encode $x_2$.

By considering all permutations of $[2\Nsf/\Ksf]$ whose first element is $i$,
we obtain (by the similar derivation of~\eqref{eq:proof of first approach bound 3}),
 % \begin{subequations}
\begin{align}
H (X^{\prime}_1) &\geq \frac{\Nsf\Bsf}{2}- y \Bsf -\frac{2}{h+2} \left\{\left(\frac{2\Nsf}{\Ksf}-1 \right) \frac{2y\Bsf}{\Nsf} \frac{\Ksf}{2} \right\} \nonumber\\& +\frac{2}{h+2}\sum_{p\in\left[\frac{2\Nsf}{\Ksf} \right]\setminus \{i\}} q^i_{1,p} \nonumber\\
 &-\sum_{p\in\left[\frac{2\Nsf}{\Ksf} \right]\setminus \{i\}} \left\{  \frac{ \frac{2\Nsf}{\Ksf} -2}{(h+1)(h+2)}\left(\frac{2y\Bsf}{\Nsf}\frac{\Ksf}{2}-q^i_{1,p} \right) \right. \nonumber\\& \left.+ \frac{h}{h+2}\left(\frac{\Bsf}{2}-\frac{y\Bsf}{\Nsf} \right)\frac{\Ksf}{2}       \right\}, \ \forall h\in \left[0:\frac{2\Nsf}{\Ksf}-3 \right]. \label{eq:proof of first approach bound 3 k user}
\end{align}
%\label{eq:proof of first approach k user}
% \end{subequations}  
  
By considering all $i\in \left[\frac{2\Nsf}{\Ksf} \right]$ to bound $H(X^{\prime}_1)$  and all   $j\in \left[\frac{2\Nsf}{\Ksf} \right]$ to bound $H(X^{\prime}_2)$, we sum all inequalities as in~\eqref{eq:proof of first approach bound 3 k user} to obtain (by the similar derivation of~\eqref{eq:proof of first approach bound 3 step 3}),
  %\begin{subequations}
\begin{align}
  \Rsf^{\star}_{\mathrm{g}} \Bsf  &\geq  \Nsf \Bsf- 2y \Bsf -\frac{4}{h+2} \left\{\left(\frac{2\Nsf}{\Ksf}-1 \right) \frac{2y\Bsf}{\Nsf} \frac{\Ksf}{2} \right\} \nonumber\\& -\frac{\frac{2\Nsf}{\Ksf}-2}{(h+1)(h+2) } \frac{4y (\frac{2\Nsf}{\Ksf}-1)\Bsf }{\Nsf}\frac{\Ksf}{2}  \nonumber\\&-\frac{h\left(\frac{2\Nsf}{\Ksf}-1 \right)}{(h+2) }\left( \Bsf -\frac{2y\Bsf}{\Nsf} \right)\frac{\Ksf}{2} \nonumber\\& +\left(\frac{2}{(h+2)\frac{2\Nsf}{\Ksf}}+\frac{\frac{2\Nsf}{\Ksf}-2}{(h+1)(h+2)(2\Nsf/\Ksf) } \right)\nonumber\\
& \left( \sum_{i\in \left[\frac{2\Nsf}{\Ksf}\right]}\sum_{p\in\left[\frac{2\Nsf}{\Ksf}\right] \setminus \{i\}} q^i_{1,p}+\sum_{j\in \left[\frac{2\Nsf}{\Ksf}\right]}\sum_{p\in\left[\frac{2\Nsf}{\Ksf}\right]\setminus \{j\}} q^j_{2,p} \right), \nonumber\\&  \forall h\in \left[0:\frac{2\Nsf}{\Ksf}-3 \right]. 
\label{eq:proof of first approach case 3 k user}
\end{align}
 %\label{eq:proof of first approach final k user}
%\end{subequations} 
  
Similarly, by the second approach of constructing converse bound described in Section~\ref{sub:two user converse}, 
when we consider  $X^{\prime}_2$ and the same permutation as the one to derive~\eqref{eq:proof of first approach bound 1 k user} and~\eqref{eq:proof of first approach bound 2 k user}, we obtain (by the similar derivations of~\eqref{eq:proof of second approach bound 1} and~\eqref{eq:proof of second approach bound 2}),
\begin{align}
 H (X^{\prime}_2)& \geq \left(\frac{\Bsf}{2}-\frac{y \Bsf}{\Nsf}\right) \frac{\Ksf}{2}; \label{eq:proof of second approach bound 1 k user}\\
 H (X^{\prime}_2) &\geq \frac{\Ksf}{2}\Bsf-\frac{\Ksf}{2}\frac{2 y \Bsf}{\Nsf}-  q^i_{1,u_2} . \label{eq:proof of second approach bound 2 k user}
\end{align}
By summing~\eqref{eq:proof of first approach bound 1 k user} and~\eqref{eq:proof of second approach bound 1 k user}, we prove~\eqref{eq:K user converse third segment}. By summing~\eqref{eq:proof of first approach bound 2 k user} and~\eqref{eq:proof of second approach bound 2 k user}, we prove~\eqref{eq:K user converse second segment}.

In addition, by the second approach of constructing converse bound described in Section~\ref{sub:two user converse},  after considering all permutations to   bound $H(X^{\prime}_1)$  and all permutations to bound $H(X^{\prime}_2)$,  we    obtain (by the similar derivation of~\eqref{eq:proof of second approach bound 3 step 3}),
  %  \begin{subequations}
\begin{align}
   \Rsf^{\star}_{\mathrm{g}} \Bsf &\geq \Nsf\Bsf -2y  \Bsf-\frac{h\left(\frac{2\Nsf}{\Ksf} -1\right)}{(h+2) }\left( \Bsf -\frac{2y\Bsf}{\Nsf} \right) \frac{\Ksf}{2}\nonumber\\& - \left( \frac{2}{(h+2)\frac{2\Nsf}{\Ksf}}+\frac{  2\Nsf/\Ksf -2 }{(h+1)(h+2)\frac{2\Nsf}{\Ksf} } \right)\nonumber\\
   &
  \left(\sum_{i\in \left[\frac{2\Nsf}{\Ksf} \right]} \sum_{p\in\left[\frac{2\Nsf}{\Ksf} \right]\setminus \{i\}} q^i_{1,p}+\sum_{j\in \left[\frac{2\Nsf}{\Ksf} \right]} \sum_{p\in\left[\frac{2\Nsf}{\Ksf} \right]\setminus \{j\}} q^j_{2,p} \right),  \nonumber\\& \forall h\in \left[0:\frac{2\Nsf}{\Ksf}-3\right].
\label{eq:proof of second approach case 3 k user}
\end{align}
 %  \label{eq:proof of second approach final k user}
 %\end{subequations} 
 
By summing~\eqref{eq:proof of first approach case 3 k user} and~\eqref{eq:proof of second approach case 3 k user}, we prove~\eqref{eq:K user converse first segment}.

\section{Conclusions}
\label{sec:conclusion} 
We introduced a new D2D private caching model, which aims to preserve the privacy of the users' demands. 
We proposed new D2D private coded caching schemes, which were proved to be order optimal by matching a new converse bound under the constraint of uncoded cache placement and user collusion to within a constant gap.
Further works include proving %even tighter converse bounds than the extension from the two-user system, 
 new converse bounds for any cache placement, and investigating the decentralized D2D private coded caching problem.

\appendices

\section{Proofs of~\eqref{eq:proof of first approach bound 1},~\eqref{eq:proof of first approach bound 2}, and~\eqref{eq:proof of first approach bound 3}} 
 \label{sec:proof of first approach}
Recall that by considering a   permutation of $[\Nsf]$, assumed to be  $\uv=(u_1,\ldots,u_{\Nsf})$, where $u_1=i$, we can derive~\eqref{eq:approache 1 step 2},
\begin{align}
 H(F_1,\ldots, F_{\Nsf}) & \leq   H(X_1)+  H(Z^{(i ,i)}_{2})\nonumber\\& +\sum_{p\in [2:\Nsf]}H\left( Z^{(i ,u_2)}_{2,u_{p}} ,\ldots, Z^{(i ,u_{p})}_{2,u_{p}}|   Z^{(i ,i)}_{2,u_{p}}  \right). \label{eq:recall approache 1 step 2}
\end{align}
For each $p\in [2:\Nsf]$, since $| Z^{(i ,i)}_{2,u_{p}}|=\frac{\Bsf}{2}+\frac{y\Bsf}{ \Nsf}$, we have 
\begin{align}
H\left( Z^{(i ,u_2)}_{2,u_{p}} ,\ldots, Z^{(i ,u_{p})}_{2,u_{p}}|   Z^{(i ,i)}_{2,u_{p}}  \right) \leq H(F_p|  Z^{(i ,i)}_{2,u_{p}})=\frac{\Bsf}{2}-\frac{y\Bsf}{ \Nsf}.\label{eq:outer bound per term approache 1}
\end{align}

 \subsection{Proof of~\eqref{eq:proof of first approach bound 1}}
 \label{sub:proof of first approach bound 1}
 Now we bound each term $H\left( Z^{(i ,u_2)}_{2,u_{p}} ,\ldots, Z^{(i ,u_{p})}_{2,u_{p}}|   Z^{(i ,i)}_{2,u_{p}}  \right)$ where $p\in [2:\Nsf]$ in~\eqref{eq:recall approache 1 step 2} by $\frac{\Bsf}{2}-\frac{y\Bsf}{ \Nsf}$, to obtain
 \begin{subequations}
 \begin{align}
 & H(F_1,\ldots, F_{\Nsf})  \leq   H(X_1)+  H(Z^{(i ,i)}_{2}) \nonumber\\& +\sum_{p\in [2:\Nsf]}H\left( Z^{(i ,u_2)}_{2,u_{p}} ,\ldots, Z^{(i ,u_{p})}_{2,u_{p}}|   Z^{(i ,i)}_{2,u_{p}}  \right) \\
 &\leq  H(X_1)+  H(Z^{(i ,i)}_{2})+ (\Nsf-1)\left(\frac{\Bsf}{2}-\frac{y\Bsf}{ \Nsf} \right) \\
 &= H(X_1)+\frac{\Nsf \Bsf}{2}+ y\Bsf + (\Nsf-1)\left(\frac{\Bsf}{2}-\frac{y\Bsf}{ \Nsf} \right).
 \end{align}
 \end{subequations}
 Hence, we have 
 \begin{align}
 H(X_1) \geq \frac{\Bsf}{2}-\frac{y\Bsf}{ \Nsf},\label{eq:first approach case 1 one perm}
 \end{align}
 which proves~\eqref{eq:proof of first approach bound 1}.
 %By considering all permutations of $[\Nsf]$ where the first element is $i$ and summing all inequalities as~\eqref{eq:first approach case 1 one perm}, we can obtain~\eqref{eq:proof of first approach bound 1}.
  \subsection{Proof of~\eqref{eq:proof of first approach bound 2}}
 \label{sub:proof of first approach bound 2}
We first prove for each $i\in [\Nsf]$ and $n,p\in [\Nsf]\setminus \{i\}$, we have 
 \begin{subequations}
\begin{align}
  &H\left( Z^{(i ,n)}_{2,p}  |   Z^{(i ,i)}_{2,p}  \right)=  H\left( Z^{(i ,n)}_{2,p}  |   Z^{(i ,i)}_{2,p}, F_{p}\setminus Z^i_{1,p}  \right)   \label{eq:contain unknown of Z_1}\\
 &=  H\left( Z^{(i ,n)}_{2,p} \cap Z^i_{1,p} |   Z^{(i ,i)}_{2,p}, F_{p}\setminus Z^i_{1,p}  \right)   \label{eq:Z_1 indep 1}\\
 &=  H\left( Z^{(i ,n)}_{2,p} \cap Z^i_{1,p} |   Z^{(i ,i)}_{2,p}  \right) \label{eq:Z_1 indep 2}\\
&\leq H\left(  Z^{(i ,n)}_{2,p} \cap Z^i_{1,p}   \right)-q^i_{1,p}  \label{eq:bound by q_1 u2}\\
& =  \frac{2y\Bsf}{\Nsf} -q^i_{1,p} ,\label{eq:sy/n-q}
\end{align} 
 \end{subequations}
 where~\eqref{eq:contain unknown of Z_1} follows since $ Z^{(i ,i)}_{2,p} \cup Z^i_{1,p} =  Z^{(i ,n)}_{2,p} \cup Z^i_{1,p} =F_{p} $ and thus  $  (F_{p}\setminus Z^i_{1,p}) \subseteq Z^{(i ,i)}_{2,p} $,
  \eqref{eq:Z_1 indep 1} and~\eqref{eq:Z_1 indep 2} follow since all bits in the library are independent, 
   \eqref{eq:bound by q_1 u2} comes from~\eqref{eq:Qi1p},~\eqref{eq:sy/n-q} comes from~\eqref{eq:all known size i}.
 
 Now we bound each term $H\left( Z^{(i ,u_2)}_{2,u_{p}} ,\ldots, Z^{(i ,u_{p})}_{2,u_{p}}|   Z^{(i ,i)}_{2,u_{p}}  \right)$ where $p\in [3:\Nsf]$ in~\eqref{eq:recall approache 1 step 2} by $\frac{\Bsf}{2}-\frac{y\Bsf}{ \Nsf}$, to obtain
 \begin{subequations}
 \begin{align}
  &H(F_1,\ldots, F_{\Nsf})  \leq   H(X_1)+  H(Z^{(i ,i)}_{2})\nonumber\\&+\sum_{p\in [2:\Nsf]}H\left( Z^{(i ,u_2)}_{2,u_{p}} ,\ldots, Z^{(i ,u_{p})}_{2,u_{p}}|   Z^{(i ,i)}_{2,u_{p}}  \right) \\
 &  \leq   H(X_1)+ \frac{\Nsf \Bsf}{2}+ y\Bsf + H\left( Z^{(i ,u_2)}_{2,u_{2}}  |   Z^{(i ,i)}_{2,u_{2}}  \right) \nonumber\\& + (\Nsf-2) \left(\frac{\Bsf}{2}-\frac{y\Bsf}{ \Nsf} \right) \label{eq:before derivation first approach}\\
 &\leq H(X_1)+  \frac{\Nsf \Bsf}{2}+ y\Bsf+  \frac{2y\Bsf}{\Nsf} -q^i_{1,u_2} \nonumber\\& + (\Nsf-2) \left(\frac{\Bsf}{2}-\frac{y\Bsf}{ \Nsf} \right),\label{eq:common term less than}
 \end{align}
 \end{subequations}
 where~\eqref{eq:common term less than} comes from~\eqref{eq:sy/n-q}.
    
   Hence, we have 
    \begin{subequations}
   \begin{align}
  & H(X_1) \geq \frac{\Nsf \Bsf}{2}- y\Bsf-\frac{2y\Bsf}{\Nsf} +q^i_{1,u_2}-(\Nsf-2) \left(\frac{\Bsf}{2}-\frac{y\Bsf}{ \Nsf} \right)\\
  &=\Bsf-\frac{4y\Bsf}{\Nsf}+q^i_{1,u_2},\label{eq:first approach case 2 one perm}
   \end{align}
    \end{subequations}
    which proves~\eqref{eq:proof of first approach bound 2}.
  %   By considering all permutations of $[\Nsf]$ where the first element is $i$ and summing all inequalities as~\eqref{eq:first approach case 2 one perm}, we can obtain~\eqref{eq:proof of first approach bound 2}.
  \subsection{Proof of~\eqref{eq:proof of first approach bound 3}}
 \label{sub:proof of first approach bound 3}
 Note that when $\Nsf=2$,~\eqref{eq:proof of first approach bound 3} does not exist. Hence, in the following we consider $\Nsf \geq 3$ to prove~\eqref{eq:proof of first approach bound 3}.   
 
 From~\eqref{eq:recall approache 1 step 2}, we have 
\begin{subequations} 
\begin{align}
 &H(F_1,\ldots, F_{\Nsf})  \leq   H(X_1)+  H(Z^{(i ,i)}_{2}) \nonumber\\& +\sum_{p\in [2:\Nsf]}H\left( Z^{(i ,u_2)}_{2,u_{p}} ,\ldots, Z^{(i ,u_{p})}_{2,u_{p}}|   Z^{(i ,i)}_{2,u_{p}}  \right)\\
 &=H(X_1)+ H(Z^{(i ,i)}_{2}) + \sum_{p\in [2:\Nsf]}\left\{ H\left(  Z^{(i ,u_{p})}_{2,u_{p}}|   Z^{(i ,i)}_{2,u_{p}}  \right) \right. \nonumber\\& \left. + H\left( Z^{(i ,u_2)}_{2,u_{p}} ,\ldots, Z^{(i ,u_{p-1})}_{2,u_{p}}|   Z^{(i ,i)}_{2,u_{p}},Z^{(i ,u_{p})}_{2,u_{p}}  \right) \right\}
\\
  &=H(X_1)+ \frac{\Nsf \Bsf}{2}+ y\Bsf + \sum_{p\in [2:\Nsf]}  H\left(  Z^{(i ,u_{p})}_{2,u_{p}}|   Z^{(i ,i)}_{2,u_{p}}  \right) \nonumber\\& +  \sum_{p\in [2:\Nsf]} H\left( Z^{(i ,u_2)}_{2,u_{p}} ,\ldots, Z^{(i ,u_{p-1})}_{2,u_{p}}|   Z^{(i ,i)}_{2,u_{p}},Z^{(i ,u_{p})}_{2,u_{p}}  \right).  \label{eq:from recall approache 1 step 2} 
\end{align}
 \end{subequations} 
By considering all permutations of $[\Nsf]$ where the first element is $i$ and summing all inequalities as~\eqref{eq:from recall approache 1 step 2}, we have 
\begin{subequations}
\begin{align}
&H(X_1)\geq \frac{\Nsf \Bsf}{2}- y\Bsf - \frac{1}{(\Nsf-1)!} \sum_{\uv:u_1=i} \sum_{p\in [2:\Nsf]} \nonumber\\& H\left(  Z^{(i ,u_{p})}_{2,u_{p}}|   Z^{(i ,i)}_{2,u_{p}}  \right)- \frac{1}{(\Nsf-1)!} \sum_{\uv: u_1=i}
\sum_{p\in [2:\Nsf]} \nonumber\\
& H\left( Z^{(i ,u_2)}_{2,u_{p}} ,\ldots, Z^{(i ,u_{p-1})}_{2,u_{p}}|   Z^{(i ,i)}_{2,u_{p}},Z^{(i ,u_{p})}_{2,u_{p}}  \right)\\
%&=\frac{\Nsf \Bsf}{2}- y\Bsf - \negmedspace\negmedspace \sum_{p\in [2:\Nsf]} \negmedspace\negmedspace H\left(  Z^{(i ,p)}_{2,p}|   Z^{(i ,i)}_{2,p}  \right) - \frac{1}{(\Nsf-1)!} \sum_{\uv: u_1=i}
%\sum_{p\in [2:\Nsf]}  H\left( Z^{(i ,u_2)}_{2,u_{p}} ,\ldots, Z^{(i ,u_{p-1})}_{2,u_{p}}|   Z^{(i ,i)}_{2,u_{p}},Z^{(i ,u_{p})}_{2,u_{p}}  \right)\label{eq:first approach rearrange 1}\\
&=\frac{\Nsf \Bsf}{2}- y\Bsf - \negmedspace\negmedspace \sum_{p\in [ \Nsf]\setminus\{i\}} \negmedspace\negmedspace  H\left(  Z^{(i ,p)}_{2,p}|   Z^{(i ,i)}_{2,p}  \right)  - \frac{1}{(\Nsf-1)!} \sum_{p\in [ \Nsf]\setminus\{i\}} \nonumber\\& \sum_{r\in [2:\Nsf]} \sum_{\uv: u_1=i,u_r=p} \negmedspace\negmedspace\negmedspace\negmedspace H\left( Z^{(i ,u_2)}_{2,p} ,\ldots, Z^{(i ,u_{r-1})}_{2,p}|   Z^{(i ,i)}_{2,p},Z^{(i ,p)}_{2,p}  \right),\label{eq:first approach rearrange 2} 
\end{align}
 \end{subequations} 
 where~\eqref{eq:first approach rearrange 2} comes from the re-arrangements on the summations.

To bound the last term in~\eqref{eq:first approach rearrange 2}, we now focus on    one file $F_p$ where $p\in [ \Nsf]\setminus\{i\}$ and bound the following term
\begin{align}
  \sum_{r\in [2:\Nsf]} \sum_{\uv: u_1=i,u_r=p} H\left( Z^{(i ,u_2)}_{2,p} ,\ldots, Z^{(i ,u_{r-1})}_{2,p}|   Z^{(i ,i)}_{2,p},Z^{(i ,p)}_{2,p}  \right).\label{eq:bound term first approach}
\end{align}
 Note that the conditional entropies in~\eqref{eq:bound term first approach} are conditioned on the same term, which is $ Z^{(i ,i)}_{2,p}\cup Z^{(i ,p)}_{2,p}$.
 In addition, for any $n\in [\Nsf]\setminus \{i,p\}$, we have 
 $$
 Z^{(i ,n)}_{2,p} \setminus ( Z^{(i ,i)}_{2,p}\cup Z^{(i ,p)}_{2,p} ) \subseteq F_p \setminus ( Z^{(i ,i)}_{2,p}\cup Z^{(i ,p)}_{2,p} ).
 $$
 Hence, we divide the bits in $F_p \setminus ( Z^{(i ,i)}_{2,p}\cup Z^{(i ,p)}_{2,p} )$ into sub-pieces, and denote (with a slight abuse of notation)
\begin{align}
& F_p \setminus ( Z^{(i ,i)}_{2,p}\cup Z^{(i ,p)}_{2,p} )=\{\Fc_{p,\Sc}: \Sc\subseteq ([\Nsf]\setminus \{i,p\}) \},  
\end{align}
where
\begin{align}
\Fc_{p,\Sc} =  &\left(F_p \setminus ( Z^{(i ,i)}_{2,p}\cup Z^{(i ,p)}_{2,p} ) \right) \cap  \left(\cap_{n\in \Sc}  Z^{(i ,n)}_{2,p}\right)\nonumber\\& \setminus \left(\cup_{n_1\notin \Sc}  Z^{(i ,n_1)}_{2,p}\right). \label{eq:def of fps}
\end{align}
In other words,  $\Fc_{p,\Sc} $ represents the    bits in $F_p \setminus ( Z^{(i ,i)}_{2,p}\cup Z^{(i ,p)}_{2,p} )$ which are exclusively in $ Z^{(i ,n)}_{2,p}$ where $n\in \Sc$.
 
 We then define 
\begin{align}
f_{t}:= \sum_{ \Sc\subseteq ([\Nsf]\setminus \{i,p\}): |\Sc|=t} |\Fc_{p,\Sc}|, \ \forall t\in [0:\Nsf-2], \label{eq:def of fpt}
\end{align} 
as the total length of sub-pieces $\Fc_{p,\Sc}$ where $|\Sc|=t$. 
 
 In~\eqref{eq:sy/n-q}, we proved that for each $n\in [\Nsf]\setminus \{i,p\}$, we have 
$
 H ( Z^{(i ,n)}_{2,p}  |   Z^{(i ,i)}_{2,p}   ) \leq \frac{2y\Bsf}{\Nsf} -q^i_{1,p}.
$
 Hence, we also have  $ H ( Z^{(i ,n)}_{2,p}  |   Z^{(i ,i)}_{2,p}, Z^{(i ,p)}_{2,p}  )\leq  H ( Z^{(i ,n)}_{2,p}  |   Z^{(i ,i)}_{2,p}   ) \leq \frac{2y\Bsf}{\Nsf} -q^i_{1,p}$. In other words,
 \begin{align}
\sum_{\Sc \subseteq [\Nsf]\setminus \{i,p\}: n\in \Sc } |\Fc_{p,\Sc}|\leq \frac{2y\Bsf}{\Nsf} -q^i_{1,p}.\label{eq:first approach cache size one n}
 \end{align}
 By summing~\eqref{eq:first approach cache size one n} over all $n\in [\Nsf]\setminus \{i,p\}$, we have 
 \begin{subequations}
 \begin{align}
 \sum_{t\in [0:\Nsf-2]}t f_t &=\sum_{n\in [\Nsf]\setminus \{i,p\}} \ \sum_{\Sc \subseteq [\Nsf]\setminus \{i,p\}: n\in \Sc } |\Fc_{p,\Sc}| \\& \leq (\Nsf-2) \left(\frac{2y\Bsf}{\Nsf} -q^i_{1,p}\right).\label{eq:first approach cache size all n}
 \end{align}
 \end{subequations}  
  
 In addition, since $F_p \setminus ( Z^{(i ,i)}_{2,p}\cup Z^{(i ,p)}_{2,p} )= (F_p \setminus Z^{(i ,i)}_{2,p}) \setminus (Z^{(i ,p)}_{2,p} \setminus  Z^{(i ,i)}_{2,p})$, we have 
  \begin{subequations}
\begin{align}
 |F_p \setminus ( Z^{(i ,i)}_{2,p}\cup Z^{(i ,p)}_{2,p} )|&=|F_p \setminus   Z^{(i ,i)}_{2,p}|- |Z^{(i ,p)}_{2,p} \setminus  Z^{(i ,i)}_{2,p}| \\& 
=\frac{\Bsf}{2}-\frac{y\Bsf}{\Nsf}-H(Z^{(i ,p)}_{2,p} |  Z^{(i ,i)}_{2,p}).
\end{align}
 \end{subequations}  
 Hence, we have 
   \begin{subequations}
 \begin{align}
 \sum_{t\in [0:\Nsf-2]} f_t& =\sum_{\Sc \subseteq [\Nsf]\setminus \{i,p\} } |\Fc_{p,\Sc}| \\& =\frac{\Bsf}{2}-\frac{y\Bsf}{\Nsf}-H(Z^{(i ,p)}_{2,p} |  Z^{(i ,i)}_{2,p}).\label{eq:first approach file size all n}
 \end{align}
  \end{subequations}

 From the above definitions, we can re-write~\eqref{eq:bound term first approach} as follows,
  \begin{align}
 & \sum_{r\in [2:\Nsf]} \sum_{\substack{  \uv:\\ u_1=i, u_r=p}} H\left( Z^{(i ,u_2)}_{2,p} ,\ldots, Z^{(i ,u_{r-1})}_{2,p}|   Z^{(i ,i)}_{2,p},Z^{(i ,p)}_{2,p}  \right) \nonumber\\& = \sum_{r\in [2:\Nsf]} \sum_{\substack{  \uv:\\ u_1=i, u_r=p}} \sum_{\substack{ \Sc\subseteq ([\Nsf]\setminus \{i,p\}): \\ \Sc\cap \{u_2,\ldots,u_{r-1}\}\neq \emptyset}} |\Fc_{p,\Sc}|.\label{eq:before all r}
  \end{align}

In~\eqref{eq:before all r}, for each $r\in [2:\Nsf]$, we can compute
 %\begin{subequations}
\begin{align}
&\sum_{\substack{  \uv:\\ u_1=i, u_r=p}} \sum_{\substack{ \Sc\subseteq ([\Nsf]\setminus \{i,p\}): \\ \Sc\cap \{u_2,\ldots,u_{r-1}\}\neq \emptyset}} |\Fc_{p,\Sc}| \nonumber\\&  =\sum_{t\in [0:\Nsf-2]} (\Nsf-2)!\frac{\binom{\Nsf-2}{t}-\binom{\Nsf-r-1}{t}}{\binom{\Nsf-2}{t}}f_{t}. \label{eq:replace fps}
\end{align}
 %\end{subequations}
This is because in $ \sum_{\substack{ \Sc\subseteq ([\Nsf]\setminus \{i,p\}): \\ \Sc\cap \{u_2,\ldots,u_{r-1}\}\neq \emptyset}} |\Fc_{p,\Sc}| $, there are  $\binom{\Nsf-2}{t}-\binom{\Nsf-2-(r-1)}{t}$ sub-pieces whose $\Sc$ has $t$ elements. Considering all permutations $  \uv$ where  $u_1=i$ and  $u_r=p$, by the symmetry, the coefficient of each $|\Fc_{p,\Sc}|$ where $\Sc=t$ should be the same. In addition, there are in total $\binom{\Nsf-2}{t}$
 sub-pieces whose $\Sc$ has $t$ elements. Hence, we obtain~\eqref{eq:replace fps}.
 
 Considering all $r\in [2:\Nsf-2]$, from~\eqref{eq:replace fps} we have 
 \begin{subequations}
 \begin{align}
  & \sum_{r\in [2:\Nsf]} \sum_{\substack{  \uv:\\ u_1=i, u_r=p}} H\left( Z^{(i ,u_2)}_{2,p} ,\ldots, Z^{(i ,u_{r-1})}_{2,p}|   Z^{(i ,i)}_{2,p},Z^{(i ,p)}_{2,p}  \right)\nonumber\\& =\sum_{r\in[2:\Nsf]}\sum_{t\in [0:\Nsf-2]} (\Nsf-2)!\frac{\binom{\Nsf-2}{t}-\binom{\Nsf-r-1}{t}}{\binom{\Nsf-2}{t}}f_{t}\\
  &= (\Nsf-2)! \sum_{t\in [0:\Nsf-2]}\sum_{r\in[2:\Nsf]}\frac{\binom{\Nsf-2}{t}-\binom{\Nsf-r-1}{t}}{\binom{\Nsf-2}{t}}f_{t}\\
  &=(\Nsf-2)! \sum_{t\in [0:\Nsf-2]} \left( \frac{(\Nsf-2)\binom{\Nsf-2}{t}-\binom{\Nsf-2}{t+1}}{\binom{\Nsf-2}{t}}\right)f_{t}\label{eq:from pascal}\\
  &=(\Nsf-1)! \sum_{t\in [0:\Nsf-2]} \frac{t}{t+1} f_{t},\label{eq:after pascal}
 \end{align}
\end{subequations} 
 where~\eqref{eq:from pascal} comes from the Pascal's Triangle, $\binom{\Nsf-3}{t}+\cdots+\binom{t}{t}=\binom{\Nsf-2}{t+1}$.
 
 The next step is to use Fourier-Motzkin elimination on $f_t$ where $t\in [0:\Nsf-2]$ in~\eqref{eq:after pascal} (as we did in~\cite{indexcodingcaching2020}) with the help of~\eqref{eq:first approach cache size all n} and~\eqref{eq:first approach file size all n}. More precisely,  we fix one integer $h\in[0:\Nsf-3]$.
We multiply~\eqref{eq:first approach cache size all n} by $\frac{(\Nsf-1)!}{(h+1)(h+2)}$ and multiply~\eqref{eq:first approach file size all n} by $\frac{(\Nsf-1)! h}{h+2}$, and sum them to obtain
\begin{align}
&\sum_{t\in [0:\Nsf-2]}\left( t \frac{(\Nsf-1)!}{(h+1)(h+2)}+ \frac{(\Nsf-1)! h}{h+2} \right) f_{t} \nonumber\\& \leq  \frac{(\Nsf-1)! (\Nsf-2)}{(h+1)(h+2)} \left(\frac{2 y \Bsf}{\Nsf} -q^{i}_{1,p}\right) \nonumber\\& + \frac{(\Nsf-1)! h}{h+2} \left(\frac{\Bsf}{2}-\frac{y \Bsf}{\Nsf} -H(Z^{(i ,p)}_{2,p} |  Z^{(i ,i)}_{2,p}) \right).\label{eq:after multiply}
\end{align} 
From~\eqref{eq:after multiply}, we have 
\begin{align}
&\frac{(\Nsf-1)! h}{h+1 }  f_{ h}+\frac{(\Nsf-1)! (h+1)}{h+2 }  f_{ h+1} \nonumber\\&\leq \frac{(\Nsf-1)! (\Nsf-2)}{(h+1)(h+2)} \left(\frac{2 y \Bsf}{\Nsf} -q^{i}_{p}\right) \nonumber\\& + \frac{(\Nsf-1)! h}{h+2} \left(\frac{\Bsf}{2}-\frac{y\Bsf}{\Nsf} -H(Z^{(i ,p)}_{2,p} |  Z^{(i ,i)}_{2,p}) \right)\nonumber\\
&-\sum_{t\in [0:\Nsf-2]:t\notin \{h, h+1\}}\left( t \frac{(\Nsf-1)!}{(h+1)(h+2)}+ \frac{(\Nsf-1)! h}{h+2} \right) f_{ t} .\label{eq:after multiply2}
\end{align}
We then take~\eqref{eq:after multiply2} into~\eqref{eq:after pascal} to obtain,
 \begin{subequations}
\begin{align}
& \sum_{r\in [2:\Nsf]} \sum_{\substack{  \uv:\\ u_1=i, u_r=p}} H\left( Z^{(i ,u_2)}_{2,p} ,\ldots, Z^{(i ,u_{r-1})}_{2,p}|   Z^{(i ,i)}_{2,p},Z^{(i ,p)}_{2,p}  \right) \nonumber\\
&\leq \frac{(\Nsf-1)! (\Nsf-2)}{(h+1)(h+2)} \left(\frac{2 y \Bsf}{\Nsf} -q^{i}_{1,p}\right) \nonumber\\& + \frac{(\Nsf-1)! h}{h+2} \left(\frac{\Bsf}{2}-\frac{y\Bsf}{\Nsf} -H(Z^{(i ,p)}_{2,p} |  Z^{(i ,i)}_{2,p}) \right)\nonumber\\
& -\sum_{t\in [0:\Nsf-2]} (\Nsf-1)!\frac{(h-t)(h+1-t)}{(h+1)(h+2)(t+1)}f_{ t} \\
&\leq\frac{(\Nsf-1)! (\Nsf-2)}{(h+1)(h+2)} \left(\frac{2 y \Bsf}{\Nsf} -q^{i}_{1,p}\right) \nonumber\\& + \frac{(\Nsf-1)! h}{h+2} \left(\frac{\Bsf}{2}-\frac{y\Bsf}{\Nsf} -H(Z^{(i ,p)}_{2,p} |  Z^{(i ,i)}_{2,p}) \right).\label{eq:after elimination}
\end{align}
 \end{subequations}

Finally, we take~\eqref{eq:after elimination} into~\eqref{eq:first approach rearrange 2} to obtain, for each $h\in [0:\Nsf-3]$,
 \begin{subequations}
\begin{align}
&H(X_1)\geq\frac{\Nsf \Bsf}{2}- y\Bsf - \negmedspace\negmedspace \sum_{p\in [ \Nsf]\setminus\{i\}} \negmedspace\negmedspace  H\left(  Z^{(i ,p)}_{2,p}|   Z^{(i ,i)}_{2,p}  \right) \nonumber\\&  -\frac{1}{(\Nsf-1)!}  \negmedspace\negmedspace \sum_{p\in [ \Nsf]\setminus\{i\}} \sum_{r\in [2:\Nsf]} \sum_{\substack{\uv:\\ u_1=i,u_r=p}} \nonumber\\&  H\left( Z^{(i ,u_2)}_{2,p} ,\ldots, Z^{(i ,u_{r-1})}_{2,p}|   Z^{(i ,i)}_{2,p},Z^{(i ,p)}_{2,p}  \right)\\
&\geq \frac{\Nsf \Bsf}{2}- y\Bsf -   \sum_{p\in [ \Nsf]\setminus\{i\}}   H\left(  Z^{(i ,p)}_{2,p}|   Z^{(i ,i)}_{2,p}  \right) \nonumber\\
&-\frac{1}{(\Nsf-1)!}\sum_{p\in [ \Nsf]\setminus\{i\}}\left\{ \frac{(\Nsf-1)! (\Nsf-2)}{(h+1)(h+2)} \left(\frac{2 y \Bsf}{\Nsf} -q^{i}_{1,p}\right) \right.\nonumber\\& \left.+ \frac{(\Nsf-1)! h}{h+2} \left(\frac{\Bsf}{2}-\frac{y\Bsf}{\Nsf} -H(Z^{(i ,p)}_{2,p} |  Z^{(i ,i)}_{2,p}) \right)\right\}\\
&=\frac{\Nsf \Bsf}{2}- y\Bsf -  \frac{2}{h+2} \negmedspace\negmedspace \sum_{p\in [ \Nsf]\setminus\{i\}} \negmedspace\negmedspace    H\left(  Z^{(i ,p)}_{2,p}|   Z^{(i ,i)}_{2,p}  \right) - \sum_{p\in [ \Nsf]\setminus\{i\}} \nonumber\\& \left\{ \frac{  (\Nsf-2)}{(h+1)(h+2)} \left(\frac{2 y \Bsf}{\Nsf} -q^{i}_{1,p}\right)+ \frac{  h}{h+2} \left(\frac{\Bsf}{2}-\frac{y\Bsf}{\Nsf}  \right)\right\} \label{eq:before final first approach}\\
&\geq \frac{\Nsf \Bsf}{2}- y\Bsf -  \frac{2}{h+2} \negmedspace\negmedspace  \sum_{p\in [ \Nsf]\setminus\{i\}} \negmedspace\negmedspace  \left(   \frac{2y\Bsf}{\Nsf} -q^i_{1,p} \right) - \sum_{p\in [ \Nsf]\setminus\{i\}} \nonumber\\& \left\{ \frac{  (\Nsf-2)}{(h+1)(h+2)} \left(\frac{2 y \Bsf}{\Nsf} -q^{i}_{1,p}\right)+ \frac{  h}{h+2} \left(\frac{\Bsf}{2}-\frac{y\Bsf}{\Nsf}  \right)\right\}, \label{eq:final first approach}
\end{align}
 \end{subequations}
where~\eqref{eq:final first approach} comes from~\eqref{eq:sy/n-q}. Hence, we prove~\eqref{eq:proof of first approach bound 3}.

 \section{Proofs of~\eqref{eq:proof of second approach bound 1},~\eqref{eq:proof of second approach bound 2}, and~\eqref{eq:proof of second approach bound 3}} 
 \label{sec:proof of second approach}
 The proofs of~\eqref{eq:proof of second approach bound 1} \eqref{eq:proof of second approach bound 2} \eqref{eq:proof of second approach bound 3} come  from a similar strategy used in Appendix~\ref{sec:proof of first approach}. Hence, in the following, we briefly describe the proofs of~\eqref{eq:proof of second approach bound 1} \eqref{eq:proof of second approach bound 2} \eqref{eq:proof of second approach bound 3}.
  
  Recall from~\eqref{eq:approache 2 step 1} that by considering a   permutation of $[\Nsf]$, assumed to be  $\uv=(u_1,\ldots,u_{\Nsf})$, where $u_1=i$, we can derive 
\begin{align}
 &H(X_2)\geq  \left(H(F_{i})-H(Z^{i}_{1,i})\right)+  \nonumber\\& \sum_{p\in [2:\Nsf]} \left(H(F_{u_{p}})-H(Z^{u_{p}}_{1,u_{p}},  \Qc^{i}_{1,u_{p}},\Qc^{u_2}_{1,u_{p}},\ldots,  \Qc^{u_{p-1}}_{1,u_{p}} ) \right).\label{eq:recall approache 2}
\end{align}
For each $p\in [2:\Nsf]$,   we have 
\begin{align}
H(F_{u_{p}})-H(Z^{u_{p}}_{1,u_{p}},   \Qc^{i}_{1,u_{p}},\Qc^{u_2}_{1,u_{p}}, \ldots,  \Qc^{u_{p-1}}_{1,u_{p}} )\geq 0.\label{eq:outer bound per term approache 2}
\end{align}
 
 \subsection{Proof of~\eqref{eq:proof of second approach bound 1}}
 \label{sub:proof of second approach bound 1}
Now we bound each term $H(F_{u_{p}})-H(Z^{u_{p}}_{1,u_{p}}, \Qc^{i}_{1,u_{p}},\Qc^{u_2}_{1,u_{p}},\ldots,  \Qc^{u_{p-1}}_{1,u_{p}} )$ where $p\in [2:\Nsf]$ in~\eqref{eq:recall approache 2} by $0$, to obtain 
 \begin{align}
 &H(X_2)\geq   H(F_{i})-H(Z^{i}_{1,i})=\frac{\Bsf}{2}-\frac{y\Bsf}{\Nsf},\label{eq:second approach case 1 j}
 \end{align}
 which proves~\eqref{eq:proof of second approach bound 1}.
%By considering all permutations of $[\Nsf]$ where the first element is $j$ and summing all inequalities as~\eqref{eq:second approach case 1 j}, we can obtain~\eqref{eq:proof of second approach bound 1}.
 
  \subsection{Proof of~\eqref{eq:proof of second approach bound 2}}
 \label{sub:proof of second approach bound 2}
 Now we bound each term $H(F_{u_{p}})-H(Z^{u_{p}}_{1,u_{p}},   \Qc^{i}_{1,u_{p}},\Qc^{u_2}_{1,u_{p}},\ldots,  \Qc^{u_{p-1}}_{1,u_{p}} )$ where $p\in [3:\Nsf]$ in~\eqref{eq:recall approache 2} by $0$, to obtain
  \begin{subequations}
 \begin{align}
 &H(X_2)\geq  \left(H(F_{i})-H(Z^{i}_{1,i})\right) \nonumber\\& + \sum_{p\in [2:\Nsf]} \left(H(F_{u_{p}})-H(Z^{u_{p}}_{1,u_{p}},   \Qc^{i}_{1,u_{p}},\Qc^{u_2}_{1,u_{p}},\ldots,  \Qc^{u_{p-1}}_{1,u_{p}} ) \right)\\
 &\geq  \left(H(F_{i})-H(Z^{i}_{1,i})\right)+\left(H(F_{u_{2}})-H(Z^{u_{2}}_{1,u_{2}},   \Qc^{i}_{1,u_{2}} ) \right)\\
 &\geq H(F_{i})-H(Z^{i}_{1,i}+H(F_{u_{2}})-H(Z^{u_{2}}_{1,u_{2}})-H(\Qc^{i}_{1,u_{2}})\\
 &=\Bsf -\frac{2y\Bsf}{\Nsf}-q^{i}_{1,u_{2}}.\label{eq:second approach case 2 j}
 \end{align}
  \end{subequations}
  which proves~\eqref{eq:proof of second approach bound 2}.
%       By considering all permutations of $[\Nsf]$ where the first element is $j$ and summing all inequalities as~\eqref{eq:second approach case 2 j}, we can obtain~\eqref{eq:proof of second approach bound 2}.
  \subsection{Proof of~\eqref{eq:proof of second approach bound 3}}
 \label{sub:proof of second approach bound 3}
 Note that when $\Nsf=2$,~\eqref{eq:proof of second approach bound 3} does not exist. Hence, in the following we consider $\Nsf \geq 3$ to prove~\eqref{eq:proof of second approach bound 3}. 
 
 From~\eqref{eq:recall approache 2}, we have 
   \begin{subequations}
 \begin{align}
  &H(X_2)\geq  \left(H(F_{i})-H(Z^{i}_{1,i})\right)  \nonumber\\& + \sum_{p\in [2:\Nsf]} \left(H(F_{u_{p}})-H(Z^{u_{p}}_{1,u_{p}},  \Qc^{i}_{1,u_{p}},\Qc^{u_2}_{1,u_{p}},\ldots,  \Qc^{u_{p-1}}_{1,u_{p}} ) \right)\\
  &=\left(H(F_{i})-H(Z^{i}_{1,i})\right)+ \negmedspace\negmedspace \sum_{p\in [2:\Nsf]} \negmedspace\negmedspace \left(H(F_{u_{p}})-H(Z^{u_{p}}_{1,u_{p}}) \right.\nonumber\\& \left. -H(\Qc^{i}_{1,u_{p}}|Z^{u_{p}}_{1,u_{p}})-H( \Qc^{u_2}_{1,u_{p}},\ldots,  \Qc^{u_{p-1}}_{1,u_{p}} | Z^{u_{p}}_{1,u_{p}},  \Qc^{i}_{1,u_{p}} )\right)\\ 
  &=\Nsf\left(\frac{\Bsf}{2}-\frac{y\Bsf}{\Nsf}\right)-\sum_{p\in [2:\Nsf]}  H(\Qc^{i}_{1,u_{p}}|Z^{u_{p}}_{1,u_{p}}) \nonumber\\& - \sum_{p\in [2:\Nsf]}  H( \Qc^{u_2}_{1,u_{p}},\ldots,  \Qc^{u_{p-1}}_{1,u_{p}}|Z^{u_{p}}_{1,u_{p}},  \Qc^{i}_{1,u_{p}}).  \label{eq:from recall approache 2 step 2} 
 \end{align}
   \end{subequations}
  By considering all permutations of $[\Nsf]$ where the first element is $i$ and summing all inequalities as~\eqref{eq:from recall approache 2 step 2}, we can obtain
     \begin{subequations}
  \begin{align}
   &H(X_2)\geq \Nsf\left(\frac{\Bsf}{2}-\frac{y\Bsf}{\Nsf}\right) \nonumber\\& -\frac{1}{(\Nsf-1)!}\sum_{\uv:u_1=i} \sum_{p\in [2:\Nsf]}  H(\Qc^{i}_{1,u_{p}}|Z^{u_{p}}_{1,u_{p}})\nonumber\\&-\frac{1}{(\Nsf-1)!}\sum_{\uv:u_1=i}\sum_{p\in [2:\Nsf]}    H( \Qc^{u_2}_{1,u_{p}},\ldots,  \Qc^{u_{p-1}}_{1,u_{p}}|Z^{u_{p}}_{1,u_{p}},  \Qc^{i}_{1,u_{p}})\\
   &=\Nsf\left(\frac{\Bsf}{2}-\frac{y\Bsf}{\Nsf}\right)- \negmedspace\negmedspace\sum_{p\in [ \Nsf]\setminus\{i\}} \negmedspace\negmedspace H(\Qc^{i}_{1,p}|Z^{p}_{1,p}) - \frac{1}{(\Nsf-1)!}  \nonumber\\& \sum_{p\in [ \Nsf]\setminus\{i\}}\sum_{r\in [2:\Nsf]} \sum_{\uv: u_1=i,u_r=p} \negmedspace\negmedspace\negmedspace\negmedspace H( \Qc^{u_2}_{1,p},\ldots,  \Qc^{u_{r-1}}_{1,p}|Z^{p}_{1,p},  \Qc^{i}_{1,p}),\label{eq:second approach rearrange 2} 
\end{align}
 \end{subequations} 
 where~\eqref{eq:second approach rearrange 2} comes from the re-arrangements on the summations.

To bound the last term in~\eqref{eq:second approach rearrange 2}, we now focus on    one file $F_p$ where $ p\in [ \Nsf]\setminus\{i\}$ and bound the following term
\begin{align}
  \sum_{r\in [2:\Nsf]} \sum_{\uv: u_1=i,u_r=p} H( \Qc^{u_2}_{1,p},\ldots,  \Qc^{u_{r-1}}_{1,p}|Z^{p}_{1,p},  \Qc^{i}_{1,p}).\label{eq:bound term second approach}
\end{align}
We divide the bits in $F_p \setminus (Z^{p}_{1,p} \cup \Qc^{i}_{1,p})$ into sub-pieces, and denote 
\begin{align}
& F_p \setminus ( Z^{p}_{1,p} \cup \Qc^{i}_{1,p})=\{\Gc_{p,\Sc}: \Sc\subseteq ([\Nsf]\setminus \{i,p\}) \}, 
\end{align}
where 
\begin{align}
\Gc_{p,\Sc}=  \left(F_p \setminus ( Z^{p}_{1,p} \cup \Qc^{i}_{1,p}) \right) \cap  \left(\cap_{n\in \Sc}  \Qc^{n}_{1,p}\right) \setminus \left(\cup_{n_1\notin \Sc}  \Qc^{n_1}_{1,p}\right). \label{eq:def of gps}
\end{align}
 We then define 
\begin{align}
g_{t}:= \sum_{ \Sc\subseteq ([\Nsf]\setminus \{i,p\}): |\Sc|=t} |\Gc_{p,\Sc}|, \ \forall t\in [0:\Nsf-2]. \label{eq:def of gpt}
\end{align}  

 For each $n\in [\Nsf]\setminus \{i,p\}$, we have $H( \Qc^{n}_{1,p}| Z^{p}_{1,p},  \Qc^{i}_{1,p})\leq H( \Qc^{n}_{1,p} )$. Hence, we have 
 \begin{align}
 \sum_{t\in [0:\Nsf-2]}t g_t \leq \sum_{n\in [\Nsf]\setminus \{i,p\}} q^{n}_{1,p}.\label{eq:second approach cache size}
 \end{align}
 
  In addition, since $F_p \setminus  ( Z^{p}_{1,p} \cup \Qc^{i}_{1,p})= (F_p \setminus Z^{p}_{1,p}) \setminus ( \Qc^{i}_{1,p}  \setminus   Z^{p}_{1,p})$, we have 
 \begin{align}
 \sum_{t\in [0:\Nsf-2]} g_t=\sum_{\Sc \subseteq [\Nsf]\setminus \{i,p\} } |\Gc_{p,\Sc}|=\frac{\Bsf}{2}-\frac{y\Bsf}{\Nsf}-H(\Qc^{i}_{1,p}  |   Z^{p}_{1,p}).\label{eq:second approach file size}
 \end{align} 
 
  From the above definitions, we can re-write~\eqref{eq:second approach rearrange 2} (as we did to obtain~\eqref{eq:after pascal}),
  \begin{subequations}
  \begin{align}
  & \sum_{r\in [2:\Nsf]} \sum_{\substack{ \uv:\\ u_1=i,u_r=p}} H( \Qc^{u_2}_{1,p},\ldots,  \Qc^{u_{r-1}}_{1,p}|Z^{p}_{1,p},  \Qc^{i}_{1,p})\nonumber\\& = \sum_{r\in [2:\Nsf]} \sum_{\substack{ \uv:\\ u_1=i,u_r=p}}\sum_{\substack{ \Sc\subseteq ([\Nsf]\setminus \{i,p\}): \\ \Sc\cap \{u_2,\ldots,u_{r-1}\}\neq \emptyset}} |\Gc_{p,\Sc}|\\
&  =\sum_{r\in[2:\Nsf]}\sum_{t\in [0:\Nsf-2]} (\Nsf-2)!\frac{\binom{\Nsf-2}{t}-\binom{\Nsf-r-1}{t}}{\binom{\Nsf-2}{t}}g_{t}\\
    &=(\Nsf-1)! \sum_{t\in [0:\Nsf-2]} \frac{t}{t+1} g_{t}.
  \label{eq:second after pascal}  
  \end{align}
\end{subequations}

By  Fourier-Motzkin elimination on $g_t$ where $t\in [0:\Nsf-2]$ in~\eqref{eq:second after pascal}  with the help of~\eqref{eq:second approach cache size} and~\eqref{eq:second approach file size}, we obtain for each $h\in [0:\Nsf-3]$,
 %\begin{subequations}
\begin{align}
& \sum_{r\in [2:\Nsf]} \sum_{\substack{ \uv:\\ u_1=i,u_r=p}} H( \Qc^{u_2}_{1,p},\ldots,  \Qc^{u_{r-1}}_{1,p}|Z^{p}_{1,p},  \Qc^{i}_{1,p}) \nonumber\\ 
 &\leq \frac{(\Nsf-1)!  }{(h+1)(h+2)}  \sum_{n\in [\Nsf]\setminus \{p,i\}}q^{n}_{1,p} \nonumber\\& + \frac{(\Nsf-1)! h}{h+2} \left(\frac{\Bsf}{2}-\frac{y\Bsf}{\Nsf} -H(\Qc^{i}_{1,p}  |   Z^{p}_{1,p})\right).\label{eq:second after elimination}
\end{align}
  %\end{subequations} 
 
 Finally, by taking~\eqref{eq:second after elimination} into~\eqref{eq:second approach rearrange 2}, we obtain for each $h\in [0:\Nsf-3]$,
 \begin{subequations}
 \begin{align}
  &H(X_2)\geq \Nsf\left(\frac{\Bsf}{2}-\frac{y\Bsf}{\Nsf}\right)- \frac{2}{h+2}\sum_{p\in [ \Nsf]\setminus\{i\}} \negmedspace\negmedspace H(\Qc^{i}_{1,p}|Z^{p}_{1,p})\nonumber\\& -\sum_{p\in [ \Nsf]\setminus\{i\}}\left\{ \frac{ \sum_{n\in [\Nsf]\setminus \{p,i\}}q^{n}_{1,p} }{(h+1)(h+2)}  - \frac{  h}{h+2} \left(\frac{\Bsf}{2}-\frac{y\Bsf}{\Nsf}  \right) \right\} \\
  &\geq \Nsf\left(\frac{\Bsf}{2}-\frac{y\Bsf}{\Nsf}\right)- \frac{2}{h+2}\sum_{p\in [ \Nsf]\setminus\{i\}} q^{i}_{1,p} \nonumber\\& -\sum_{p\in [ \Nsf]\setminus\{i\}}\left\{ \frac{ \sum_{n\in [\Nsf]\setminus \{p,i\}}q^{n}_{1,p} }{(h+1)(h+2)}  - \frac{  h}{h+2} \left(\frac{\Bsf}{2}-\frac{y\Bsf}{\Nsf}  \right) \right\},\label{eq:second approach last one}
 \end{align}
 \end{subequations}
 where~\eqref{eq:second approach last one} follows since $H(\Qc^{i}_{1,p}|Z^{p}_{1,p})\leq H(\Qc^{i}_{1,p})= q^{i}_{1,p}$.
 Hence, we prove~\eqref{eq:proof of second approach bound 3}.

\section{Generalization of the Proof in Section~\ref{sub:K user converse}} 
\label{sec:general k,n}  
In Section~\ref{sub:K user converse}, we prove Theorem~\ref{thm:K user converse} for the case  where $\Ksf/2$ is an integer and $2\Nsf/\Ksf$ is also an integer. In the following, we only consider the   case  where $\Ksf/2$ is not integer and $\frac{\Nsf}{\left\lfloor \Ksf/2\right\rfloor }$ is not an integer neither. 
 The proof for the case where $\Ksf/2$ is an integer  and $2\Nsf/\Ksf$ is not an integer, or $\Ksf/2$ is not an integer  and $\frac{\Nsf}{\left\lfloor \Ksf/2\right\rfloor }$ is   an integer, can be directly derived from the following proof.

Recall $\Msf=\frac{\Nsf}{\Ksf}+\frac{2 y }{\Ksf}$, where $y\in \left[0,\frac{\Nsf}{2}\right]$.
We first fix one user $k\in [\Ksf]$ (assuming now $k=\Ksf$). We can divide the users in $[\Ksf]\setminus \{k\}$ into two groups, and generate an aggregate user for each group. Denoted
the  two aggregate users    by $k_1$ and $k_2$, respectively.  The cache size of each aggregate user is $\Msf \Bsf \times \frac{\Ksf-1}{2}$.
In addition, the demanded files of aggregate users $k_1$ and $k_2$ are the union sets of the demanded files of users in $[(\Ksf-1)/2]$ and   of users in $[(\Ksf+1)/2:\Ksf-1]$,   respectively.
  
 % Assume $a$ is the maximum integer such that $a\left\lfloor \Ksf/2\right\rfloor \leq \Nsf$ and denote $\Nsf_1= a \left\lfloor \Ksf/2\right\rfloor$. 
 By denoting    $\Nsf_1:= \left\lfloor 2\Nsf/\Ksf \right\rfloor \left\lfloor \Ksf/2\right\rfloor $, we  divide  files in $\left[ \Nsf_1 \right]$ into $\left\lfloor 2\Nsf/\Ksf \right\rfloor$  non-overlapping groups, each of which contains $\left\lfloor \Ksf/2\right\rfloor$ files. 
Each aggregate user demands one group of files. 

 We   assume  that the caches of aggregate users $k_1$ and $k_2$ are  $A^1_1$ and $A^1_2$.
The transmitted packets by aggregate users $k_1$ and $k_2$ are denoted by $X^{\prime}_1$ and $X^{\prime}_2$,  and   the transmitted packets by user $k=\Ksf$ are denoted by $X_{k}$, such that from $(X^{\prime}_2,X_k,A^1_1)$ aggregate user $k_1$ can decode the files in group $1$ and from $(X^{\prime}_1,X_k,A^1_2)$ aggregate user $k_2$ can also  decode the files in group $1$.
We then   construct   the cache configurations of aggregate users $k_1$ and $k_2$ by a $ \left\lfloor 2\Nsf/\Ksf \right\rfloor$-ary  tree, as we did in Section~\ref{sub:two user converse}.

   In the first approach, when we consider $A^i_1$ where $i\in [\left\lfloor 2\Nsf/\Ksf \right\rfloor]$ (cache of aggregate user $k_1$ where from $(X^{\prime}_2,X_k,A^i_1)$,  the files in group $i$ can be decoded),  by constructing a genie-aided  super-user as in~\eqref{eq:virtual user cache i1} (the cache of this super-user is denoted by $A$), by Fano's inequality,  
   \begin{align}
&   H(F_1,\ldots, F_{\Nsf}|\{F_{\ell}:\ell \in [\Nsf_1+1: \Nsf] \} ) \nonumber\\& \leq H(X^{\prime}_1) +H(X_k)  + H(A|\{F_{\ell}:\ell \in [\Nsf_1+1: \Nsf] \} ).\label{eq:fano}
   \end{align}
   
By considering one    permutation  of $[\left\lfloor 2\Nsf/\Ksf \right\rfloor]$, denoted by $\uv=(u_1,\ldots,u_{\left\lfloor 2\Nsf/\Ksf \right\rfloor})$ where  $u_1=i$,
by the similar derivations of~\eqref{eq:proof of first approach bound 1 k user} and~\eqref{eq:proof of first approach bound 2 k user}, we obtain
\begin{align}
 & H (X^{\prime}_1)+H(X_k) \geq \left(\frac{\Bsf}{2}-\frac{y \Bsf}{\Nsf}\right) \left\lfloor \Ksf/2\right\rfloor ; \label{eq:proof of first approach bound 1 k user no inte}\\
 & H (X^{\prime}_1)+H(X_k) \geq \left\lfloor \Ksf/2\right\rfloor\Bsf-\left\lfloor \Ksf/2\right\rfloor\frac{4 y \Bsf}{\Nsf}+  q^i_{1,u_2}.  \label{eq:proof of first approach bound 2 k user no inte} 
\end{align}

By considering all permutations of $[\left\lfloor 2\Nsf/\Ksf \right\rfloor]$ where the first element is $i$ to develop~\eqref{eq:fano} as we did in~\eqref{eq:approache 1 step 2}, %and sum  all inequalities as in~\eqref{eq:fano}. From the similar derivation of~\eqref{eq:proof of first approach bound 3 k user}, we obtain
 and by the similar derivation of~\eqref{eq:proof of first approach bound 3 k user}, we obtain
   \begin{subequations}
\begin{align}
   & H (X^{\prime}_1)+H(X_k) \nonumber\\& \geq 
  \left(\frac{ \Bsf}{2}- \frac{y  \Bsf}{\Nsf}  \right)  \Nsf_1  -\frac{2}{h+2} \left( (\left\lfloor 2\Nsf/\Ksf \right\rfloor-1  ) \frac{2y\Bsf}{\Nsf} \left\lfloor \Ksf/2\right\rfloor \right) \nonumber\\& +\frac{2}{h+2}\sum_{p\in\left[\left\lfloor 2\Nsf/\Ksf \right\rfloor \right]\setminus \{i\}} q^i_{1,p}  \nonumber\\
&-  \frac{ (\left\lfloor 2\Nsf/\Ksf \right\rfloor-1)(\left\lfloor 2\Nsf/\Ksf \right\rfloor-2)}{(h+1)(h+2)} \frac{2y\Bsf}{\Nsf}\left\lfloor \Ksf/2\right\rfloor \nonumber\\
& -\frac{(\left\lfloor 2\Nsf/\Ksf \right\rfloor-1)h}{h+2}\left(\frac{\Bsf}{2}-\frac{y\Bsf}{\Nsf} \right)\left\lfloor \Ksf/2\right\rfloor  \nonumber\\
&+ \frac{  \left\lfloor 2\Nsf/\Ksf \right\rfloor-2 }{(h+1)(h+2)} \sum_{p\in\left[\left\lfloor 2\Nsf/\Ksf \right\rfloor \right]\setminus \{i\}}q^i_{1,p}\nonumber\\
&\geq \frac{\Nsf_1}{\Nsf}\left\{ \left(\frac{ \Bsf}{2}- \frac{y  \Bsf}{\Nsf}  \right)  \Nsf   -\frac{2}{h+2} \left( (2\Nsf/\Ksf -1  ) \frac{2y\Bsf}{\Nsf} \frac{\Ksf}{2} \right)  \right. \nonumber\\
&\left. -  \frac{ (  2\Nsf/\Ksf  -1)(  2\Nsf/\Ksf  -2)}{(h+1)(h+2)} \frac{2y\Bsf}{\Nsf} \frac{\Ksf}{2}  - \frac{(  2\Nsf/\Ksf  -1)h}{h+2}\right. \nonumber\\& \left. \left(\frac{\Bsf}{2}-\frac{y\Bsf}{\Nsf} \right) \frac{\Ksf}{2}  \right\} +\left( \frac{2}{h+2} + \frac{  \left\lfloor 2\Nsf/\Ksf \right\rfloor-2 }{(h+1)(h+2)}\right)\nonumber\\
& \sum_{p\in\left[\left\lfloor 2\Nsf/\Ksf \right\rfloor \right]\setminus \{i\}}q^i_{1,p},
  \ \forall h\in \left[0:\left\lfloor 2\Nsf/\Ksf \right\rfloor -3 \right],  
   \label{eq:proof of first approach bound 3 k user no inte}
\end{align}
%\label{eq:proof of first approach k user no inte}
 \end{subequations} 
  where~\eqref{eq:proof of first approach bound 3 k user no inte} follows since 
\begin{align}
 \frac{\Nsf}{\Nsf_1} (\left\lfloor 2\Nsf/\Ksf \right\rfloor-1)\left\lfloor \Ksf/2\right\rfloor= \Nsf-\frac{\Nsf}{ \left\lfloor 2\Nsf/\Ksf \right\rfloor} \leq   (  2\Nsf/\Ksf  -1) \frac{\Ksf}{2}.
\end{align}

   By considering all $i\in [\left\lfloor 2\Nsf/\Ksf \right\rfloor]$ to bound $H(X^{\prime}_1)+H(X_k)$, and all   $j\in [\left\lfloor 2\Nsf/\Ksf \right\rfloor]$ to bound $H(X^{\prime}_2)+H(X_k)$, we sum all inequalities as~\eqref{eq:proof of first approach bound 3 k user no inte} to obtain (by the similar derivation of~\eqref{eq:proof of first approach case 3 k user}),
 % \begin{subequations}
\begin{align}
  &\Rsf^{\star}_{\mathrm{u,c}} \Bsf+H(X_k) \geq \frac{\Nsf_1}{\Nsf}\left\{ \left(\Bsf- \frac{2 y  \Bsf}{\Nsf}  \right)  \Nsf  -\frac{4}{h+2} \right. \nonumber\\&\left.  \left( ( 2\Nsf/\Ksf -1 ) \frac{2y\Bsf}{\Nsf} \frac{\Ksf}{2} \right)   -\frac{( 2\Nsf/\Ksf-1)( 2\Nsf/\Ksf-2)}{(h+1)(h+2) } \frac{4y  \Bsf }{\Nsf}\frac{\Ksf}{2} \right. \nonumber\\& \left.-\frac{h ( 2\Nsf/\Ksf-1 )}{(h+2) }\left( \Bsf -\frac{2y\Bsf}{\Nsf} \right)\frac{\Ksf}{2} \right\}
 +\left( \frac{2}{(h+2)\left\lfloor 2\Nsf/\Ksf \right\rfloor} \right. \nonumber\\ & \left.  
 +\frac{  \left\lfloor 2\Nsf/\Ksf \right\rfloor-2 }{(\left\lfloor 2\Nsf/\Ksf \right\rfloor)(h+1)(h+2)}\right)
 \left( \sum_{i\in  [\left\lfloor 2\Nsf/\Ksf \right\rfloor]}\sum_{p\in [\left\lfloor 2\Nsf/\Ksf \right\rfloor] \setminus \{i\}} q^i_{1,p} \right. \nonumber\\& \left. +\sum_{j\in  [\left\lfloor 2\Nsf/\Ksf \right\rfloor]}\sum_{p\in [\left\lfloor 2\Nsf/\Ksf \right\rfloor]\setminus \{j\}} q^j_{2,p} \right) , \   \forall h\in \left[0:\left\lfloor 2\Nsf/\Ksf \right\rfloor -3 \right].
\label{eq:proof of first approach case 3 k user no inte} 
 \end{align}
% \label{eq:proof of first approach final k user no inte}
%\end{subequations} 

  Similarly,  in the second approach, when we consider $(X^{\prime}_2,X_k)$ and the same permutation as the one to derive~\eqref{eq:proof of first approach bound 1 k user no inte} and~\eqref{eq:proof of first approach bound 2 k user no inte},  by constructing a genie-aided super-user as in~\eqref{eq:virtual user cache j2} (the cache of this super-user is denoted by $A^{\prime}$), by Fano's inequality,  
   \begin{align}
   &H(F_1,\ldots, F_{\Nsf}|\{F_{\ell}:\ell \in [\Nsf_1+1: \Nsf] \} ) \nonumber\\&  \leq H(X^{\prime}_2) +H(X_k)+H(A^{\prime}|\{F_{\ell}:\ell \in [\Nsf_1+1: \Nsf] \} ).\label{eq:fano 2}
   \end{align}
  By the similar derivations of~\eqref{eq:proof of second approach bound 1 k user} and~\eqref{eq:proof of second approach bound 2 k user}, we obtain
\begin{align}
 & H (X^{\prime}_2)+H(X_k) \geq \left(\frac{\Bsf}{2}-\frac{y \Bsf}{\Nsf}\right) \left\lfloor \Ksf/2\right\rfloor ; \label{eq:proof of second approach bound 1 k user no inte}\\
 & H (X^{\prime}_2)+H(X_k) \geq \left\lfloor \Ksf/2\right\rfloor\Bsf-\left\lfloor \Ksf/2\right\rfloor\frac{2 y \Bsf}{\Nsf}- q^i_{1,u_2}.  \label{eq:proof of second approach bound 2 k user no inte} 
\end{align}

 In addition, by   considering all permutations to   bound $H(X^{\prime}_1)+H(X_k)$  and all permutations to bound $H(X^{\prime}_2)+H(X_k)$,  we  sum all inequalities 
 %as in~\eqref{eq:fano 2} 
  to  obtain (by the similar derivation of~\eqref{eq:proof of second approach case 3 k user}),
  \begin{align}
&\Rsf^{\star}_{\mathrm{u,c}} \Bsf+H(X_k) \nonumber\\&   \geq \frac{\Nsf_1}{\Nsf} \left\{ \Nsf\Bsf -2y  \Bsf -\frac{h\left(\frac{2\Nsf}{\Ksf} -1\right)}{(h+2) }\left( \Bsf -\frac{2y\Bsf}{\Nsf} \right) \frac{\Ksf}{2} \right\} \nonumber\\& -\left(  \frac{2}{(h+2)\left\lfloor 2\Nsf/\Ksf \right\rfloor}  + \frac{\left\lfloor 2\Nsf/\Ksf \right\rfloor-2}{(h+1)(h+2)\left\lfloor 2\Nsf/\Ksf \right\rfloor } \right) \nonumber\\& \left(\sum_{j\in \left[\frac{2\Nsf}{\Ksf} \right]} \sum_{p\in\left[\frac{2\Nsf}{\Ksf} \right]\setminus \{j\}} q^j_{2,p}+\sum_{i\in \left[\frac{2\Nsf}{\Ksf} \right]} \sum_{p\in\left[\frac{2\Nsf}{\Ksf} \right]\setminus \{i\}} q^i_{1,p} \right),\nonumber\\&  \forall h\in \left[0:\left\lfloor 2\Nsf/\Ksf \right\rfloor-3\right].
\label{eq:proof of second approach case 3 k user no inte}
  \end{align}

   By summing~\eqref{eq:proof of first approach bound 1 k user no inte} and~\eqref{eq:proof of second approach bound 1 k user no inte}, summing~\eqref{eq:proof of first approach bound 2 k user no inte} and~\eqref{eq:proof of second approach bound 2 k user no inte}, and
   summing~\eqref{eq:proof of first approach case 3 k user no inte} and~\eqref{eq:proof of second approach case 3 k user no inte}, we obtain
      \begin{subequations}
  \begin{align}
  &\Rsf^{\star}_{\mathrm{u,c}} \Bsf+H(X_k)  \geq   \left( \Bsf -\frac{2y \Bsf}{\Nsf}\right) \left\lfloor \Ksf/2\right\rfloor; \label{eq:proof of sum case 1 k user no inte}\\  
  &\Rsf^{\star}_{\mathrm{u,c}} \Bsf+H(X_k)  \geq  \left( 2\Bsf-\frac{6 y \Bsf}{\Nsf} \right) \left\lfloor \Ksf/2\right\rfloor ; \label{eq:proof of sum case 2 k user no inte}\\
   &\Rsf^{\star}_{\mathrm{u,c}} \Bsf+H(X_k)  \nonumber\\& \geq \frac{\Nsf_1}{\Nsf}\left\{ \Nsf\Bsf -2y  \Bsf -\frac{2}{h+2} \left( ( 2\Nsf/\Ksf -1 ) \frac{2y\Bsf}{\Nsf} \frac{\Ksf}{2} \right) \right.\nonumber\\
&\left. -\frac{( 2\Nsf/\Ksf-1)( 2\Nsf/\Ksf-2)}{(h+1)(h+2) } \frac{2y  \Bsf }{\Nsf}\frac{\Ksf}{2} \right. \nonumber\\& \left. -\frac{h ( 2\Nsf/\Ksf-1 )}{(h+2) }\left( \Bsf -\frac{2y\Bsf}{\Nsf} \right)\frac{\Ksf}{2} \right\},
 \ \forall h\in \left[0:\left\lfloor 2\Nsf/\Ksf \right\rfloor-3\right].
\label{eq:proof of sum case 3 k user no inte}
  \end{align}
  \label{eq:proof of sum k user no inte}
     \end{subequations}

Finally we consider all       $k\in [\Ksf]$ and sum inequalities as~\eqref{eq:proof of sum k user no inte}, to obtain (recall that $\Rsf^{\star}_{\mathrm{u,c}} \Bsf \geq \sum_{k\in [\Ksf]} H(X_k)$),
  \begin{subequations}
\begin{align}
& \Rsf^{\star}_{\mathrm{u,c}} \Bsf   \geq  \frac{ \Ksf}{2 \left\lceil\Ksf/2 \right\rceil } \left( \Bsf -\frac{2y \Bsf}{\Nsf}\right) \left\lfloor \Ksf/2\right\rfloor \nonumber\\& =  \frac{\left\lfloor \Ksf/2 \right\rfloor }{ \left\lceil\Ksf/2 \right\rceil  } \left( \Bsf -\frac{2y \Bsf}{\Nsf}\right) \frac{\Ksf}{2}; \label{eq:proof of final sum case 1 k user no inte}\\  
  &\Rsf^{\star}_{\mathrm{u,c}} \Bsf   \geq \frac{ \Ksf}{2 \left\lceil\Ksf/2 \right\rceil }  \left( 2\Bsf-\frac{6 y \Bsf}{\Nsf} \right) \left\lfloor \Ksf/2\right\rfloor \nonumber\\& = \frac{\left\lfloor \Ksf/2 \right\rfloor }{ \left\lceil\Ksf/2 \right\rceil  } \left( 2\Bsf-\frac{6 y \Bsf}{\Nsf} \right) \frac{\Ksf}{2}; \label{eq:proof of final sum case 2 k user no inte}\\  
  &\Rsf^{\star}_{\mathrm{u,c}} \Bsf   \geq  \frac{ \Ksf}{2 \left\lceil\Ksf/2 \right\rceil } 
  \frac{\Nsf_1}{\Nsf}\left\{ \Nsf\Bsf -2y \Bsf -\frac{2}{h+2} \right. \nonumber\\ & \left. \left( ( 2\Nsf/\Ksf -1 ) \frac{2y\Bsf}{\Nsf} \frac{\Ksf}{2} \right)  -\frac{( 2\Nsf/\Ksf-1)( 2\Nsf/\Ksf-2)}{(h+1)(h+2) } \frac{2y  \Bsf }{\Nsf}\frac{\Ksf}{2} \right. \nonumber\\& \left. -\frac{h ( 2\Nsf/\Ksf-1 )}{(h+2) }\left( \Bsf -\frac{2y\Bsf}{\Nsf} \right)\frac{\Ksf}{2} \right\} \nonumber\\
&=  \frac{\left\lfloor \Ksf/2 \right\rfloor }{ \left\lceil\Ksf/2 \right\rceil  }\frac{\left\lfloor 2\Nsf/\Ksf \right\rfloor  }{  2\Nsf/\Ksf } \left\{ \Nsf\Bsf -2y \Bsf -\frac{2}{h+2} \right.\nonumber\\
&\left. \left( ( 2\Nsf/\Ksf -1 ) \frac{2y\Bsf}{\Nsf} \frac{\Ksf}{2} \right)  -\frac{( 2\Nsf/\Ksf-1)( 2\Nsf/\Ksf-2)}{(h+1)(h+2) } \frac{2y  \Bsf }{\Nsf}\frac{\Ksf}{2} \right.\nonumber\\
&\left. -\frac{h ( 2\Nsf/\Ksf-1 )}{(h+2) }\left( \Bsf -\frac{2y\Bsf}{\Nsf} \right)\frac{\Ksf}{2} \right\},
 \ \forall h\in \left[0:\left\lfloor 2\Nsf/\Ksf \right\rfloor-3\right],\label{eq:proof of final sum case 3 k user no inte}
\end{align}
\label{eq:proof of final sum k user no inte}
 \end{subequations}  
where~\eqref{eq:proof of final sum case 3 k user no inte} comes from (recall that $\Nsf_1:=\left\lfloor 2\Nsf/\Ksf \right\rfloor \left\lfloor \Ksf/2\right\rfloor$),
\begin{align}
 \frac{ \Ksf}{2 \left\lceil\Ksf/2 \right\rceil } 
  \frac{\Nsf_1}{\Nsf}=\frac{ \Ksf}{2 \left\lceil\Ksf/2 \right\rceil }  \frac{\left\lfloor 2\Nsf/\Ksf \right\rfloor \left\lfloor \Ksf/2\right\rfloor}{\Nsf}= \frac{\left\lfloor \Ksf/2 \right\rfloor }{ \left\lceil\Ksf/2 \right\rceil  }\frac{\left\lfloor 2\Nsf/\Ksf \right\rfloor  }{  2\Nsf/\Ksf }.
\end{align}
Hence, we prove Theorem~\ref{thm:K user converse}.

\section{Proof of Theorem~\ref{thm:OrderOptimalityA}}
\label{sec:proof of order optimality} 
We first provide a direct upper bound of the achieved load of Scheme A in Theorem~\ref{thm:SchemeA}, since $\frac{\binom{\Usf}{t}-\binom{\Usf-\Nsf}{t}}{ \binom{\Usf}{t-1}} \leq \frac{\binom{\Usf}{t} }{ \binom{\Usf}{t-1}}=\frac{\Usf-t+1}{t}$.
\begin{lem}
\label{lem:upper bound}
The achieved load of Scheme A in Theorem~\ref{thm:SchemeA} is upper bound by the lower convex envelop of $(\Nsf/\Ksf, \Nsf)$ and 
\begin{align}
& \left( \frac{\Nsf+t-1}{\Ksf},
\frac{\Usf-t+1}{t} \right), \ \forall  t\in [ \Usf+1].
\label{eq:extended scheme A}
\end{align}
\end{lem}
 
We then introduce the following lemma, whose proof is in Appendix~\ref{sec:proof of lemma1}.
\begin{lem}
\label{lem:multiplicative gap}
The multiplicative gap between the lower convex envelop of the memory-load tradeoff $\left(\frac{\Nsf+t_1-1 }{\Ksf},
\frac{\Usf-t_1+1}{t_1} \right)$ where $t_1\in [ \Usf]$,   and the lower convex envelop of the memory-load tradeoff $ \left(\frac{\Nsf t}{\Ksf},
\frac{\Ksf-t}{t+1} \right)$ where $t\in [2: \Ksf]$, is at most $3$ when $\Msf \geq \frac{2\Nsf}{\Ksf}$.
\end{lem}
 
 We then prove the two cases in Theorem~\ref{thm:OrderOptimalityA}, where $\Nsf\geq \Ksf$ and $\Nsf< \Ksf$.
 \subsection{$\Nsf \geq \Ksf$}
 \label{sub:N>K} 
{\it Converse.}
It was proved in~\cite{yufactor2TIT2018} that for the shared-link caching model with $\Nsf \geq \Ksf$, the   lower  convex envelope of the corner points $ \left( \frac{\Nsf t }{\Ksf }, \frac{\Ksf-t}{  t+1 } \right)$, where $t\in [0:\Ksf]$,  achieved by the MAN caching scheme in~\cite{dvbt2fundamental} is order optimal to within a factor of $2$.
 In addition, it was proved in~\cite{indexcodingcaching2020} that these corner points are successively convex. Hence, when $\Msf \geq 2\Nsf/\Ksf$, the lower convex envelop of $ \left( \frac{\Nsf t }{\Ksf }, \frac{\Ksf-t}{  t+1 } \right)$, where $t\in [2:\Ksf]$ is order optimal to  within a factor of $2$.
 We will also use this converse in our model. Hence, 
 for $\Msf \in [2\Nsf/\Ksf,\Nsf]$,
 $\Rsf^{\star}$ is lower bounded by  the   lower  convex envelope $ \left( \frac{\Nsf t }{\Ksf }, \frac{\Ksf-t}{ 2(t+1) } \right)$, where $t\in [2: \Ksf]$. 
 
 {\it Achievability.}
 From Lemma~\ref{lem:multiplicative gap}, it can be seen that from the proposed scheme in Theorem~\ref{thm:SchemeA}, we can achieve the lower convex envelop of the memory-load tradeoff $ \left(\frac{\Nsf t}{\Ksf},
\frac{3(\Ksf-t)}{t+1} \right)$ where $t\in [2: \Ksf]$.

As a result, the proposed scheme in Theorem~\ref{thm:SchemeA} is order optimal to  within a factor of $6$  when $\Nsf\geq \Ksf$  and $\Msf \geq \frac{2\Nsf}{\Ksf}$.
 \subsection{$\Nsf< \Ksf$}
 \label{sub:N<K} 
 {\it Converse.}
It was proved in~\cite{improvedlower2017Ghasemi} that for the shared-link caching model with $\Nsf<\Ksf$, the   lower  convex envelope of the corner points $(0,\Nsf)$ and $ \left( \frac{\Nsf t }{\Ksf }, \frac{\Ksf-t}{  t+1 } \right)$, where $t\in [\Ksf]$,  achieved by the MAN caching scheme in~\cite{dvbt2fundamental} is order optimal to  within a factor of $4$.

Since  the corner points $ \left( \frac{\Nsf t }{\Ksf }, \frac{\Ksf-t}{  t+1 } \right)$ where $t\in [\Ksf]$, are successively convex, the lower convex envelop of the MAN caching scheme for $\Nsf<\Ksf$ is as follows. There exists one $t_2 \in [\Ksf]$, such that the lower convex envelop of the MAN caching scheme for $\Msf\in [0,\Nsf t_2/\Ksf]$ is the memory-sharing between $(0,\Nsf)$ and  $ \left( \frac{\Nsf t_2 }{\Ksf }, \frac{\Ksf-t_2}{  t_2+1 } \right)$, while the lower convex envelop for $\Msf \in [\Nsf t_2/\Ksf, \Nsf]$ is the lower convex envelop of the successive corner points  $ \left( \frac{\Nsf t }{\Ksf }, \frac{\Ksf-t}{  t+1 } \right)$  where $t\in [t_2:\Ksf]$. 
In addition, it is obvious that $t_2$ is the maximum value among $x\in [\Ksf]$ such that the memory-sharing between $(0,\Nsf)$ and $ \left( \frac{\Nsf x}{\Ksf }, \frac{\Ksf-x}{  x+1 } \right)$ at the memory $\Msf^{\prime}=\frac{\Nsf(x-1)}{\Ksf}$ leads to a lower load than $\frac{\Ksf-x+1}{  x}$. More precisely, if we    interpolate $(0,\Nsf)$ and $ \left( \frac{\Nsf x }{\Ksf }, \frac{\Ksf-x}{ x+1 } \right)$ where $x\in [\Ksf]$ to match $\Msf^{\prime}=\frac{\Nsf(x-1)}{\Ksf}$, the achieved load is 
$$
-\frac{\Nsf-\frac{\Ksf-x}{x+1}}{\frac{\Nsf x}{\Ksf}} \frac{\Nsf (x-1)}{\Ksf}+\Nsf=\frac{(\Ksf-x)(x-1)}{x(x+1)}+\frac{\Nsf}{x}.
$$ 
Hence, we have 
\begin{subequations}
\begin{align}
t_2 &:= \arg\max_{x\in [\Ksf]} \left\{\frac{(\Ksf-x)(x-1)}{x(x+1)}+\frac{\Nsf}{x} \leq \frac{\Ksf-x+1}{  x} \right\} \\& = \left\lfloor \frac{2\Ksf-\Nsf+1}{\Nsf+1} \right\rfloor. \label{eq:def of t1} 
\end{align}
\end{subequations}

We then interpolate $(0,\Nsf)$ and $ \left( \frac{\Nsf t_2 }{\Ksf }, \frac{\Ksf-t_2}{ t_2+1 } \right)$ to match $\Msf_1=\Nsf/\Ksf$, to get the memory-load tradeoff 
\begin{align}
(\Msf_1,\Rsf_1)=\left(\frac{\Nsf}{\Ksf}, \Nsf-\frac{\Nsf-\frac{\Ksf-t_2}{t_2+1}}{t_2} \right).\label{eq:start point}
\end{align}
Hence, it is equivalent to say that the lower convex envelop of the achieved memory-load tradeoffs by the MAN caching scheme for $\Msf \geq \Nsf/\Ksf$ also has two regimes.
\begin{enumerate}
\item $\Msf\in \left[\frac{\Nsf}{\Ksf}, \frac{\Nsf t_2}{\Ksf}\right]$. The lower convex envelop is the memory-sharing between $(\Msf_1,\Rsf_1)$ and $\left(\frac{\Nsf t_2}{\Ksf},\frac{\Ksf-t_2}{  t_2+1 } \right)$.
\item $\Msf \in \left[ \frac{\Nsf t_2}{\Ksf} ,\Nsf\right]$. The lower convex envelop of the MAN scheme is the lower convex envelop  of  the corner points $ \left( \frac{\Nsf t }{\Ksf }, \frac{\Ksf-t}{  t+1 } \right)$, where $t\in [t_2:\Ksf]$. 
\end{enumerate}
Since the MAN scheme is order optimal to  within a factor of $4$,  $\Rsf^{\star}$ is lower bounded by  the   lower  convex envelope of the corner points  $\left(\Msf_1,\frac{\Rsf_1}{4}\right)$ and  $ \left( \frac{\Nsf t }{\Ksf }, \frac{\Ksf-t}{ 4(t+1) } \right)$, where $t\in [t_2: \Ksf]$.   

 {\it Achievability.}
Let us first focus on $\Msf=\Nsf/\Ksf$. The achieved load by  the proposed scheme in Theorem~\ref{thm:SchemeA} is $\Nsf$. In the following, we will prove $\Nsf \leq 2\Rsf_1$. More precisely,
\begin{align}
\Nsf-2\Rsf_1&=2\frac{\Nsf-\frac{\Ksf-t_2}{t_2+1}}{t_2} -\Nsf\nonumber\\
&=\frac{2\Nsf(t_2+1)- 2(\Ksf-t_2)-\Nsf t_2(t_2+1) }{t_2(t_2+1)}\nonumber\\
&=\frac{-\Nsf t_2^2+(\Nsf+2)t_2-2(\Ksf-\Nsf)  }{t_2(t_2+1)}\nonumber\\
&=\frac{-t_2(\Nsf t_2 -\Nsf -2)-2(\Ksf-\Nsf) }{t_2(t_2+1)}\nonumber\\
&=\frac{- (\Nsf t_2 -\Nsf -2)-\frac{2(\Ksf-\Nsf)}{t_2} }{ (t_2+1)}.\label{eq:discuss t1}
\end{align}
 We consider  the following two cases.
 \begin{enumerate}
 \item $t_2=1$. From~\eqref{eq:discuss t1}, we have 
 \begin{align}
 \Nsf-2\Rsf_1=\frac{2 - 2(\Ksf-\Nsf)  }{ 2} \leq  0,
 \end{align}
 which follows $\Ksf>\Nsf$.
 \item $t_2>1$.   From~\eqref{eq:discuss t1}, we have 
 \begin{align}
 \Nsf-2\Rsf_1 &\leq \frac{- (2 \Nsf -\Nsf -2)-\frac{2(\Ksf-\Nsf)}{t_2} }{  t_2+1 }<0,
 \end{align}
 which follows $\Nsf\geq 2$ and $\Ksf>\Nsf$.
 \end{enumerate}
 Hence, from the proposed scheme in Theorem~\ref{thm:SchemeA},  we can achieve $(\Msf_1,2\Rsf_1)$.
 In addition, from Lemma~\ref{lem:multiplicative gap}, it can be seen that from the proposed scheme in Theorem~\ref{thm:SchemeA}, we can achieve the lower convex envelop of the memory-load tradeoff $ \left(\frac{\Nsf t}{\Ksf},
\frac{3(\Ksf-t)}{t+1} \right)$ where $t\in [t_2: \Ksf]$.

 As a result, the proposed scheme in Theorem~\ref{thm:SchemeA} is order optimal to  within a factor of $12$  when $\Nsf<\Ksf$.

\section{Proof of Lemma~\ref{lem:multiplicative gap}}
\label{sec:proof of lemma1} 
It  was proved in~\cite{indexcodingcaching2020} that the  corner points  $ \left(\frac{\Nsf t}{\Ksf}, \frac{\Ksf-t}{t+1} \right)$ where $t\in [0: \Ksf]$ are successively convex, i.e., for each memory size $\Msf \in \left[ \frac{\Nsf t }{\Ksf },  \frac{\Nsf (t+1) }{\Ksf }\right] $ where $t\in [0:\Ksf-1]$, the lower convex envelop is obtained by memory-sharing between $\left(\frac{\Nsf t }{\Ksf },  \frac{\Ksf-t}{  t+1 }\right)$ and $\left(\frac{\Nsf(t+1) }{\Ksf },  \frac{\Ksf-t-1}{  t+2 }\right)$. Hence, in order to prove Lemma~\ref{lem:multiplicative gap}, in the following we prove from %$ \left(\frac{(\Ksf-1)(t^{\prime}-1)+\Usf }{\Ksf\Usf}\Nsf,
$\left(\frac{\Nsf+t_1-1 }{\Ksf},
\frac{\Usf-t_1+1}{t_1} \right)$ where $t_1\in [ \Usf]$, we can achieve  $ \left(\frac{\Nsf t}{\Ksf},
3\frac{\Ksf-t}{(t+1)} \right)$ for each $t\in [2: \Ksf]$.

We now focus on one $t\in [2:\Ksf]$. We let $t_1=\Nsf(t-1)+1$ such that  the memory size is 
\begin{align}
\frac{\Nsf+t_1-1 }{\Ksf} =\frac{\Nsf+\Nsf(t-1)+1-1 }{\Ksf} =\frac{\Nsf t}{\Ksf}.
\end{align}
The achieved load is 
\begin{align}
&\frac{\Usf-t_1+1}{t_1}=\frac{\Usf-\frac{\Usf(t-1)}{\Ksf-1}}{\frac{\Usf(t-1)}{\Ksf-1}+1}\nonumber\\
&=\frac{\Usf(\Ksf-1)-\Usf(t-1)}{\Usf(t-1)+(\Ksf-1)}\nonumber\\
&=\frac{\Ksf-t}{t-1+\frac{\Ksf-1}{\Nsf}}\nonumber\\
&\leq \frac{\Ksf-t}{t-1}\nonumber\\
&\leq 3\frac{\Ksf-t}{t+1},\label{eq:integer case}
\end{align}
where~\eqref{eq:integer case} comes from $t\geq 2$.
  Hence, we prove the proof of Lemma~\ref{lem:multiplicative gap}.

\section{Proof of Corollary~\ref{cor:proof of strictly better}}
\label{sec:proof of strictly better}
Recall that for the two-user system, the achieved corner points of Scheme~A are $\left(\frac{\Nsf+t-1}{2},\frac{\Nsf-t+1}{t} \right)$, where $t\in [ \Nsf+1]$. The achieved corner points of Scheme~B are $\left(\frac{\Nsf}{2}+\frac{\Nsf t^{\prime}}{2(\Nsf+t^{\prime}-1)}, \frac{\Nsf(\Nsf-1)}{(t^{\prime}+1)(\Nsf+t^{\prime}-1)} \right)$ and $(\Nsf,0)$, where $t^{\prime} \in [0:\Nsf-1]$.

To prove Scheme~B is  better than Scheme~A for the two-user system, we prove that for each $t\in [ \Nsf]$, by memory-sharing between  $\left(\frac{\Nsf}{2}+\frac{\Nsf t^{\prime} }{2(\Nsf+t^{\prime}  -1)}, \frac{\Nsf(\Nsf-1)}{(t^{\prime}  +1)(\Nsf+t^{\prime}  -1)} \right)$ and $(\Nsf,0)$, where $t^{\prime} =t-1$, we can obtain  $\left(\frac{\Nsf+t-1}{2},\frac{\Nsf-t+1}{t} \right)$. More precisely, we let $\alpha=\frac{(\Nsf+t^{\prime} -1)(\Nsf-t^{\prime} )}{\Nsf(\Nsf-1)}$. We have 
\begin{align}
&\alpha \left(\frac{\Nsf}{2}+\frac{\Nsf t^{\prime}  }{2(\Nsf+t^{\prime}  -1)} \right)+(1-\alpha) \Nsf \nonumber\\& =\frac{(\Nsf+t^{\prime} -1)(\Nsf-t^{\prime} )}{\Nsf(\Nsf-1)}\frac{\Nsf(\Nsf+2 t^{\prime}-1)}{2(\Nsf+  t^{\prime}-1)}+\frac{t^{\prime} (t^{\prime}-1)}{\Nsf(\Nsf-1)} \Nsf \nonumber\\
&=\frac{(\Nsf+2 t^{\prime} -1)(\Nsf-t^{\prime})}{2(\Nsf-1)}+\frac{t^{\prime} (t^{\prime}-1)}{ \Nsf-1 }\nonumber\\
&=\frac{(\Nsf-1)(\Nsf-t^{\prime})}{2(\Nsf-1)}\nonumber\\
&=\frac{\Nsf-t+1}{2};\label{eq:shared memory}\\
& \alpha \frac{\Nsf(\Nsf-1)}{(t^{\prime}  +1)(\Nsf+t^{\prime}  -1)} +(1-\alpha)\times 0 = \frac{\Nsf-t^{\prime}}{t^{\prime}+1}=\frac{\Nsf-t+1}{t}.\label{eq:shared load}
\end{align}

\section{Proof of Theorem~\ref{thm:optimality K2}}
\label{sec:optimality K2}   
\subsection{Optimality in  Theorem~\ref{thm:optimality K2}}
\label{sub:opmality of Scheme B}
When  $\Nsf=2$, it can be easily checked that   the converse bound in Theorem~\ref{thm:two user converse} is a piecewise curve with corner points $\left( \frac{\Nsf}{2}, \Nsf \right)$, $\left( \frac{3\Nsf }{4}, \frac{1}{2}\right)$, and $(\Nsf, 0)$, which can be achieved by Scheme~B in~\eqref{eq:second scheme}. Hence, in the following, we focus on $\Nsf>2$.

   Recall that  $\Msf=\frac{\Nsf}{2}+y$.
For $0\leq y \leq \frac{1}{2}$, from the converse bound in~\eqref{eq:K2 converse 1} with $h=0$,    we have 
\begin{align}
\Rsf^{\star}_{\mathrm{u}} &\geq \Nsf-2y-\frac{4y+(\Nsf-1)h}{h+2} \nonumber\\& +\frac{h^2(\Nsf-1)-\Nsf(\Nsf-3)+h(\Nsf+1) }{(h+1)(h+2)}\frac{2y}{\Nsf} \nonumber\\
&= \Nsf-2y-2y-y(\Nsf-3)\nonumber\\
&=\Nsf- y(\Nsf+1).
\label{eq:K2 converse 1 first seg}
\end{align}  
In other words, when $\frac{\Nsf}{2} \leq \Msf \leq \frac{\Nsf+1}{2}$, the converse bound on $\Rsf^{\star}_{\mathrm{u}}$ in~\eqref{eq:K2 converse 1 first seg} is a straight line  between $\left(\frac{\Nsf}{2}, \Nsf \right)$ and $\left(\frac{\Nsf+1}{2}, \frac{\Nsf-1}{2} \right)$. In addition, Scheme~B in~\eqref{eq:second scheme} achieves $\left(\frac{\Nsf}{2}, \Nsf \right)$ with $t^{\prime}=0$, and   $\left(\frac{\Nsf+1}{2}, \frac{\Nsf-1}{2} \right)$ with $ t^{\prime}=1$.
Hence, we prove Scheme~B is optimal under the constraint of uncoded cache placement when $\frac{\Nsf}{2} \leq \Msf \leq \frac{\Nsf+1}{2}$.

%We then focus on the converse bound in~\eqref{eq:K2 converse 2} for $\frac{2\Nsf}{3} \leq \Msf \leq \frac{3\Nsf}{4}$,
For $\frac{2\Nsf}{3} \leq \Msf \leq \frac{3\Nsf}{4}$ (i.e., $\frac{\Nsf}{6}\leq y \leq \frac{\Nsf}{4}$), from   the converse bound in~\eqref{eq:K2 converse 2}
\begin{align}
\Rsf^{\star}_{\mathrm{u}} \geq 2-\frac{6y}{\Nsf}=5-\frac{6\Msf}{\Nsf}.\label{eq:K2 converse 1 sec seg}
\end{align}
By noticing that  $\frac{\Nsf(3\Nsf-5)}{2(2\Nsf-3)} \geq \frac{2 \Nsf}{3}$ when $\Nsf \geq 3$,
from~\eqref{eq:K2 converse 1 sec seg}, it can be seen that when $\Msf=\frac{\Nsf(3\Nsf-5)}{2(2\Nsf-3)}$, $\Rsf^{\star}_{\mathrm{u}} \geq \frac{\Nsf}{2\Nsf-3}$, coinciding with Scheme~B in~\eqref{eq:second scheme}  with $t^{\prime}= \Nsf-2$. When $\Msf= \frac{3\Nsf}{4}$, $\Rsf^{\star}_{\mathrm{u}} \geq \frac{1}{2}$, coinciding with Scheme~B in~\eqref{eq:second scheme}  with $t^{\prime}= \Nsf-1$.
Hence, we prove that Scheme~B is optimal under the constraint of uncoded cache placement when $\frac{\Nsf(3\Nsf-5)}{2(2\Nsf-3)} \leq \Msf \leq \frac{3\Nsf}{4}$.

Finally, for $\frac{3\Nsf}{4} \leq  \Msf \leq\Nsf$ (i.e., $\frac{\Nsf}{4} \leq  y \leq \frac{\Nsf}{2}$),
from the converse bound in~\eqref{eq:K2 converse 3}, we have  
\begin{align}
\Rsf^{\star}_{\mathrm{u}} \geq 1-\frac{2y}{\Nsf}=2-\frac{2\Msf}{\Nsf}.\label{eq:K2 converse 1 thir seg}
\end{align} 
From~\eqref{eq:K2 converse 1 thir seg}, it can be seen that when $\Msf=\frac{3\Nsf}{4}$, $\Rsf^{\star}_{\mathrm{u}} \geq \frac{1}{2 }$, coinciding with Scheme~B in~\eqref{eq:second scheme}  with $t^{\prime}= \Nsf-1$. When $\Msf= \Nsf$, $\Rsf^{\star}_{\mathrm{u}} \geq 0$, which can be also achieved by Scheme~B.
Hence, we prove that Scheme~B is optimal under the constraint of uncoded cache placement when  $\frac{3\Nsf}{4} \leq  \Msf \leq\Nsf$.

% In conclusion, we prove  the first part of Theorem~\ref{thm:optimality K2}.

\subsection{Order optimality in  Theorem~\ref{thm:optimality K2}}
\label{sub:order opmality of Scheme B }
From Theorem~\ref{thm:two user converse}, we can compute that the proposed converse bound is a piecewise curve with the corner points  
\begin{align}
&\left( \frac{\Nsf}{2}+\frac{\Nsf h^{\prime}}{2(\Nsf+2h^{\prime}-2)}, \frac{(h^{\prime}-1)(\Nsf+h^{\prime})+(\Nsf-1)\Nsf }{ (h^{\prime}+1)(\Nsf +2h^{\prime}-2)}  \right), \nonumber\\&  \forall h^{\prime}\in [0:\Nsf-2],\label{eq:corner points converse K 2}
\end{align}
$\left(\frac{3\Nsf}{4}, \frac{1}{2} \right)$, and $(\Nsf,0)$.\footnote{\label{foot:intersection two user}The first corner point in~\eqref{eq:corner points converse K 2} is $\left( \frac{\Nsf}{2}, \Nsf \right)$ with $h^{\prime}=0$, and the last corner point is $ (\Nsf,0)$. For each $h^{\prime} \in [\Nsf-3]$, we obtain the corner point in~\eqref{eq:corner points converse K 2} by taking the intersection between the converse bounds in~\eqref{eq:K2 converse 1} with $h=h^{\prime}-1$ and $h=h^{\prime}$. The corner point in~\eqref{eq:corner points converse K 2} with $h^{\prime}=\Nsf-2$, is obtained by taking the intersection between   the converse bounds in~\eqref{eq:K2 converse 1} with $h=\Nsf-3$ and the converse bound in~\eqref{eq:K2 converse 2}. The corner point  $\left(\frac{3\Nsf}{4}, \frac{1}{2} \right)$ is obtained by taking the intersection between the converse bounds in~\eqref{eq:K2 converse 2} and~\eqref{eq:K2 converse 3}. }
 Note that the proposed converse bound is
a piecewise linear curve with the above corner points, and that the straight line in
the memory-load tradeoff between two achievable points is also achievable by memory-sharing.  Hence, in the following, we focus on each corner point of the converse bound,
  and characterize the multiplicative gap between Scheme~B and the converse bound.

Note that in~\eqref{eq:corner points converse K 2}, when $h^{\prime}=0$, we have $\left( \frac{\Nsf}{2}, \Nsf \right)$; when $h^{\prime}=1$, we have $\left( \frac{\Nsf+1}{2}, \frac{\Nsf-1}{2} \right)$; when $h^{\prime}=\Nsf-2$, we have $\left(\frac{2\Nsf}{3}, 1 \right)$. 
In addition, in Appendix~\ref{sub:opmality of Scheme B}, we proved the optimality of Scheme~B under the constraint of uncoded cache placement when $\Msf \leq \frac{\Nsf+1}{2}$ or when $\Msf \geq \frac{3\Nsf}{4}$. Hence, in the following, we only need to compare Scheme~B and 
the corner points in~\eqref{eq:corner points converse K 2} where $h^{\prime} \in [2: \Nsf-2]$ and $\Nsf \geq 4$.

In Corollary~\ref{cor:proof of strictly better}, we show that Scheme~B is   better than Scheme~A. We will prove the multiplicative gap between Scheme~A and
the corner points in~\eqref{eq:corner points converse K 2} where $h^{\prime} \in [2: \Nsf-2]$ and $\Nsf \geq 4$, is no more than $3$. %Thus we can prove Theorem~\ref{thm:order optimality K2}.

Recall that the achieved points of Scheme~A for the two-user system   are 
\begin{align}
\left( \frac{\Nsf+t-1}{2},\frac{\Nsf-t+1}{t} \right), \forall t \in [\Nsf+1].\label{eq:recall Scheme A case 2}
\end{align}
We  want to interpolate the achieved points of  Scheme~A to match the converse bound   
at the memory size $\Msf= \frac{\Nsf}{2}+\frac{\Nsf h^{\prime}}{2(\Nsf+2 h^{\prime}-2)}$ where $h^{\prime}\in [2:\Nsf-2]$. 
By computing
\begin{align}
\frac{\Nsf+t-1}{2}=\frac{\Nsf}{2}+\frac{\Nsf h^{\prime}}{2(\Nsf+2 h^{\prime}-2)}\nonumber\\
\Longleftrightarrow t=\frac{\Nsf h^{\prime}}{ \Nsf+ 2 h^{\prime} -2}+1,
\end{align}
and observing $\frac{\Nsf-t+1}{t}$ is non-increasing with $t$, it can be seen that the achieved load of Scheme~A at $\Msf= \frac{\Nsf}{2}+\frac{\Nsf h^{\prime}}{2(\Nsf+2 h^{\prime} -2)}$  is lower than
\begin{align}
\Rsf^{\prime}=\frac{\Nsf-\frac{\Nsf h^{\prime}}{ \Nsf+2 h^{\prime} -2} +1}{ \frac{\Nsf h}{ \Nsf+2 h^{\prime} -2}}=\frac{\Nsf^2+(\Nsf+2)(h^{\prime}-1)}{\Nsf h^{\prime}}. 
\end{align}

By comparing  $\Rsf^{\prime}$ and $\frac{(h^{\prime}-1)(\Nsf+h^{\prime})+(\Nsf-1)\Nsf }{ (h^{\prime}+1)(\Nsf +2 h^{\prime}-2)} $, we have 
\begin{align}
&\frac{\Rsf^{\prime}}{\frac{(h^{\prime}-1)(\Nsf+h^{\prime})+(\Nsf-1)\Nsf }{ (h^{\prime}+1)(\Nsf +2 h^{\prime}-2)} } \nonumber\\& = \frac{\big(\Nsf^2+(\Nsf+2)(h^{\prime}-1)\big)(h^{\prime}+1)(\Nsf+2 h^{\prime}-2) }{\Nsf h^{\prime} \big( (h^{\prime}-1)(\Nsf+h^{\prime})+(\Nsf-1)\Nsf \big) }. \label{eq:multiplicative gap two user}
\end{align}

In addition, we compute 
\begin{align}
&3 \Nsf h^{\prime} \big( (h^{\prime}-1)(\Nsf+h^{\prime})+(\Nsf-1)\Nsf \big) \nonumber\\& -  \big(\Nsf^2+(\Nsf+2)(h^{\prime}-1)\big)(h^{\prime}+1)(\Nsf+2 h^{\prime}-2) \nonumber\\
&= 2\Nsf^3 h^{\prime}- \Nsf^3 - 6 \Nsf^2 h^{\prime} -3\Nsf h^{{\prime}^2} +(\Nsf-4) h^{{\prime}^3} \nonumber\\& + 3 \Nsf^2+2\Nsf h^{\prime}+4 h^{\prime}(h^{\prime}+1)-4. \label{eq:times by 3}
\end{align} 

Now we want to prove the RHS of~\eqref{eq:times by 3} is larger than $0$ for $\Nsf \geq 4$ and $h^{\prime} \in [2:\Nsf-2]$.
More precisely, when $\Nsf=4$ and $h^{\prime}=2$, we can compute the RHS of~\eqref{eq:times by 3} is equal to $36$; when $\Nsf=5$ and $h^{\prime}=2$,    the RHS of~\eqref{eq:times by 3} is equal to $138$;    when $\Nsf=5$ and $h^{\prime}=3$,    the RHS of~\eqref{eq:times by 3} is equal to $216$. Now we only need to consider $\Nsf \geq 6$ and $h^{\prime} \in [2:\Nsf-2]$.

 When $\Nsf \geq 6$ and $h^{\prime}\in [2:\Nsf-2]$, we have 
\begin{align}
 &2\Nsf^3 h^{\prime} - \Nsf^3 - 6 \Nsf^2 h^{\prime}-3\Nsf h^{{\prime}^2} +(\Nsf-4) h^{{\prime}^3}+ 3 \Nsf^2 \nonumber\\& +2\Nsf h^{\prime}+4 h^{\prime}(h^{\prime}+1)-4 \nonumber\\ &>   2\Nsf^3 h^{\prime} - \Nsf^3 - 6 \Nsf^2 h^{\prime} -3\Nsf h^{{\prime}^2} \nonumber\\
 &=(\Nsf^3 h^{\prime} -6 \Nsf^2 h^{\prime}  )+(0.5 \Nsf^3 h^{\prime} -3\Nsf h^{{\prime}^2}  )+(0.5  \Nsf^3 h^{\prime} -  \Nsf^3)\nonumber\\
 &\geq 0. \label{eq:2 user geq 0}
\end{align}

Hence, we prove 
\begin{align}
&3 \Nsf h^{\prime} \big( (h^{\prime}-1)(\Nsf+h^{\prime})+(\Nsf-1)\Nsf \big) \nonumber\\& -  \big(\Nsf^2+(\Nsf+2)(h^{\prime}-1)\big)(h^{\prime}+1)(\Nsf+2 h^{\prime}-2) >0.\label{eq:times by 3 >0}
\end{align}
By taking~\eqref{eq:times by 3 >0} into~\eqref{eq:multiplicative gap two user}, we prove that the multiplicative gap between Scheme~A and the corner points in~\eqref{eq:corner points converse K 2} where $h^{\prime} \in [2: \Nsf-2]$ and $\Nsf \geq 4$, is no more than $3$. 

In conclusion, we prove that Scheme~B is order optimal under the constraint of uncoded cache placement to within a factor of $3$.

\section{Proof of Theorem~\ref{thm:order optimality K users}}
  \label{sec:order optimality K users} 
In this proof, for the achievability,   we consider the load in Lemma~\ref{lem:upper bound}, which is an upper bound of the achieved load of Scheme A.
  
We first focus on the case where $\Nsf \leq 6 \Ksf$, and compare Scheme~A with   
the   shared-link caching converse bound   under the constraint of uncoded cache placement (without privacy) in~\cite{indexcodingcaching2020}. 
Recall that when $\Msf \in \left[\frac{\Nsf}{\Ksf}, \Nsf \right]$,  the converse bound in~\cite{indexcodingcaching2020} is a piecewise curve with corner points $\left(\frac{\Nsf t}{\Ksf}, \frac{\Ksf-t}{t+1} \right)$, where $t\in [\Ksf]$.
It was proved in Appendix~\ref{sub:N>K} that Scheme~A can achieve the corner points $\left(\frac{\Nsf t}{\Ksf}, 3\frac{\Ksf-t}{t+1}  \right) $, where $t \in [2:\Ksf]$.
In addition, when $\Msf=\frac{\Nsf}{\Ksf}$, the converse bound in~\cite{indexcodingcaching2020}  is $\Rsf^{\star}_{\mathrm{u}} \geq \frac{\Ksf-1}{2}$, while the achieved load of Scheme~A is 
$$
\Nsf \leq 6\Ksf \leq 9 (\Ksf-1), \ \text{ when }\Ksf \geq 3.
$$
Hence, the multiplicative gap between Scheme~A and the converse bound in~\cite{indexcodingcaching2020} at $\Msf=\frac{\Nsf}{\Ksf}$ is no more than $18$.  So we prove that  $\Nsf \leq 6 \Ksf$,  Scheme~A
is order optimal  under the constraint of uncoded cache placement within a factor of $18$.

In the rest of the proof, we focus on the case where $\Nsf >6 \Ksf$.  It was proved in Theorem~\ref{thm:OrderOptimalityA} that when $\Nsf\geq \Ksf$  and $\Msf \geq  \frac{2\Nsf}{\Ksf} $, Scheme~A is order optimal to  within a factor of $6$. Hence, in the following we consider $\frac{\Nsf}{\Ksf} \leq \Msf \leq  \frac{2\Nsf}{\Ksf}$, which is then divided into
  three memory size regimes, and prove the order optimality of Scheme~A separately, 
    \begin{subequations}
\begin{align}
&\text{Regime }1: \frac{\Nsf}{\Ksf} \leq  \Msf \leq \frac{\Nsf}{\Ksf} + \frac{ \Nsf h_1 }{2(\Nsf+\Ksf h_1-\Ksf)}, \nonumber\\& \text{ where } h_1:=\left\lfloor \frac{4 (\Ksf-2)(\Nsf-\Ksf) }{\Ksf(\Nsf-4\Ksf+8)} \right\rfloor;\label{eq:k user memory regime 1}\\
& \text{Regime }2: \frac{\Nsf}{\Ksf} + \frac{ \Nsf h_1 }{2(\Nsf+\Ksf h_1-\Ksf)}  \leq  \Msf \leq \nonumber\\& \frac{\Nsf}{\Ksf} + \frac{ \Nsf h_2 }{2(\Nsf+\Ksf h_2-\Ksf)},  \text{ where } h_2:=\left\lfloor \frac{2\Nsf}{\Ksf}-2 \right\rfloor;\label{eq:k user memory regime 2}\\
&\text{Regime }3:\frac{\Nsf}{\Ksf} + \frac{ \Nsf h_2 }{2(\Nsf+\Ksf h_2-\Ksf)}  \leq  \Msf \leq  \frac{2\Nsf}{\Ksf}.\label{eq:k user memory regime 3}
\end{align}
  \end{subequations}
Note that when $\Nsf >6\Ksf$, we have $h_1:=\left\lfloor \frac{4 (\Ksf-2)(\Nsf-\Ksf) }{\Ksf(\Nsf-4\Ksf+8)} \right\rfloor<10$ and $ h_2:=\left\lfloor \frac{2\Nsf}{\Ksf}-2 \right\rfloor \geq 10$. Thus we have  $h_1<h_2$. 
In addition, we have 
\begin{align}
\frac{\Nsf}{\Ksf} + \frac{ \Nsf h_2 }{2(\Nsf+\Ksf h_2-\Ksf)} &\leq \frac{\Nsf}{\Ksf} + \frac{ \Nsf\frac{2\Nsf}{\Ksf}-2}{2\left(\Nsf+\Ksf \frac{2\Nsf}{\Ksf}-2\Ksf -\Ksf \right)} \nonumber\\ & =\frac{4\Nsf}{3\Ksf}.
\end{align}
Hence, the above memory regime division is possible.

 From the converse bound in~\eqref{eq:K user converse first segment}, for each $h\in \left[0:  \left\lfloor 2\Nsf/\Ksf -3 \right\rfloor \right]$ we have, 
  %for $  \frac{\Nsf\Ksf  h}{4(\Nsf+\Ksf h -\Ksf)} \leq y \leq \frac{ \Nsf\Ksf(h+1) }{4(\Nsf+\Ksf h)}$ where  $h\in \left[0:  \left\lfloor 2\Nsf/\Ksf -3 \right\rfloor \right]$. 
\begin{align}
&\Rsf^{\star}_{\mathrm{u,c}} \geq \frac{\left\lfloor \Ksf/2 \right\rfloor }{ \left\lceil\Ksf/2 \right\rceil  }\frac{\left\lfloor 2\Nsf/\Ksf \right\rfloor  }{  2\Nsf/\Ksf }  \left\{ \Nsf-2y-\frac{8y +h(2\Nsf-\Ksf)}{2h+4} \right. \nonumber\\& \left. +\frac{h^2 \Ksf(2\Nsf-\Ksf)-2\Nsf(2\Nsf-3\Ksf)+h\Ksf(\Ksf+2\Nsf) }{(h+1)(h+2)\Ksf\Nsf}y \right\}\nonumber\\
&\geq \frac{6}{13}   \left\{ \Nsf-2y-\frac{8y +h(2\Nsf-\Ksf)}{2h+4} \right. \nonumber\\& \left. +\frac{h^2 \Ksf(2\Nsf-\Ksf)-2\Nsf(2\Nsf-3\Ksf)+h\Ksf(\Ksf+2\Nsf) }{(h+1)(h+2)\Ksf\Nsf}y \right\},
\label{eq:K user converse first segment 9/4}
\end{align}
where~\eqref{eq:K user converse first segment 9/4} follows since $\Ksf \geq 3$ and $\Nsf >6\Ksf$.
 
 In Regimes $1$ and $2$, we will use~\eqref{eq:K user converse first segment 9/4} as the converse bound. In Regime $3$, we use the   shared-link caching converse bound   under the constraint of uncoded cache placement  in~\cite{indexcodingcaching2020}.

 \subsection{Regime $1$}
 \label{sub:regime 1}
  It can be computed that the converse bound in~\eqref{eq:K user converse first segment 9/4} for  $\frac{\Nsf}{\Ksf} \leq  \Msf \leq \frac{\Nsf}{\Ksf} + \frac{ \Nsf h_1 }{2(\Nsf+\Ksf h_1-\Ksf)}$ is a piecewise curve with the corner points 
 \begin{align}
 &\left(\frac{\Nsf}{\Ksf}+\frac{\Nsf h^{\prime}}{2(\Nsf+\Ksf h^{\prime}-\Ksf)},\right. \nonumber\\& \left.  \frac{6}{13}  \frac{\Ksf(h^{\prime}-1)(2\Nsf+\Ksf h^{\prime})+2\Nsf(2\Nsf-\Ksf)}{4(h^{\prime}+1)(\Nsf+\Ksf h^{\prime} - \Ksf)} \right), \ \forall h^{\prime}\in [0: h_1],\label{eq:converse corner points regime 1}
 \end{align}
 where $h^{\prime}=0$  represents the first corner point where $\Msf=\Nsf/2$,   and each corner point in~\eqref{eq:converse corner points regime 1} with $h^{\prime}$ is obtained by taking the intersection of the converse bounds in~\eqref{eq:K user converse first segment 9/4} between $h=h^{\prime}-1$ and $h=h^{\prime}$.
 
 For the achievability, we take the memory-sharing between $\left( \frac{\Nsf}{\Ksf}, \Nsf \right)$ and $\left(\frac{\Nsf+ t_3-1}{\Ksf}, \frac{\Usf-t_3+1}{t_3}  \right)$, where $t_3= 2\Ksf-3$. 
 Notice  that 
 \begin{align}
& \frac{\Nsf+ t_3-1}{\Ksf}=\frac{\Nsf+2\Ksf-4}{\Ksf}=\frac{\Nsf}{\Ksf}+\frac{2\Ksf-4}{\Ksf}.\label{eq:t3 memory} 
\end{align}  
In addition, we have 
    \begin{subequations}
\begin{align} 
&\frac{\Nsf}{\Ksf}+\frac{\Nsf h_1}{2(\Nsf+\Ksf h_1-\Ksf)}=\frac{\Nsf}{\Ksf}+\frac{\Nsf h_1}{2(\Nsf+\Ksf h_1-\Ksf)} \\
&\leq \frac{\Nsf}{\Ksf}+\frac{\Nsf \frac{4 (\Ksf-2)(\Nsf-\Ksf) }{\Ksf(\Nsf-4\Ksf+8)}}{2(\Nsf+\Ksf \frac{4 (\Ksf-2)(\Nsf-\Ksf) }{\Ksf(\Nsf-4\Ksf+8)}-\Ksf)}\label{eq:remove floor h1}\\
&=\frac{\Nsf}{\Ksf}+\frac{4\Nsf  (\Ksf-2)(\Nsf-\Ksf)  }{2\big((\Nsf-\Ksf)\Ksf(\Nsf-4\Ksf+8) + 4\Ksf   (\Ksf-2)(\Nsf-\Ksf)  \big)} \\
&=\frac{\Nsf}{\Ksf}+\frac{4\Nsf  (\Ksf-2)(\Nsf-\Ksf)}{2\Ksf \Nsf ( \Nsf-\Ksf)} \\
&=\frac{\Nsf}{\Ksf} +\frac{2\Ksf-4}{\Ksf},\label{eq:remove floor h1 memory}
\end{align}   
   \end{subequations}
   where~\eqref{eq:remove floor h1} comes from $\frac{\Nsf h_1}{2(\Nsf+\Ksf h_1-\Ksf)}$ is  increasing with $h_1$ and $h_1 \leq \frac{4 (\Ksf-2)(\Nsf-\Ksf) }{\Ksf(\Nsf-4\Ksf+8)} $.
   From~\eqref{eq:t3 memory} and~\eqref{eq:remove floor h1 memory}, we can see that this memory-sharing can cover all memory sizes in regime $1$.
   
   When $h^{\prime}=0$, we have the corner point in~\eqref{eq:converse corner points regime 1} is $\left(\frac{\Nsf}{2},\frac{ 6\Nsf}{13} \right)$, while Scheme~A achieves  $\left(\frac{\Nsf}{2}, \Nsf\right)$. Hence, the multiplicative gap between Scheme~A and the converse is $\frac{13}{6}$.
   
   For each $h^{\prime}\in [ h_1]$, we now interpolate  Scheme~A between $(\Msf_1,\Rsf_1)=\left( \frac{\Nsf}{\Ksf}, \Nsf \right)$ and $(\Msf_2,\Rsf_2)=\left(\frac{\Nsf+ t_3-1}{\Ksf}, \frac{\Usf-t_3+1}{t_3}  \right)$ to match the corner point in the converse bound\\ $(\Msf_3,\Rsf_3)= \left(\frac{\Nsf}{\Ksf}+\frac{\Nsf h^{\prime}}{2(\Nsf+\Ksf h^{\prime}-\Ksf)}, \frac{6}{13}   \frac{\Ksf(h^{\prime}-1)(2\Nsf+\Ksf h^{\prime})+2\Nsf(2\Nsf-\Ksf)}{4(h^{\prime}+1)(\Nsf+\Ksf h^{\prime} - \Ksf)} \right)$. More precisely,  by memory-sharing between $(\Msf_1,\Rsf_1)$ and $(\Msf_2,\Rsf_2)$ with coefficient 
\begin{align}
  \alpha=\frac{\Msf_2-\Msf_3}{\Msf_2-\Msf}= \frac{\Nsf(4\Ksf -h^{\prime} \Ksf-8)+ 4 \Ksf (h^{\prime}-1)(\Ksf-2)}{4(\Ksf-2)(\Nsf+h^{\prime}\Ksf-\Ksf)} \label{eq:alpha regime 1}
\end{align}
   such that $\alpha \Msf_1+ (1-\alpha) \Msf_2=\Msf_3$,
   we get  at $\Msf_3$ Scheme~A can achieve,
\begin{align}
\Rsf^{\prime}&=\alpha \Rsf_1+ (1-\alpha) \Rsf_2 \nonumber\\& =\Nsf\frac{-12 \Nsf +8\Ksf^2(h^{\prime}-1)+\Ksf\big( \Nsf (8-h^{\prime})-14 h^{\prime}+12 \big)}{4(2\Ksf-3)(\Nsf+h^{\prime}\Ksf-\Ksf)}.
\label{eq:regime 1 achieved load}
\end{align}
 
 In the following, we compare $\Rsf^{\prime}$ and $\Rsf_3$ to obtain
 \begin{align}
 \frac{\Rsf^{\prime}}{\Rsf_3}&=   \frac{13\Nsf(h^{\prime}+1)}{6(2\Ksf-3)\big(\Ksf(h^{\prime}-1)(2\Nsf+\Ksf h^{\prime})+2\Nsf(2\Nsf-\Ksf) \big)} \nonumber\\ & \left( -12 \Nsf +8\Ksf^2(h^{\prime}-1)+\Ksf\big( \Nsf (8-h^{\prime})-14 h^{\prime}+12 \big) \right).
 \label{eq:regime 1 gap}
\end{align} 
 
Finally, we will prove  
\begin{align}
\frac{6 \Rsf^{\prime}}{13 \Rsf_3}& =\frac{\Nsf(h^{\prime}+1)}{(2\Ksf-3)\big(\Ksf(h^{\prime}-1)(2\Nsf+\Ksf h^{\prime})+2\Nsf(2\Nsf-\Ksf) \big)} \nonumber\\& \left( -12 \Nsf +8\Ksf^2(h^{\prime}-1)+\Ksf\big( \Nsf (8-h^{\prime})-14 h^{\prime}+12 \big) \right) \nonumber\\& < 8.\label{eq:regime 1 less than 8}
\end{align}

We can compute that
    \begin{subequations}
\begin{align}
&8(2\Ksf\negmedspace -\negmedspace 3)\big(\Ksf(h^{\prime}-1)(2\Nsf\negmedspace +\negmedspace \Ksf h^{\prime})\negmedspace + \negmedspace 2\Nsf(2\Nsf-\Ksf) \big)\negmedspace   -\negmedspace  \Nsf(h^{\prime}+1)\nonumber\\&  \left( -12 \Nsf \negmedspace + \negmedspace 8\Ksf^2(h^{\prime}-1)\negmedspace + \negmedspace \Ksf\big( \Nsf (8-h^{\prime})-14 h^{\prime}+12 \big)   \right)\nonumber\\
%&\geq 8(2\Ksf-3)\big( 4\Nsf^2+2\Ksf \Nsf (h^{\prime}-2)\big)-\Nsf(h^{\prime}+1)\left\{ -12 \Nsf +8\Ksf^2(h^{\prime}-1)+\Ksf\big( \Nsf (8-h^{\prime})-14 h^{\prime}+12 \big) \right\}\\
&\geq 8(2\Ksf\negmedspace -\negmedspace 3)\big(\Ksf(h^{\prime}-1)(2\Nsf\negmedspace +\negmedspace \Ksf h^{\prime})\negmedspace + \negmedspace 2\Nsf(2\Nsf-\Ksf) \big) \nonumber\\&  -\Nsf(h^{\prime}+1)\left( -12 \Nsf +8\Ksf^2(h^{\prime}-1)+\Ksf  \Nsf (8-h^{\prime})  \right) \label{eq:from h >=1 regime 1}\\
&=\big( 32 (2\Ksf-3) +12(h^{\prime}+1)-\Ksf(8-h^{\prime})(h^{\prime}+1)\big) \Nsf^2 \nonumber\\& - \big(8\Ksf (h^{\prime}+1)(h^{\prime}-1)-16(2\Ksf-3)(h^{\prime}-2) \big)\Ksf \Nsf \nonumber \\ & +8 (2\Ksf-3) \Ksf^2 h^{\prime}(h^{\prime}-1)\label{eq:from k>=3 regime 1}\\
&\geq \big( 32 (2\Ksf-3) +12(h^{\prime}+1)-\Ksf(8-h^{\prime})(h^{\prime}+1)\big) \Nsf^2 \nonumber\\& -  \big(8  (h^{\prime}+1)(h^{\prime}-1)-16 (h^{\prime}-2) \big)\Ksf^2 \Nsf \nonumber \\ & +8 (2\Ksf-3) \Ksf^2 h^{\prime}(h^{\prime}-1), \label{eq:regime to final check}
\end{align}
    \end{subequations}
where~\eqref{eq:from h >=1 regime 1} comes from $h^{\prime} \geq 1$ and~\eqref{eq:from k>=3 regime 1} comes from $\Ksf \geq 3$.   

Recall that $\Nsf > 6\Ksf$, and  that $h^{\prime} \leq h_1=\left\lfloor \frac{4 (\Ksf-2)(\Nsf-\Ksf) }{\Ksf(\Nsf-4\Ksf+8)} \right\rfloor <10$. 

We first focus on $h^{\prime}=9$.  If  $h^{\prime}=9$, it can be seen that $6\Ksf<\Nsf < \frac{32}{5} \Ksf$. Hence, we have 
\begin{align}
&8 (2\Ksf-3) \Ksf^2 h^{\prime}(h^{\prime}-1)> \frac{5}{4}(2\Ksf-3) \Ksf\Nsf h^{\prime}(h^{\prime}-1) \nonumber\\& \geq\frac{5}{4}  \Ksf^2 \Nsf h^{\prime}(h^{\prime}-1)=90 \Ksf^2 \Nsf .\label{eq:h=9}
\end{align}
We take $h^{\prime}=9$ and~\eqref{eq:h=9} into~\eqref{eq:regime to final check} to obtain
   \begin{subequations}
\begin{align}
&8(2\Ksf\negmedspace -\negmedspace 3)\big(\Ksf(h^{\prime}-1)(2\Nsf\negmedspace +\negmedspace \Ksf h^{\prime})\negmedspace + \negmedspace 2\Nsf(2\Nsf-\Ksf) \big)\negmedspace   -\negmedspace  \Nsf(h^{\prime}+1) \nonumber\\&  \left( -12 \Nsf \negmedspace + \negmedspace 8\Ksf^2(h^{\prime}-1)\negmedspace + \negmedspace \Ksf\big( \Nsf (8-h^{\prime})-14 h^{\prime}+12 \big)   \right) \nonumber\\
&> (74 \Ksf +24) \Nsf^2-( 640-112-90) \Ksf^2 \Nsf \\
&>74 \Ksf\Nsf^2 - 438 \Ksf^2 \Nsf \\
&>0,\label{eq:regime 1 h9} 
\end{align}
    \end{subequations}
where~\eqref{eq:regime 1 h9} comes from $\Nsf >6\Ksf$. %From~\eqref{eq:regime 1 h9}, we   prove that when $h^{\prime}=9$, we have  $\frac{6 \Rsf^{\prime}}{13 \Rsf_3} < 8$, and thus $\frac{  \Rsf^{\prime}}{  \Rsf_3}<18$.
 
 We then focus on $h^{\prime}= 8$. If $\Ksf=3$, from~\eqref{eq:regime to final check}, we have the RHS of~\eqref{eq:regime to final check} becomes $204\Nsf(\Nsf-18)+12096$, which is larger than $0$ since $\Nsf>6\Ksf \geq 18$.
 Now we consider $\Ksf \geq 4$.
 From~\eqref{eq:from k>=3 regime 1}, we have 
  \begin{subequations}
 \begin{align}
 &8(2\Ksf\negmedspace -\negmedspace 3)\big(\Ksf(h^{\prime}-1)(2\Nsf\negmedspace +\negmedspace \Ksf h^{\prime})\negmedspace + \negmedspace 2\Nsf(2\Nsf-\Ksf) \big)\negmedspace -\negmedspace  \Nsf(h^{\prime}+1)\nonumber\\&  \left( -12 \Nsf \negmedspace + \negmedspace 8\Ksf^2(h^{\prime}-1)\negmedspace + \negmedspace \Ksf\big( \Nsf (8-h^{\prime})-14 h^{\prime}+12 \big)   \right) \nonumber\\
  %&\geq \big( 32 (2\Ksf-3) +12(h^{\prime}+1)-\Ksf(8-h^{\prime})(h^{\prime}+1)\big) \Nsf^2- \big(8\Ksf (h^{\prime}+1)(h^{\prime}-1)-16(2\Ksf-3)(h^{\prime}-2) \big)\Ksf \Nsf \nonumber \\ & +8 (2\Ksf-3) \Ksf^2 h^{\prime}(h^{\prime}-1)\\
&>\big( 32 (2\Ksf-3) +12(h^{\prime}+1)-\Ksf(8-h^{\prime})(h^{\prime}+1)\big) \Nsf^2 \nonumber\\& - \big(8\Ksf (h^{\prime}+1)(h^{\prime}-1)-16(2\Ksf-3)(h^{\prime}-2) \big)\Ksf \Nsf  \\
&\geq   \big( 32 (2\Ksf-3) +12(h^{\prime}+1)-\Ksf(8-h^{\prime})(h^{\prime}+1)\big) \Nsf^2 \nonumber\\& - \big(8\Ksf (h^{\prime}+1)(h^{\prime}-1)-20\Ksf (h^{\prime}-2) \big)\Ksf \Nsf  \label{eq:from k>=4}\\
&= \big((56+h^{{\prime}^2}-7 h^{\prime})  \Ksf+12 h^{\prime}- 84 \big) \Nsf^2 \nonumber\\& -(32 + 8 h^{{\prime}^2} - 20 h^{\prime})\Ksf^2 \Nsf  \\
&\geq  (56+h^{{\prime}^2}-7 h^{\prime})  \Ksf\Nsf^2-(32 + 8 h^{{\prime}^2} - 20 h^{\prime})\Ksf^2 \Nsf \\
& > 6 (56+h^{{\prime}^2}-7 h^{\prime})  \Ksf^2\Nsf -(32 + 8 h^{{\prime}^2} - 20 h^{\prime})\Ksf^2 \Nsf  \label{eq:h=8 N>6k}\\
&=0,
 \label{eq:regime to final check h<=8}
 \end{align}
  \end{subequations}
where~\eqref{eq:from k>=4} comes from $\Ksf \geq 4$ and thus $\frac{2\Ksf-3}{\Ksf} \geq \frac{5}{4}$, and~\eqref{eq:h=8 N>6k} comes from $\Nsf >6\Ksf$.

Lastly, we consider  $h^{\prime} \in [7]$. From~\eqref{eq:regime to final check}, we have 
  \begin{subequations}
\begin{align}
 &8(2\Ksf\negmedspace -\negmedspace 3)\big(\Ksf(h^{\prime}-1)(2\Nsf\negmedspace +\negmedspace \Ksf h^{\prime})\negmedspace + \negmedspace 2\Nsf(2\Nsf-\Ksf) \big)\negmedspace -\negmedspace  \Nsf(h^{\prime}+1)\nonumber\\&  \left( -12 \Nsf \negmedspace + \negmedspace 8\Ksf^2(h^{\prime}-1)\negmedspace + \negmedspace \Ksf\big( \Nsf (8-h^{\prime})-14 h^{\prime}+12 \big)  \right) \nonumber\\
  &> \big( 32 (2\Ksf-3) +12(h^{\prime}+1)-\Ksf(8-h^{\prime})(h^{\prime}+1)\big) \Nsf^2\nonumber\\&- \big(8  (h^{\prime}+1)(h^{\prime}-1)-16 (h^{\prime}-2) \big)\Ksf^2 \Nsf\\ 
  &= \big((56+h^{{\prime}^2}-7 h^{\prime})  \Ksf+12 h^{\prime}- 84 \big) \Nsf^2\nonumber\\& -(24 + 8 h^{{\prime}^2} - 16 h^{\prime})\Ksf^2 \Nsf \\
  &\geq  (56+h^{{\prime}^2}-7 h^{\prime} +4 h^{\prime}- 28)  \Ksf  \Nsf^2 \nonumber\\& -(24 + 8 h^{{\prime}^2} - 16 h^{\prime})\Ksf^2 \Nsf \label{eq:h=7 k>3}\\
  &> 6(28+h^{{\prime}^2}-3 h^{\prime})  \Ksf^2  \Nsf-(24 + 8 h^{{\prime}^2} - 16 h^{\prime})\Ksf^2 \Nsf \label{eq:h=7 N>6k}\\
  &=(144-2 h^{{\prime}^2}-2 h^{\prime})  \Ksf^2  \Nsf \\
  &>0  \label{eq:regime to final check h<=7}
\end{align}
  \end{subequations}
  where~\eqref{eq:h=7 k>3} comes from $  h^{\prime} \leq 7$ and $\Ksf \geq 3$, which lead to $12 h^{\prime}- 84 \geq  (4 h^{\prime}- 28)\Ksf$, and
  \eqref{eq:h=7 N>6k} comes from $\Nsf >6\Ksf$, and~\eqref{eq:regime to final check h<=7} comes from $h^{\prime} \in [7]$.
  
  In conclusion, we     prove~\eqref{eq:regime 1 less than 8}. In other words, under the constraint of uncoded cache placement and user collusion, Scheme~A is order optimal to  within a factor of $\frac{13}{6}\times 8<18$ for the memory size Regime $1$.
\subsection{Regime $2$}
 \label{sub:regime 2}
Similar  to the converse bound for Regime  $1$,  it can be computed that the converse bound in~\eqref{eq:K user converse first segment 9/4} for  $\frac{\Nsf}{\Ksf} + \frac{ \Nsf h_1 }{2(\Nsf+\Ksf h_1-\Ksf)}  \leq  \Msf \leq \frac{\Nsf}{\Ksf} + \frac{ \Nsf h_2 }{2(\Nsf+\Ksf h_2-\Ksf)}$ is a piecewise curve with the corner points 
 \begin{align}
 &\left(\frac{\Nsf}{\Ksf}+\frac{\Nsf h^{\prime}}{2(\Nsf+\Ksf h^{\prime}-\Ksf)}, \right. \nonumber\\& \left. \frac{6}{13}  \frac{\Ksf(h^{\prime}-1)(2\Nsf+\Ksf h^{\prime})+2\Nsf(2\Nsf-\Ksf)}{4(h^{\prime}+1)(\Nsf+\Ksf h^{\prime} - \Ksf)} \right), \ \forall h^{\prime}\in [h_1: h_2].\label{eq:converse corner points regime 2}
 \end{align} 
 
 For the achievability, we take the memory-sharing  among the  achieved points  in~\eqref{eq:extended scheme A},   $\left(\frac{\Nsf+ t -1}{\Ksf}, \frac{\Usf-t +1}{t }  \right)$, where $t \in [\Usf+1]$.  
 We  want to interpolate the achieved points of  Scheme~A to match the converse bound   
at the memory size $\Msf=\frac{\Nsf}{\Ksf}+\frac{\Nsf h^{\prime}}{2(\Nsf+\Ksf h^{\prime}-\Ksf)}$ where $h^{\prime}\in [h_1: h_2]$. 
By computing
\begin{align}
\frac{\Nsf+t-1}{\Ksf}=\frac{\Nsf}{\Ksf}+\frac{\Nsf h^{\prime}}{2(\Nsf+\Ksf h^{\prime}-\Ksf)}\nonumber\\
\Longleftrightarrow t=\frac{\Nsf h^{\prime} \Ksf }{ 2(\Nsf+ \Ksf h^{\prime} -\Ksf)}+1,
\end{align}
and observing $\frac{\Usf-t+1}{t}$ is non-increasing with $t$, it can be seen that the achieved load of Scheme~A at $\Msf= \frac{\Nsf}{\Ksf}+\frac{\Nsf h^{\prime}}{2(\Nsf+\Ksf h^{\prime}-\Ksf)}$  is lower than
\begin{align}
\Rsf^{\prime}=\frac{\Usf-\frac{\Nsf h^{\prime} \Ksf }{ 2(\Nsf+ \Ksf h^{\prime} -\Ksf)} +1}{ \frac{\Nsf h^{\prime} \Ksf }{ 2(\Nsf+ \Ksf h^{\prime} -\Ksf)} }.\label{eq:kuser 2regime load} 
\end{align}
 By comparing $\Rsf^{\prime}$ and $\frac{6}{13}  \frac{\Ksf(h^{\prime}-1)(2\Nsf+\Ksf h^{\prime})+2\Nsf(2\Nsf-\Ksf)}{4(h^{\prime}+1)(\Nsf+\Ksf h^{\prime} - \Ksf)}$, we have~\eqref{eq:compare R and 6/13} (at the top of the next page).
 \begin{figure*}
 \begin{align}
  \frac{\Rsf^{\prime}}{\Rsf_3}=\frac{13}{6} \frac{4(\Nsf+\Ksf h^{\prime}-\Ksf)(h^{\prime}+1)\big(2 \Ksf^2 \Nsf (h^{\prime}-1)+\Ksf (2 \Nsf^2  +2 \Nsf  + 2 h^{\prime} - 3 \Nsf h^{\prime}-2 )-2\Nsf(\Nsf-1) \big)}{\Ksf \Nsf h^{\prime}\big(\Ksf(h^{\prime}-1)(2\Nsf+\Ksf h^{\prime})+2\Nsf(2\Nsf-\Ksf) \big)}.
  \label{eq:compare R and 6/13}
 \end{align}
 \end{figure*} 
 
Since $\Ksf \geq 3$, we have 
  \begin{subequations}
\begin{align}
&h^{\prime} \geq h_1=\left\lfloor \frac{4 (\Ksf-2)(\Nsf-\Ksf) }{\Ksf(\Nsf-4\Ksf+8)} \right\rfloor \geq \left\lfloor \frac{2 (\Nsf-\Ksf) }{\Nsf-4\Ksf+8} \right\rfloor >2; \label{eq:h>2}\\
&h^{\prime} \leq h_2=\left\lfloor \frac{2\Nsf}{\Ksf}-2 \right\rfloor <\frac{2\Nsf}{\Ksf}.\label{eq:h<2n} 
\end{align}
\label{eq:h range}
  \end{subequations}

  In the following, we will use~\eqref{eq:h range} and  $\Nsf >6\Ksf \geq 18$ to prove~\eqref{eq:task regime 2} (at the top of the next page).
    \begin{figure*}
 \begin{align}
   \frac{6 \Rsf^{\prime}}{13 \Rsf_3}= \frac{4(\Nsf+\Ksf h^{\prime}-\Ksf)(h^{\prime}+1)\big(2 \Ksf^2 \Nsf (h^{\prime}-1)+\Ksf (2 \Nsf^2  +2 \Nsf  + 2 h^{\prime} - 3 \Nsf h^{\prime}-2 )-2\Nsf(\Nsf-1) \big)}{\Ksf \Nsf h^{\prime}\big(\Ksf(h^{\prime}-1)(2\Nsf+\Ksf h^{\prime})+2\Nsf(2\Nsf-\Ksf) \big)} <8.\label{eq:task regime 2}
 \end{align}
 \end{figure*}  
 
 We can compute that 
   \begin{subequations}
 \begin{align}
 &8  \Ksf \Nsf h^{\prime}\big(\Ksf(h^{\prime}-1)(2\Nsf+\Ksf h^{\prime})+2\Nsf(2\Nsf-\Ksf) \big)\nonumber\\
 &-4(\Nsf+\Ksf h^{\prime}-\Ksf)(h^{\prime}+1)  \big(2 \Ksf^2 \Nsf (h^{\prime}-1) \nonumber\\& +\Ksf (2 \Nsf^2  +2 \Nsf  + 2 h^{\prime} - 3 \Nsf h^{\prime}-2 )-2\Nsf(\Nsf-1) \big) \nonumber\\
 &\geq 8  \Ksf \Nsf h^{\prime}\big(\Ksf(h^{\prime}-1)(2\Nsf+\Ksf h^{\prime})+2\Nsf(2\Nsf-\Ksf) \big)\nonumber\\
 &-4(\Nsf+\Ksf h^{\prime}-\Ksf)(h^{\prime}+1)\big(2 \Ksf^2 \Nsf (h^{\prime}-1) \nonumber\\& +\Ksf (2 \Nsf^2  +2 \Nsf  + 2 h^{\prime} - 3 \Nsf h^{\prime}-2 ) \big) \\
 &=8\Ksf(\Nsf-\Ksf)+8 \Ksf^3 \Nsf (h^{\prime} -1)+4\Ksf \Nsf^2(h^{\prime}-2) \nonumber\\& + 8\Ksf \Nsf(3 \Nsf^2 h^{\prime}- 4\Ksf \Nsf h^{\prime}-\Nsf^2) \nonumber\\
 & +4 \Ksf h^{\prime}(3 \Ksf \Nsf h^{\prime^2} -3\Ksf \Nsf-2 \Ksf h^{\prime^{2}})  +4\Ksf \Nsf h^{\prime^2} (3\Nsf-2\Ksf-2) \nonumber\\&  + 8\Ksf^2 \Nsf+ 16 \Ksf^2 \Nsf^2+8\Ksf^2 h^{\prime}+8\Ksf^2 h^{\prime^{2}}\\
 &>8\Ksf(\Nsf-\Ksf)+8 \Ksf^3 \Nsf (h^{\prime} -1)+4\Ksf \Nsf^2(h^{\prime}-2) \nonumber\\& + 8\Ksf \Nsf(3 \Nsf^2 h^{\prime}- 4\Ksf \Nsf h^{\prime}-\Nsf^2) \nonumber\\
 & +4 \Ksf h^{\prime}(3 \Ksf \Nsf h^{\prime^2} -3\Ksf \Nsf-2 \Ksf h^{\prime^{2}}) +4\Ksf \Nsf h^{\prime^2} (3\Nsf-2\Ksf-2) \\
 &>8\Ksf \Nsf(3 \Nsf^2 h^{\prime}- 4\Ksf \Nsf h^{\prime}-\Nsf^2)  \nonumber\\& +4 \Ksf h^{\prime}(3 \Ksf \Nsf h^{\prime^2} -3\Ksf \Nsf-2 \Ksf h^{\prime^{2}}) \label{eq:regime 2 remove 1}\\
 &=8\Ksf \Nsf(\Nsf^2 h^{\prime}- \Nsf^2)+8\Ksf \Nsf(2 \Nsf^2 h^{\prime} - 4\Ksf \Nsf h^{\prime}) \nonumber\\& + 4 \Ksf h^{\prime} (  \Ksf \Nsf h^{\prime^2} -3\Ksf \Nsf)+4 \Ksf h^{\prime}(2 \Ksf \Nsf h^{\prime^2}-2 \Ksf h^{\prime^{2}}) \\
 &>0,\label{eq:regime 2 final}
 \end{align}
   \end{subequations}
   where~\eqref{eq:regime 2 remove 1} and~\eqref{eq:regime 2 final} come from $\Nsf >6\Ksf$ and $h^{\prime}>2$.
   
     In conclusion, we     prove~\eqref{eq:task regime 2}. In other words, under the constraint of uncoded cache placement and user collusion, Scheme~A is order optimal to  within a factor of $\frac{13}{6}\times 8<18$ for the memory size Regime $2$.
 \subsection{Regime $3$}
 \label{sub:regime 3}
 When $\frac{\Nsf}{\Ksf} \leq \Msf \leq \frac{2\Nsf}{\Ksf}$, the converse bound in~\cite{indexcodingcaching2020} is a straight line between $\left(\frac{\Nsf}{\Ksf}, \frac{\Ksf-1}{2} \right)$ and $\left(\frac{2\Nsf}{\Ksf}, \frac{\Ksf-2}{3} \right)$, which is denoted by $\Rsf_{\text{\cite{indexcodingcaching2020}}}(\Msf)$.
 Hence, the converse bound in~\cite{indexcodingcaching2020} for Regime $3$ where $\frac{\Nsf}{\Ksf} + \frac{ \Nsf h_2 }{2(\Nsf+\Ksf h_2-\Ksf)}  \leq  \Msf \leq  \frac{2\Nsf}{\Ksf}$ is   a straight line. 
When  $\Msf =  \frac{2\Nsf}{\Ksf}$, we proved in Appendix~\ref{sub:N>K} that the multiplicative gap between Scheme~A and the converse bound in~\cite{indexcodingcaching2020} is no more than $6$. Hence, in the rest of this proof, we focus on the memory size   $\Msf=\frac{\Nsf}{\Ksf} + \frac{ \Nsf h_2 }{2(\Nsf+\Ksf h_2-\Ksf)}  \leq  \Msf \leq  \frac{2\Nsf}{\Ksf}$.

 Recall that $h_2:= \left\lfloor \frac{2\Nsf}{\Ksf}-2 \right\rfloor \leq\frac{2\Nsf}{\Ksf}-2 $, we note that 
 \begin{align}
 \frac{\Nsf}{\Ksf} + \frac{ \Nsf h_2 }{2(\Nsf+\Ksf h_2-\Ksf)}& \leq  \frac{\Nsf}{\Ksf} + \frac{ \Nsf \left(\frac{2\Nsf}{\Ksf}-2 \right) }{2\{\Nsf+\Ksf \left(\frac{2\Nsf}{\Ksf}-2 \right)-\Ksf\}} \nonumber\\& = \frac{4 \Nsf}{3 \Ksf}.
 \end{align}
 Hence, the load of the converse bound in~\cite{indexcodingcaching2020} at $\Msf=  \frac{\Nsf}{\Ksf} + \frac{ \Nsf h_2 }{2(\Nsf+\Ksf h_2-\Ksf)}$ is strictly higher than the one at $\Msf^{\prime}=\frac{4 \Nsf}{3 \Ksf}$. By computing the converse bound in~\cite{indexcodingcaching2020} at $\Msf^{\prime}=\frac{4 \Nsf}{3 \Ksf}$ is 
 \begin{align}
 \Rsf_{\text{\cite{indexcodingcaching2020}}}(\Msf^{\prime})=\frac{2}{3} \frac{\Ksf-1}{2}+\frac{1}{3}  \frac{\Ksf-2}{3}=\frac{4\Ksf -5}{9},\label{eq:R M prime}
 \end{align}
 at $\Msf=  \frac{\Nsf}{\Ksf} + \frac{ \Nsf h_2 }{2(\Nsf+\Ksf h_2-\Ksf)}$,  we have  
 \begin{align}
 \Rsf^{\star}_{\mathrm{u,c}} \geq  \Rsf_{\text{\cite{indexcodingcaching2020}}}(\Msf) >\Rsf_{\text{\cite{indexcodingcaching2020}}}(\Msf^{\prime})=\frac{4\Ksf -5}{9}.\label{eq:R M}
 \end{align}

For the achievability,  it was proved in~\eqref{eq:kuser 2regime load} that the achieved load of Scheme~A at $\Msf= \frac{\Nsf}{\Ksf}+\frac{\Nsf h_2 }{2(\Nsf+\Ksf h_2 -\Ksf)}$  is lower than
   \begin{subequations}
\begin{align}
\Rsf^{\prime}&=\frac{\Usf-\frac{\Nsf h_2  \Ksf }{ 2(\Nsf+ \Ksf h_2  -\Ksf)} +1}{ \frac{\Nsf h_2\Ksf }{ 2(\Nsf+ \Ksf h_2 -\Ksf)} } \\
&\leq  \frac{\Usf-\frac{\Nsf (2\Nsf/\Ksf-3)  \Ksf }{ 2\big(\Nsf+ \Ksf (2\Nsf/\Ksf-3)  -\Ksf \big)} +1}{ \frac{\Nsf (2\Nsf/\Ksf-3) \Ksf }{ 2\big(\Nsf+ \Ksf (2\Nsf/\Ksf-3) -\Ksf \big)} }\label{eq:kuser 3regime load 1} \\
&= \frac{(6 \Ksf-8) \Nsf^2 -(8\Ksf - 11) \Ksf \Nsf+6\Nsf   -8 \Ksf   }{2 \Nsf^2- 3\Ksf \Nsf} \label{eq:kuser 3regime load 2}, 
\end{align} 
   \end{subequations}
where~\eqref{eq:kuser 3regime load 1} comes that $\frac{\Usf-t+1}{t}$ is non-increasing with $t$, and that $h_2 \leq 2\Nsf/\Ksf-3$.
 
 Finally, we compare $\Rsf^{\prime}$ and $\frac{4\Ksf -5}{9}$ to obtain,
 \begin{align}
 \frac{\Rsf^{\prime}}{\frac{4\Ksf -5}{9}}= 9\frac{(6 \Ksf-8) \Nsf^2 -(8\Ksf - 11) \Ksf \Nsf+6\Nsf   -8 \Ksf   }{(2 \Nsf^2- 3\Ksf \Nsf)(4\Ksf -5)}.\label{eq:regime 3 gap}
 \end{align}
In addition, we compute
   \begin{subequations}
\begin{align}
&2 (2 \Nsf^2- 3\Ksf \Nsf)(4\Ksf -5) \nonumber\\& - \left( (6 \Ksf-8) \Nsf^2 -(8\Ksf - 11) \Ksf \Nsf+6\Nsf   -8 \Ksf \right) \nonumber\\
&=2 \Nsf (5 \Ksf \Nsf - 6 \Nsf  - 8\Ksf^2) +(  19 \Ksf  \Nsf-6\Nsf)+8\Ksf \\
& > 2 \Nsf (5 \Ksf \Nsf - 6 \Nsf  - 8\Ksf^2) \label{eq:regime 3 k >3 1}\\
 &\geq  2 \Nsf (3 \Ksf \Nsf    - 8\Ksf^2) \label{eq:regime 3 k >3 2}\\
 &>0, \label{eq:regime 3 n >6k} 
\end{align}  
   \end{subequations}
 where~\eqref{eq:regime 3 k >3 1} and~\eqref{eq:regime 3 k >3 2} come from $\Ksf \geq 3$,  and~\eqref{eq:regime 3 n >6k} comes from $\Nsf >6\Ksf$.
By taking~\eqref{eq:regime 3 n >6k} into~\eqref{eq:regime 3 gap}, it can be seen that the multiplicative gap between Scheme~A and the converse bound in~\cite{indexcodingcaching2020} at $\Msf=  \frac{\Nsf}{\Ksf} + \frac{ \Nsf h_2 }{2(\Nsf+\Ksf h_2-\Ksf)}$ is less than $18$.

In conclusion, we prove that  under the constraint of uncoded cache placement and user collusion, Scheme~A is order optimal to  within a factor of $ 18$ for the memory size Regime $3$.
\bibliographystyle{IEEEtran}
\bibliography{IEEEabrv,IEEEexample}

\begin{IEEEbiographynophoto}
						{Kai Wan} (S '15 -- M '18)
						received  the B.E. degree in    Optoelectronics from  Huazhong University of Science and Technology, China, in 2012, the   M.Sc. and Ph.D. degrees in Communications from Universit{\'e}  Paris-Saclay, France, in 2014 and 2018.  He is currently a post-doctoral    researcher with the Communications and Information Theory Chair   (CommIT) at Technische Universit\"at Berlin, Berlin, Germany. His   research interests include information theory, coding techniques, and   their applications on coded caching,  index coding, distributed storage,  distributed computing, wireless communications,   privacy and security. He has served as an Associate Editor of IEEE Communications Letters from Aug. 2021.
					\end{IEEEbiographynophoto}
					
					\begin{IEEEbiographynophoto}{Hua Sun} (S '12 -- M '17) received the B.E. degree in Communications Engineering from Beijing University of Posts and Telecommunications, China, in 2011, and the M.S. degree in Electrical and Computer Engineering and the Ph.D. degree in Electrical Engineering from University of California Irvine, USA, in 2013 and 2017, respectively. He is an Assistant Professor in the Department of Electrical Engineering at the University of North Texas, USA. His research interests include information theory and its applications to communications, privacy, security, and storage. 

Dr. Sun is a recipient of the NSF CAREER award in 2021, and the UNT College of Engineering Distinguished Faculty Fellowship in 2021. His co-authored papers received the IEEE Jack Keil Wolf ISIT Student Paper Award in 2016, and an IEEE GLOBECOM Best Paper Award in 2016.
					\end{IEEEbiographynophoto}
					
										\begin{IEEEbiographynophoto}{Mingyue Ji}
(S '09 -- M '15) received the B.E. in Communication Engineering from Beijing University of Posts and Telecommunications (China), in 2006, the M.Sc. degrees in Electrical Engineering from Royal Institute of Technology (Sweden) and from University of California, Santa Cruz, in 2008 and 2010, respectively, and the PhD from the Ming Hsieh Department of Electrical Engineering at University of Southern California in 2015. He subsequently was a Staff II System Design Scientist with Broadcom Corporation (Broadcom Limited) in 2015-2016. He is now an Assistant Professor of Electrical and Computer Engineering Department and an Adjunct Assistant Professor of School of Computing at the University of Utah. He received the NSF CAREER Award in 2022, the IEEE Communications Society Leonard G. Abraham Prize for the best IEEE JSAC paper in 2019, the best paper awards at 2021 IEEE Globecom conference and at 2015  IEEE ICC conference, the best student paper award at 2010 IEEE European Wireless conference and USC Annenberg Fellowship from 2010 to 2014. He has served as an Associate Editor of IEEE Transactions on Communications from 2020. He is interested the broad area of information theory, coding theory, concentration of measure and statistics with the applications of caching networks, wireless communications, distributed storage and computing systems, distributed machine learning, and (statistical) signal processing.
					\end{IEEEbiographynophoto}
					
						\begin{IEEEbiographynophoto}{Daniela Tuninetti}  (M '98 -- SM '13 -- F '21)
is currently a Professor within the Department of Electrical
and Computer Engineering at the University of Illinois at Chicago (UIC),
which she joined in 2005. Dr. Tuninetti got her Ph.D. in Electrical Engineering
in 2002 from ENST/T{\'e}l{\'e}com ParisTech (Paris, France, with work done at the
Eurecom Institute in Sophia Antipolis, France), and she was a postdoctoral
research associate at the School of Communication and Computer Science
at the Swiss Federal Institute of Technology in Lausanne (EPFL, Lausanne,
Switzerland) from 2002 to 2004. Dr. Tuninetti is a recipient of a best paper
award at the European Wireless Conference in 2002, of an NSF CAREER
award in 2007, and named University of Illinois Scholar in 2015. Dr. Tuninetti
was the editor-in-chief of the IEEE Information Theory Society Newsletter
from 2006 to 2008, an editor for IEEE COMMUNICATION LETTERS from
2006 to 2009, for IEEE TRANSACTIONS ON WIRELESS COMMUNICATIONS
from 2011 to 2014; and for IEEE TRANSACTIONS ON INFORMATION
THEORY from 2014 to 2017. She is currently a distinguished lecturer for the
Information Theory society. She is also currently an editor for IEEE Transactions on Communications.
 Dr. Tuninetti's research interests are in the
ultimate performance limits of wireless interference networks (with special
emphasis on cognition and user cooperation), coexistence between radar and
communication systems, multi-relay networks, content-type coding, cache-aided
systems and distributed private coded computing.

					\end{IEEEbiographynophoto}

										\begin{IEEEbiographynophoto}{Giuseppe Caire}
 (S '92 -- M '94 -- SM '03 -- F '05) 
was born in Torino in 1965. He received the B.Sc. in Electrical Engineering  from Politecnico di Torino in 1990, 
the M.Sc. in Electrical Engineering from Princeton University in 1992, and the Ph.D. from Politecnico di Torino in 1994. 
He has been a post-doctoral research fellow with the European Space Agency (ESTEC, Noordwijk, The Netherlands) in 1994-1995,
Assistant Professor in Telecommunications at the Politecnico di Torino, Associate Professor at the University of Parma, Italy, 
Professor with the Department of Mobile Communications at the Eurecom Institute,  Sophia-Antipolis, France,
a Professor of Electrical Engineering with the Viterbi School of Engineering, University of Southern California, Los Angeles,
and he is currently an Alexander von Humboldt Professor with the Faculty of Electrical Engineering and Computer Science at the
Technical University of Berlin, Germany.

He received the Jack Neubauer Best System Paper Award from the IEEE Vehicular Technology Society in 2003,  the
IEEE Communications Society and Information Theory Society Joint Paper Award in 2004 and in 2011, 
the Okawa Research Award in 2006,   
the Alexander von Humboldt Professorship in 2014, the Vodafone Innovation Prize in 2015, an ERC Advanced Grant in 2018, 
the Leonard G. Abraham Prize for best IEEE JSAC paper in 2019, the IEEE Communications Society Edwin Howard Armstrong Achievement Award in 2020, 
and he is a recipient of the 2021 Leibinz Prize  of the German National Science Foundation (DFG). 
Giuseppe Caire is a Fellow of IEEE since 2005.  He has served in the Board of Governors of the IEEE Information Theory Society from 2004 to 2007,
and as officer from 2008 to 2013. He was President of the IEEE Information Theory Society in 2011. 
His main research interests are in the field of communications theory, information theory, channel and source coding
with particular focus on wireless communications.   
					\end{IEEEbiographynophoto}

\end{document}